\documentclass[twocolumn]{aastex7}
\usepackage{tablefootnote}

\begin{document}

\title{The low-mass and structured stellar halo of M83 argues against a merger origin for its starburst and extended neutral hydrogen disk\footnote{This research is based in part on data collected at the Subaru Telescope, which is operated by the National Astronomical Observatory of Japan. We are honored and grateful for the opportunity of observing the Universe from Maunakea, which has the cultural, historical, and natural significance in Hawaii.}}

\correspondingauthor{Eric Bell}

\author[0000-0002-5564-9873]{Eric F.\ Bell}
\affiliation{Department of Astronomy, University of Michigan, 1085 S. University Ave, Ann Arbor, MI 48109-1107, USA}
\email[show]{ericbell@umich.edu}

\author[0000-0001-6281-6724]{Benjamin Harmsen}
\affiliation{Department of Astronomy, University of Michigan, 1085 S. University Ave, Ann Arbor, MI 48109-1107, USA}
\email{benharms@umich.edu}

\author{Matthew Cosby}
\affiliation{Department of Astronomy, University of Michigan, 1085 S. University Ave, Ann Arbor, MI 48109-1107, USA}
\affiliation{L3Harris, 1919 W Cook Rd, Fort Wayne, IN 46818}
\email{mattcosby96@gmail.com}

\author[0000-0003-0511-0228]{Paul A.\ Price}
\affiliation{Department of Astrophysical Sciences, Princeton University, Princeton, NJ 08544, USA}
\email{price@astro.princeton.edu}

\author[0000-0003-0256-5446]{Sarah Pearson}
\affiliation{DARK, Niels Bohr Institute, University of Copenhagen, Denmark}
\email{sarah.pearson@nbi.ku.dk}

\author[0000-0003-2325-9616]{Antonela Monachesi}
\affiliation{Departamento de Astronom\'{i}a, Universidad de La Serena, Avda. R\'{a}ul Bitr\'{a}n 1305, La Serena, Chile}
\email{amonachesi@userena.cl}

\author[0000-0001-6982-4081]{Roelof S. de Jong}
\affiliation{Leibniz-Institut f\"{u}r Astrophysik Potsdam (AIP), An der Sternwarte 16, 14482 Potsdam, Germany}
\email{rdejong@aip.de}

\author[0000-0001-9269-8167]{Richard D'Souza}
\affiliation{Vatican Observatory, Specola Vaticana, V-00120, Vatican City State}
\email{rdsouza@specola.va}

\author[0000-0003-2294-4187]{Katya Gozman}
\affiliation{Department of Astronomy, University of Michigan, 1085 S. University Ave., Ann Arbor, MI 48109-1107, USA}
\email{kgozman@umich.edu}

\author[0000-0001-8042-5794]{Jacob Nibauer}
\affiliation{Department of Astronomy, Princeton University}
\email{jnibauer@princeton.edu}

\author[0000-0003-4961-6511]{Michael P.\ Busch}
\affiliation{National Radio Astronomy Observatory, 520 Edgemont Road, Charlottesville, VA 22903, USA}
\email{mbusch@nrao.edu}

\author[0000-0001-6380-010X]{Jeremy Bailin}
\affiliation{Department of Physics and Astronomy, University of Alabama, Box 870324, Tuscaloosa, AL 35487-0324, USA}
\email{jbailin@ua.edu}

\author[0000-0002-4884-6756]{Benne W.\ Holwerda}
\affiliation{Department of Physics and Astronomy, 102 Natural Science Building, University of Louisville, Louisville KY 40292, USA}
\email{benne.holwerda@louisville.edu}

\author[0000-0002-2502-0070]{In Sung Jang}
\affiliation{Department of Astronomy and Astrophysics, University of Chicago, Chicago, IL 60637, USA}
\email{hanlbomi@gmail.com}

\author[0000-0003-2599-7524]{Adam Smercina}\thanks{Hubble Fellow}
\affiliation{Space Telescope Science Institute, 3700 San Martin Dr., Baltimore, MD 21218, USA}
\email{asmercina@stsci.edu}

\begin{abstract}
A merger origin has been suggested for M83's massive, metal-rich extended {\sc Hi} disk and nuclear starburst. We observe M83's stellar halo to test this idea. We train nearest-neighbor star--galaxy separation on wide-area {\it Subaru} imaging with {\it Hubble Space Telescope} data to map M83's halo in resolved stars. We find that M83 has an extended, very low density smooth stellar halo of old and metal-poor [M/H]$\sim -1.15$ RGB stars with a mass between 15 and 40\,kpc of $\log_{10}M_{*,15-40,maj}/M_{\odot}=8.02\pm0.10$. In addition to M83's well-known Northern Stream, our ground-based Subaru imaging reveals a new stream to M83's south, which modeling suggests could be its trailing arm. The combined stream masses are $\log_{10}M_{stream}/M_{\odot}=7.93\pm0.10$, with metallicity [M/H]$= -1.0\pm0.2$. The stream progenitor was only recently accreted, as its stellar populations suggest that it formed stars until $2.1\pm1.3$\,Gyr ago. M83 lies on the stellar halo mass--metallicity correlation seen for other Milky Way mass galaxies, albeit with low stellar halo mass. We infer a total accreted mass of $\log_{10}M_{*,accreted}/M_{\odot}=8.78^{+0.22}_{-0.28}$, with the most massive past merger having  $\log_{10}M_{*,dom}/M_{\odot}=8.5\pm0.3$. We identify plausible M83 analogs in TNG-50 with similar stellar halos, finding that while a recent accretion can create a prominent stellar stream, such accretions do not trigger starburst activity, nor do they deliver enough gas to form M83's extended {\sc Hi} disk. We conclude that other non-merger mechanisms, such as secular evolution or accretion of gas from the IGM, are likely to be responsible for M83's remarkable properties. 
\end{abstract}



\section{Introduction} \label{sec:intro}

Among nearby Milky Way (MW)-mass galaxies, M83 (NGC 5236) is unusual for its extended, massive and metal-rich {\sc Hi} disk \citep{HuchtmeierBohnenstengel1981,Heald2016,Bresolin2009} its nuclear starburst, and its possible double nucleus \citep{Calzetti1999,Thatte2000}. While a wide range of processes might be responsible for these features, accretion and mergers are of particular interest \citep[e.g.,][]{Somerville2015}. With the development of both sensitive resolved-star halo datasets and sophisticated models of stellar halos in a cosmological context, it has become possible to quantitatively constrain a galaxy's most important merger events using stellar halo datasets (\citealt{Harmsen2017}, \citealt{Bell2017}, \citealt{DSouza2018_MNRAS}, \citealt{DSouza2018_M31}, \citealt{Smercina2020}, \citealt{Gozman2023}; see also e.g., \citealt{Zhu2022a}, \citealt{Cai2025}). M83 has resolved halo stellar populations information from the HST Galaxy Halos, Outer Discs, Substructure, Thick discs, and Star clusters (GHOSTS) survey \citep{Radburn-Smith2011,Monachesi2016_GHOSTS} and a wide-field deep Subaru survey, allowing us to pose the question: ``Is there evidence for a massive accretion/merger origin for M83's unusually massive {\sc Hi} envelope and nuclear starburst, or must other drivers be sought?"

M83 is a nearby ($D = 4.8$\,Mpc; \citealt{Radburn-Smith2011}; Galactic latitude $\sim 32^{\circ}$), nearly face-on spiral galaxy with a prominent bar \citep{KormendyKennicutt2004}. While M83's total stellar mass of $5.2\pm1.5 \times 10^{10} M_{\odot}$ and star formation rate (SFR) of $3.2\pm1.6 M_{\odot}$/yr \citep{Barnes2014} are unremarkable, M83's massive, $\geq 200$ kpc-wide {\sc Hi} envelope  \citep{HuchtmeierBohnenstengel1981, Heald2016} is unique in the Local Volume. While M83's total {\sc Hi} mass $\sim9 \times 10^9 M_{\odot}$ is only a little larger than the Milky Way's {\sc Hi} mass of $8 \times 10^9 M_{\odot}$ \citep{Kalberla2009}, most of that mass is at large radius, with $6 \times 10^9 M_{\odot}$ lying outside of the optical disc's $R_{25}$ \citep{Heald2016}. The densest parts of this {\sc Hi} disk form stars, visible as an extended UV disk \citep{Thilker2005}; this {\sc Hi} disk gives important insight into star formation at low gas densities \citep{Bruzzese19}. {\sc Hii} region spectroscopy of some of the star-forming parts of the disk show that the {\sc Hi} envelope has, remarkably, a metallicity around 1/3 solar with no detectable gradient \citep{Bresolin2009}. Furthermore, the disk is kinematically disturbed, and appears to be warped \citep{Barnes2014,Heald2016}.  

M83 also has an unusual nuclear region. Its starbursting nucleus \citep{Calzetti1999} contains tens of massive $\sim 10$\,Myr-old star clusters distributed asymmetrically \citep{Harris2001,Chandar2010}, and looks to be driving a starburst outflow \citep{DellaBruna2022}. The photometric center of M83 in the optical has a considerable mass of warm molecular hydrogen \citep{Hernandez23,Jones24}, and is offset by $\sim 60$\,pc from its kinematic center \citep{Knapen2010}. Its kinematic center is obscured by dust, and claims have been made of a possible second nucleus with red giant-like spectral features, suggestive of an older, established stellar population \citep{Thatte2000,Diaz2006}, although others argue against a double nucleus \citep[e.g.,][]{Knapen2010}. Highly-ionized neon emission has been observed in the mid-infrared, suggestive of ionization by an obscured AGN, although extreme shock emission is not completely ruled out \citep{Hernandez25}. Regardless of the exact properties of the nuclear region, it has been suggested that a single merger or interaction might be responsible for M83's unique nuclear properties and extended {\sc Hi} disk \citep[e.g.,][]{Thatte2000,Knapen2010}.  

In the debate over the origin of M83's unique features, it has been suggested that interactions with NGC 5253 are responsible. The dwarf starburst galaxy NGC 5253 (with a metallicity of $\sim 1/6$ solar; \citealt{Calzetti1999}) is $\sim 1.9$\,degrees or $\sim 160$\,kpc away in projection, and an interaction between M83 and NGC 5253 $\sim 1$\,Gyr ago was suggested as an appealing possible driver of both galaxies' starburst activity and M83's giant {\sc Hi} envelope \citep{vandenBergh1980,Calzetti1999}. In this scenario, M83's {\sc Hi} envelope might have been disturbed by (or even acquired from) NGC 5253, and while a $\sim 1$\,Gyr timeframe is considered too long for M83's starburst timescale \citep{Harris2001}, it has been tentatively suggested that the interaction might lead to, e.g., bar formation or some other process that would drive M83's present-day central starburst. Yet, this picture has a crucial problem in that NGC 5253, with $D \sim 3.55$\,Mpc \citep{Tully2013}, is $\sim$1.2\,Mpc closer than M83 along the line of sight. In order to travel 1\,Mpc in 1\,Gyr one needs a speed of $\sim 1000$\,km\,s$^{-1}$, dramatically in excess of the current velocity difference between NGC 5253 and M83 of $\sim 100$\,km\,s$^{-1}$. Consequently, we consider the origin of M83's peculiar features in an interaction with NGC 5253 to be extremely unlikely.

An accretion or merger event is another possible driver for the formation of the {\sc Hi} envelope, central starburst, and asymmetric or double nucleus. If the {\sc Hi} envelope has a merger origin,  its mass $M_{HI} \sim 6 \times 10^9 M_{\odot}$ and metallicity ($\sim 1/3$ solar) would indicate a merger with a system with at least $M_* \sim 5\times10^{9} M_{\odot}$, similar to or somewhat in excess of the Large Magellanic Cloud (LMC). Such a substantial merger would be expected to have effects on M83's SFR and nuclear regions. While M83 has a faint stellar stream at a maximum projected distance of $\sim 55$ kpc from the center of M83 (\citealt{vandenBergh1980}, \citealt{MalinHadley1997}; also known as KK 208; \citealt{KK1998}), it has a stellar mass of only $M_{*,stream} \sim 10^8 M_{\odot}$ \citep{Barnes2014}, far too small to have delivered such a large reservoir of enriched gas (but see also \citealt{Heald2016}). Instead, if M83's {\sc Hi} disk and starburst reflected a merger origin, one would expect a merger with a much larger $\sim$LMC-mass galaxy in its past.

Fortunately, a method has been developed that allows one to estimate the mass of the most massive satellite to have previously merged with a galaxy by studying its stellar halo \citep{Bell2017,DSouza2018_MNRAS}. Resolved-star investigation of the stellar halos of a number of Local Volume galaxies \citep[e.g.,][]{Radburn-Smith2011,Monachesi2016_GHOSTS} shows that the metallicity of a stellar halo correlates strongly with its mass \citep{Harmsen2017}. Such a correlation results if stellar halo masses and metallicities reflect primarily their most massive accretion --- a merger with a massive satellite drives up both the mass and metallicity of a stellar halo \citep{Deason2016,Harmsen2017,Bell2017,DSouza2018_MNRAS,Monachesi_2019}. Furthermore, if a merger was recent, stellar populations information may constrain the merger time \citep{DSouza2018_M31,Harmsen2023}. While the inference of merger masses or times requires the use of stellar halo formation simulations to calibrate the relationships between halo properties and merger information, the model features that are the most important are well-calibrated in most models (merger histories and halo occupation), and the merger inferences are quite robust. Consequently, measurements of M83's stellar halo could quantify its merger history and constrain whether merging is responsible for M83's unique features. 

The detection of a stellar halo in M83 presents a technical challenge, both owing to its nearly face-on viewing angle and its extended star-forming {\sc Hi} disc. In highly-inclined systems, stellar halos are relatively easy to separate from {\it in-situ} disk stars by studying integrated light \citep[e.g.,][]{Merritt2016} or resolved stars \citep{Radburn-Smith2011,Monachesi2013,Monachesi2016_GHOSTS} at $>$15\,kpc along the minor axes \citep{Monachesi_2016_Auriga,Monachesi_2019}. In face-on systems, the stellar disk is projected onto the stellar halo, outshining the halo's inner parts entirely, forcing one to seek more subtle signatures of the presence of a halo such as a break in the density profile or a transition in stellar populations \citep{Merritt2016,Jang19,Gozman2023}. M83's star-forming extended {\sc Hi} disk presents a further complication, as light from young stars contributes at large radii, where a halo would be most prominent. 

Accordingly, in this work we use resolved stellar populations to characterise M83's outskirts and identify and characterize its stellar halo. We combine narrow fields from the HST GHOSTS survey \citep{Radburn-Smith2011,Monachesi2016_GHOSTS} with wide-field data from HyperSuprime Cam on the 8.4-m Subaru telescope. With resolved stars, one can distinguish between the metal-rich outer disk, young stars formed in M83's giant {\sc Hi} disk \citep{Bruzzese19}, and old and metal poor RGB stars in M83's stellar halo. We use the HST data to train nearest neighbor star--galaxy separation methods for the wider field Subaru data, revealing a never-before-seen extension of M83's stellar stream to the south, and allowing quantitative measures of the halo and stream's stellar masses and metallicities. M83’s stream has a prominent AGB star population, showing that it stopped forming stars within the past 2 Gyr. We use the TNG-50 simulation to connect M83's face-on halo properties with measurements from other (usually edge-on) stellar halos with resolved-star information. These constrain M83's merger history, ruling out a large LMC-mass merger in its past. We use the TNG-50 simulation suite to seek likely analogs (in terms of merger history), allowing us to comment on the origin of M83's starburst and massive extended {\sc Hi} disk.  
We assume that the distance to M83 is $D = 4.8$\,Mpc following \citet{Radburn-Smith2011}, giving a physical scale of 1.396\,kpc/arcminute, and assume a \citet{Chabrier} IMF.

\section{Resolving stellar populations in the outskirts of M83}
\label{data}

We use two complementary datasets for our work. The first is a large-area ground-based mosaic of a roughly 2.5 square degree region of the M83 group, with the advantages of a wide area but with moderate depth and limited star-galaxy separation. This is complemented by deeper HST data, offering superior photometric quality and star-galaxy separation for some pencil beams in M83's outskirts. These HST data are critical for both providing a high-quality training set for supervised machine learning star--galaxy separation methods, and enabling accurate stellar population and surface brightness profile analyses along the particular lines of sight probed by HST. We describe first the ground-based data, then the HST data, then our star--galaxy separation methods. 

\subsection{Ground-based Subaru data}

\begin{deluxetable*}{llllllll}
\tablecaption{ Subaru observations of M83's stellar halo\label{tab:sub}} 
\tablehead{
\colhead{Field name } &
\colhead{RA } &
\colhead{Dec } &
\colhead{Passband } &
\colhead{Exposure time } &
\colhead{Av.\ FWHM } &
\colhead{Av.\ Airmass } &
\colhead{$5\sigma$ depth }}
\startdata
 M83\_1 & 13:34:40 & $-$29:21:56 & $g$ & 4560s & 0.78$"$ & 1.56 & 27.45 \\
 M83\_1 & 13:34:40 & $-$29:21:56 & $r$ & 3060s & 0.80$"$ & 1.66 & 26.62 \\
 M83\_1 & 13:34:40 & $-$29:21:56 & $i$ & 4260s & 0.85$"$ & 2.22 & 26.37 \\
 \hline
 M83\_2 & 13:34:40 & $-$30:21:56 & $g$ & 3960s & 0.83$"$ & 1.58 & 27.41 \\
 M83\_2 & 13:34:40 & $-$30:21:56 & $r$ & 3060s & 0.86$"$ & 1.70 & 26.55 \\
 M83\_2 & 13:34:40 & $-$30:21:56 & $i$ & 4860s & 1.00$"$ & 2.38 & 26.32 \\
 \enddata
\end{deluxetable*}

Our Subaru Hyper Suprime-Cam observations were taken through the Gemini--Subaru exchange program (PI: Bell, 2015A-0281). Two overlapping pointings (each $\sim$1\fdg5 FOV, overlapping by around $\sim$0\fdg5) were imaged in each of three ($g,r,i$) filters in the `classical' observing mode over the nights of March\,26--27, 2015. Data were taken for two pointings of a three pointing mosaic; the data focus on the western outskirts of M83. Given the field layout, the maximum projected distance probed from M83 is $\sim 120$\,kpc. The characteristics for each field+filter combination are given in Table \ref{tab:sub}. 
Owing to M83's southern declination, some of the data (particularly in $i$-band) were taken at high airmass where seeing was both poorer and limitations in Subaru's active mirror solution became important. This led to rapid degradation of the PSF quality which lead to a noticable degradation of the PSF quality in the southernmost parts of our mosaic; an aspect that we will explicitly account for in our star--galaxy separation. 

The data were reduced following \citet{Smercina2017}, using the HSC optical imaging pipeline hscPipe v3.6.1 \citep{Bosch2018}. The pipeline performs photometric and astrometric calibration using the Pan-STARRS1 catalog \citep{Magnier2013} reporting the final magnitudes in the HSC natural system in AB magnitudes. A particularly critical step is aggressive background subtraction using a 32 pixel $\sim 5"$ background region. This aggressive subtraction removes all diffuse light, allowing the recovery of faint point sources even in the case of mild crowding. Sources are detected in all three bands, although $i$-band is prioritized to determine reference positions for forced photometry. Forced photometry is then performed on sources in the $gri$\ co-added image stack. All magnitudes were corrected for galactic extinction using the Planck Collaboration's GNILC dust map (\citealt{Planck}; $E(B-V) \sim 0.04$), using the \texttt{dustmaps} software package as an interface \citep{dustmaps}. The version of the pipeline used for this data reduction did not incorporate the ability to insert artificial stars; we will later quantify completeness and contamination using comparison with HST data.

\subsection{HST Data}
A total of 14 fields --- spanning projected radii between $\sim 10$\,kpc and $\sim 50$\,kpc --- in M83's outer parts have been processed and analyzed by the GHOSTS survey (Table \ref{tab:obs}; these observations can be accessed via \dataset[doi:10.17909/8zcw-7w85]{https://doi.org/10.17909/8zcw-7w85}).Data reduction and photometry were carried out following \citet{Radburn-Smith2011} and \citet{Monachesi2016_GHOSTS}.
We either downloaded (ACS) or generated (WFC3) $*$\verb+_flc+ FITS files from MAST\footnote{\url{http://archive.stsci.edu}}. These data products have been bias-subtracted, cosmic
ray flagged and removed, flat fielded and corrected for charge transfer efficiency \citep[CTE;][]{Anderson10}. 
We use AstroDrizzle \citep{Gonzaga12} to combine the individual FLC images on a filter-by-filter basis and correct for geometric distortions. 

As described in \citet{Monachesi2016_GHOSTS}, we used DOLPHOT \citep{Dolphin00,DOLPHOT16} to perform simultaneous point-spread function (PSF) fitting photometry on the individual FLC exposures. Magnitudes are returned in the VEGAmag system, have been corrected for CTE loss, and include aperture corrections calculated using isolated stars. For most of our fields, M83's stars are both sparse enough and faint enough that compact background galaxies are the dominant contaminant. We select sources most likely to be stars in ACS and WFC3 catalogs following \citet{Radburn-Smith2011} and \citet{Monachesi2016_GHOSTS}, 
using diagnostic parameters returned by DOLPHOT
such as sharpness and crowding; this process (``culling") removed $\sim$95\% of the contaminants. We then clean the catalog of spurious sources in the wings of bright objects by omitting objects detected close to bright sources (as identified using SE\verb+XTRACTOR+;  \citealt{Bertin_arnouts96}).

In order to quantify photometric uncertainty and completeness, we performed artificial star tests (ASTs) as described by \citet{Radburn-Smith2011}. Around $2\times10^6$ fake stars
were injected into each exposure with a realistic range of colors and magnitudes, distributed such that they follow the observed stellar densities. DOLPHOT is then run on each fake star, one at a time. The completeness level in a given color, magnitude and spatial bin is then the fraction of injected artificial stars that were both detected and passed the culls. This completeness is used to give us the number of stars we would expect to exist for each detection. 

\begin{deluxetable*}{lllcccccc}
\tablecaption{{\it Hubble Space Telescope} Observations \label{tab:obs}} 
\tablecolumns{9}
\tablehead{
\colhead{Field ID   } &
\colhead{RA } &
\colhead{Dec } & \multicolumn{2}{c}{Exposure Time (s)} & \multicolumn{2}{l}{ F814W Completeness} & \multicolumn{2}{l}{ Distance From Galaxy Center} \\ \colhead{}
&\colhead{(h m s)}&\colhead{($^{\circ}$ m s)}&\colhead{F606W}&\colhead{F814W}&\colhead{90\%} &\colhead{50\% } & \colhead{(arcmin)}&\colhead{(kpc)}} 
\startdata
Field 01& 13 37 01.70& $-$29 43 52.00& 1900&3000 & 24.5 & 27.6 & 8.1&11.3\\  
Field 02& 13 36 55.25& $-$29 41 50.58& 1190&890 & 24.7 & 26.6 & 10.2&14.2\\
Field 03& 13 36 47.99& $-$29 36 54.93& 740&740 & 24.8 & 26.5 & 15.3&21.3\\
Field 04& 13 37 07.62& $-$29 34 57.76& 730&730 & 24.8 & 26.5 & 17.0&23.8\\
Field 05& 13 36 44.01& $-$29 33 11.54& 750&750 & 24.8 & 26.5 &19.1&26.7\\
Field 06& 13 36 30.80& $-$29 14 11.00& 1200&900 & 25.0 & 26.6 &38.3&53.5\\
Field 07& 13 36 28.07& $-$29 54 09.90& 725&725 & 23.8 & 26.2 &7.46&10.4\\
Field 08& 13 35 55.92& $-$29 57 22.37& 1190&890 & 25.4 & 26.8 &15.1&21.1\\
Field 09& 13 35 08.10& $-$30 07 04.00& 1200&900 & 25.4 & 26.8 &28.7&40.1\\
Field 10& 13 35 43.12& $-$29 43 49.45& 1190&890 & 25.4 & 26.8 &18.8&26.2\\
Field 11& 13 36 57.02& $-$30 06 41.85& 1198&898 & 25.4 & 26.8 &13.0 &18.2\\
Field 12& 13 37 00.64& $-$30 07 29.66& 1198&898 & 25.4 & 26.8 &15.6&21.8\\
Field 13& 13 36 53.36& $-$30 16 42.79& 770&680 & 25.2 & 26.5 &24.8&34.7\\
Field 14& 13 37 08.09& $-$30 21 39.29& 725&1140 & 25.2 & 27.1 &29.8&41.6\\ 
\enddata
\tablecomments{50\% and 90\% completeness limits for F606W are 1$\pm0.1$ magnitude deeper than F814W.} 
\end{deluxetable*}

\subsection{Ground-based Star--Galaxy Separation \& RGB Selection}
\label{sec:stargal}

Resolved stars at the distance of M83 have $i>24$, at brightnesses where resolved faint stars are easily confused with faint, distant galaxies. This is a fundamental limitation of wide-area ground-based survey datasets and limits their accuracy. At brighter limits, star--galaxy separation by morphology alone (choosing them to be unresolved) is often sufficient (e.g., \citealt{Bertin_arnouts96}, \citealt{Desai12}; see also \citealt{Slater20}). At these fainter limits, such a morphological selection does help but includes many faint, unresolved background galaxies \citep[e.g.,][]{Okamoto2015,Crnojevic2016}. If the dataset includes three or more passbands, color--color information can reduce contamination by background galaxies, either by using explicit color--color criteria \citep[e.g.,][]{Baldry2010,Smercina2017,Smercina2020} or supervised machine learning techniques \citep[e.g.,][]{Soumagnac2015,Bell2022}. 

The M83 halo pointings contain enough HST fields to contain a sizeable sample of stars to train supervised machine learning algorithms. Machine learning methods offer the possibility of tailored star selection that focuses on the most discriminatory features in the dataset for distinguishing between stars at M83's distance and other objects (mostly background galaxies, but also foreground MW stars that are uniformly distributed across the field). This advantage comes at a considerable cost --- we are explicitly tied to a training set of stars at M83's distance. This has two important impacts. Firstly, we will fail to recognize as stars objects that are not well-represented in our training data. Second, we are susceptible to variations in data quality across the Subaru images in areas where we have little or no training data. However, for our purposes, the gains of machine learning techniques are large enough to outweigh the costs. 

In selecting the quantities for classification, we have focused on those quantities that have discriminating power and can be generalized for areas other than the tiny areas covered by the training HST data. Recalling that the image quality is a function of declination, we focus on four types of information: magnitude, color combinations, declination, and information about how extended an object is compared to the local PSF size in multiple bands (and in different directions). The pipeline output that we use stores covariance matrices of the 2nd moment of an object's light in the $xx$, $yy$ and $xy$ directions (where $x$ is aligned with RA and $y$ is aligned with Dec). We choose to focus on the $xx$ and $yy$ extents, classifying on the {\it difference} in sizes between an object and the estimate of local PSF size from the HSC pipeline, giving six morphological features that we classify on $\sigma_{xx,g}-\sigma_{xx,g,PSF}$, $\sigma_{yy,g}-\sigma_{yy,g,PSF}$, $\sigma_{xx,r}-\sigma_{xx,r,PSF}$, $\sigma_{yy,r}-\sigma_{yy,r,PSF}$, $\sigma_{xx,i}-\sigma_{xx,i,PSF}$ and $\sigma_{yy,i}-\sigma_{yy,i,PSF}$. The difference in sizes is more robust to variations in data quality between the two fields, and within a field, than the sizes themselves. We add to these four more features: $i$-band PSF-fit magnitude, and PSF-fit derived $g-r$ and $r-i$ colors, and declination (to account for PSF degradation towards the southern parts of the mosaic). For many types of algorithm that rely on combining information from multiple features, we rescale each feature to have a `robust' standard deviation (a scaled median absolute deviation; using \texttt{astropy}'s \texttt{mad\_std}) of one.  Half of our objects comprise the training dataset; the other half are the test dataset that we use to quantify classifier performance. 

We experimented with a number of methods for using these metrics to distinguish between stars and galaxies. A wide variety of methods offered considerable gain over morphological only and morphology$+$stellar locus cuts: decision trees, random forests, a combination of principal component analysis (to compress the relevant information into fewer quantities) and kernel support vector machines, and Gaussian mixture modeling.  We found that for this dataset (and an additional training dataset in the M81 group; \citealt{Okamoto2015,Smercina2020,Bell2022}) that three methods gave somewhat better balances between completeness and contamination (recall/purity): $K$-nearest neighbors (typically with $\sim10-25$ neighbors), Gradient Boosting Machines, and Neural Networks. We could have chosen any of these three methods for our analysis in this paper without changing any of the results and conclusions in material ways. 

For simplicity, we chose a $K$-nearest neighbor classified sample, drawing from 15 nearest neighbors for classification.  
In brief, $K$-nearest neighbors methods classify datapoints by using majority votes from the $K$ nearest neighbors in multi-dimensional space. As described above, we rescale each classification metric (e.g., magnitude, difference between PSF size in one direction) by the robust standard deviation of that measurement to give each metric equal weight in driving the distance between data points. There are only two classifications. `Stars' correspond to 3902 HST-classified stars in areas with HST overlap. Our second label is `background'.  We  choose 6263 `background' objects from the Subaru data alone that are subsampled from a large region far from M83 covering a wide range in declination (which includes foreground stars --- some of which will not be classified as stars of interest --- and background galaxies). There are more `background' objects than `stars', yielding a training and test set that is somewhat unbalanced in favor of background objects, like our dataset itself\footnote{We do not use background objects in the HST fields themselves for the test or training set --- while there are enough of them to test classification performance, there are too few of them to reliably train a classifier. We thus choose background objects from the Subaru data at large distances from M83, with the added advantage of being able to explicitly control the balance of background objects and stars in the classifier training and test sets.}. The balance of `stars' and `background' impacts classifier performance in a predictable way, where classifications dominated by more background objects have lower completeness for stars, as the evidence must be more compelling to overcome the default classification as a background object. We have confirmed that even large variations in this balance do not dramatically change any of our conclusions. 

The Subaru photometric catalog, with the necessary columns for classification and star/background separation flag, is available at   \dataset[doi:10.5281/zenodo.17398573]{https://doi.org/10.5281/zenodo.17398573}.

\begin{figure*}[t]
    \centering
    \includegraphics[width=0.95\linewidth]{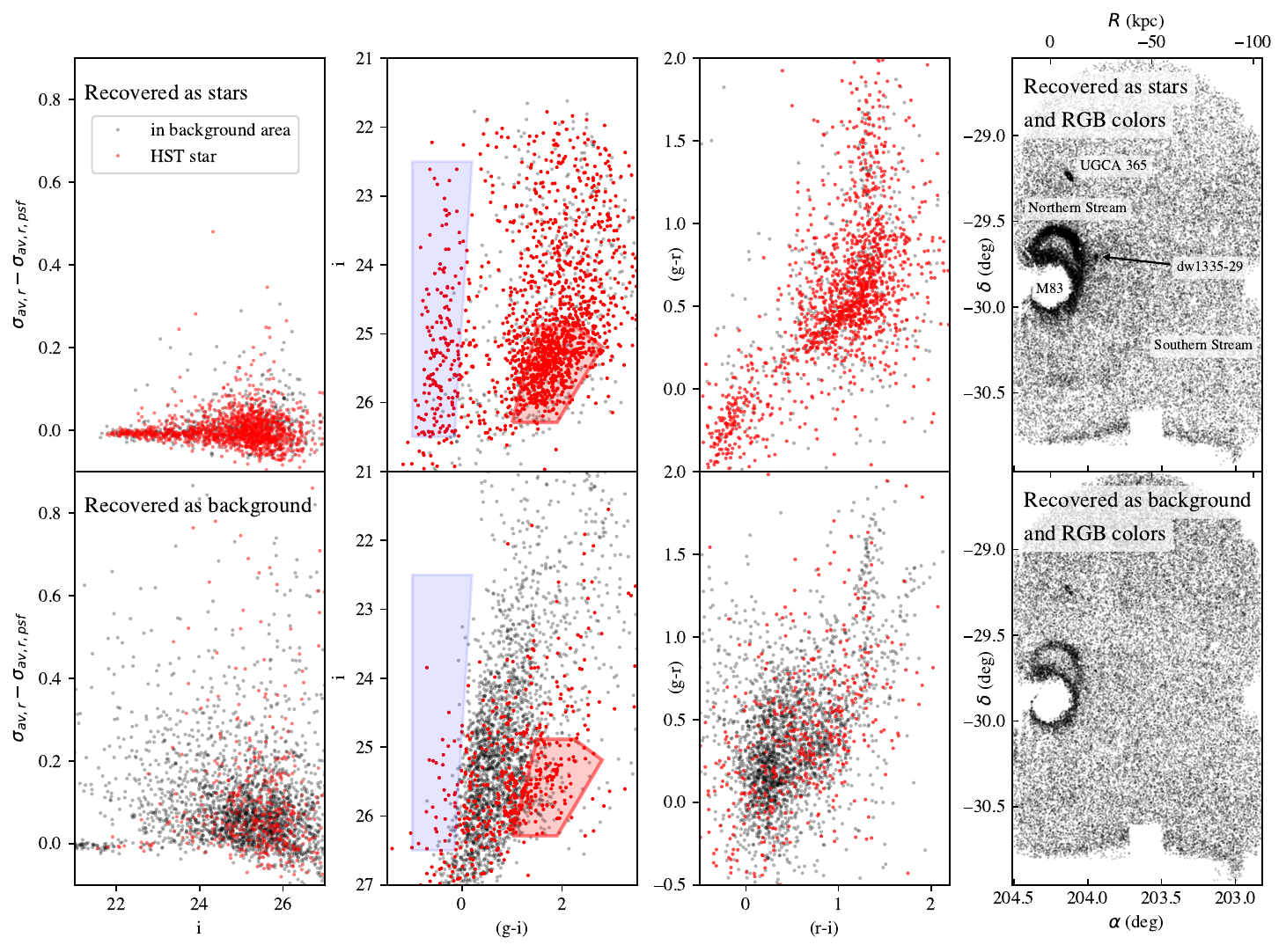}
    \caption{Results from the $K$-nearest neighbor star--galaxy classification. The top panels show objects in the test set classified as stars from the ground-based data (with HST-classified stars in red and background objects --- which include some bright foreground stars --- in black), while the bottom panels show objects classified to be background objects. The left panels show the average difference in object size and PSF size in the $xx$ and $yy$ directions in $r$-band, as a function of $i$-band PSF fit magnitude. The 2nd set of panels show $g-i$/$i$ color--magnitude diagrams, with the RGB selection region highlighted in red and the main sequence/blue helium-burning star selection shown in blue, while the 3rd set of panels show a $g-r$/$r-i$ color--color plot. The right-most panels show a map of RGB-colored objects classified as stars (top) and background (bottom). While the `background' contains overdensities in areas with large numbers of stars (M83's disk, UGCA 365, the Northern Stream), the objects classified as stars show these features much more strongly, as well as a newly discovered stream in the South and the faint satellite dw1335-29.}
    \label{fig:SGCMD}
\end{figure*}

\begin{figure}[t]
    \centering
    \includegraphics[width=0.95\columnwidth]{M83_completeness.pdf}
    \caption{Estimated point source completeness for our source detection by Subaru (black), and sources determined to be stars by our $k$-nearest neighbors classification and in our test set (dark red).  }
    \label{fig:compcont}
\end{figure}

In order to prioritize the fidelity of recovered large, low surface brightness halo features, we chose a balance between background objects and stars, and a number of nearest neighbors, that minimize the contamination of the stars without losing too much completeness in the star sample. Of the 1951 HST-identified stars in the test set (which do not include some bright foreground stars), 1464 (75\%) are recovered (this is termed the recall), but with 529 contaminating `background' objects (17\% of the background). This means that the precision (true positive/(true positive $+$ false positive) is 73\%\footnote{For known HST RGB stars specifically, the recall is 72\%, and precision is also 72\%, within the uncertainties of the classifier performance for all stars.}.
Many stars (487; 25\% of the stars) are misclassified as background, but this allows a classification that correctly classifies 2602/3131 (83\%) of the background objects, dramatically reducing contamination. 

The results of the classification on our test set are shown in Fig.\ \ref{fig:SGCMD}. The left panels show the average difference in object size and PSF size in the $xx$ and $yy$ directions in $r$-band, as a function of $i$-band PSF fit magnitude. The second set of panels show $g-i$/$i$ color--magnitude diagrams, while the 3rd set of panels show a $g-r$/$r-i$ color--color plot, which helps to separate stars (which occupy a narrow strip in color--color space --- the `stellar locus') from galaxies (which have a much wider range of colors). These panels show HST-classified stars in red and background objects (which include some bright $i<22$ foreground stars) in black. While there are misclassifications, the classifier clearly isolates many of the RGB stars $i>25$, $1<g-i<2$, showing a clear tip of the red giant branch, and very blue young stars $g-i<-0.2$. Misclassified stars tend to have properties that overlap essentially completely with background --- intermediate colors, a wider range of (often slightly larger) sizes, and a wide range of positions on the color--color plot. 

The right-most panels of Fig.\ \ref{fig:SGCMD} shows the map of RGB colored objects classified as stars (top) and background (bottom). The `background' distribution shows a smooth component (actual background) plus weak overdensities in areas with large numbers of stars (e.g., M83's disk, UGCA 365, the Northern Stream), representing actual stars misclassified as background. In contrast, the objects classified as stars show these features much more strongly, as well as a newly discovered stream in the South and the faint satellite dw1335-29 --- reflecting that the star--galaxy separation is successfully isolating many of the stars. In addition, the map of `stars' shows a considerable smooth background --- these are background sources that are indistinguishable in this 10-dimensional parameter space from actual stars. We note that the mosaics west of RA$\sim203.5$ appear essentially free of stellar subtructure; we therefore have wide areas available for background subtraction of halo features for our subsequent analyses. 

We show an alternate visualization of completeness in Fig.\ \ref{fig:compcont}. We restrict our attention to HST stars with colors similar to the RGB. For $i<25.5$, roughly 80\% or more of the HST stars are detected by Subaru (black line). Remarkably, in the test set (not used to train our star classifier), the vast majority of these are recovered by the nearest-neighbor classification as stars (dark red line). The 50\% completeness limit is $i=25.5$; in this but we focus on the TRGB region with $i<25.5$ for many of our quantitative inferences, where the average completeness is 66\%. 

In what follows, when we are discussing Subaru data, we will refer to those objects classified as stars by this $K$-nearest neighbors technique as `Subaru stars'. 

\section{Resolving M83's stellar halo and streams} \label{sec:halo}

In this section, we provide an overview of M83's stellar halo and streams from the Subaru and HST data, before focusing more quantitatively on M83's giant stellar stream and its diffuse halo in subsequent sections. 

\begin{figure*}[t]
    \centering
    \includegraphics[width=0.95\linewidth]{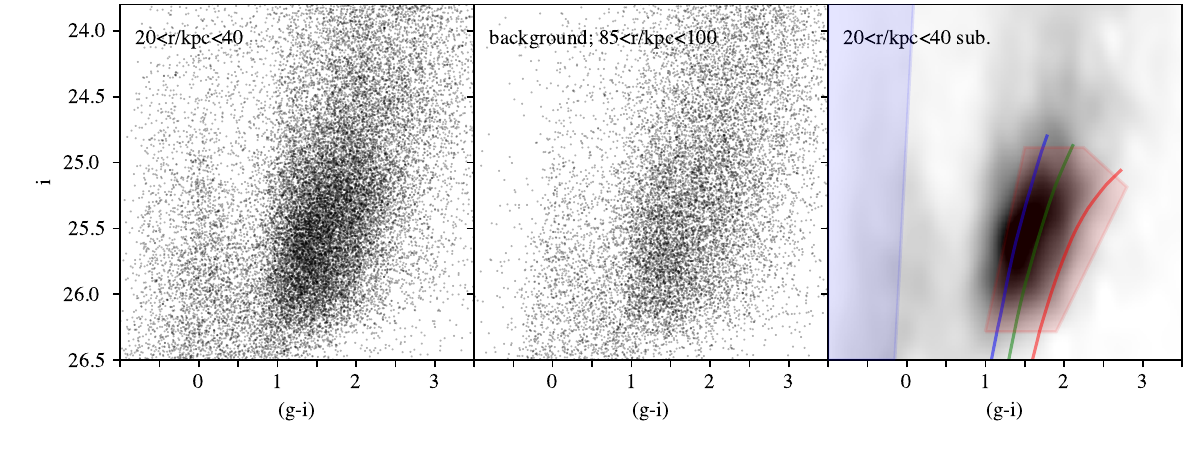}
    \caption{Color-magnitude diagrams of Subaru stars in the $20<r/kpc<40$ region, Subaru stars in a background region, and a background-subtracted Hess diagram of the $20<r/kpc<40$ region. Recall that the CMDs include only Subaru stars, which are selected in 10-dimensional space (including color and magnitude) to look similar to stars selected from the HST imaging. Overplotted are isochrones for $[M/H] = -1.5, -1, -0.5$ in blue/green/red, and the selection regions for the maps of RGB (red) and blue main sequence/blue helium-burning stars (blue) shown in Fig.\ \ref{fig:M83_pano}.  The most prominent feature is a relatively metal-poor RGB with $[M/H] \sim -1.1$, with a modest population of young upper main sequence and helium-burning stars with $g-i<0$.  }
    \label{fig:M83_2040}
\end{figure*}

\begin{figure*}[t]
    \centering
    \includegraphics[width=0.99\linewidth]{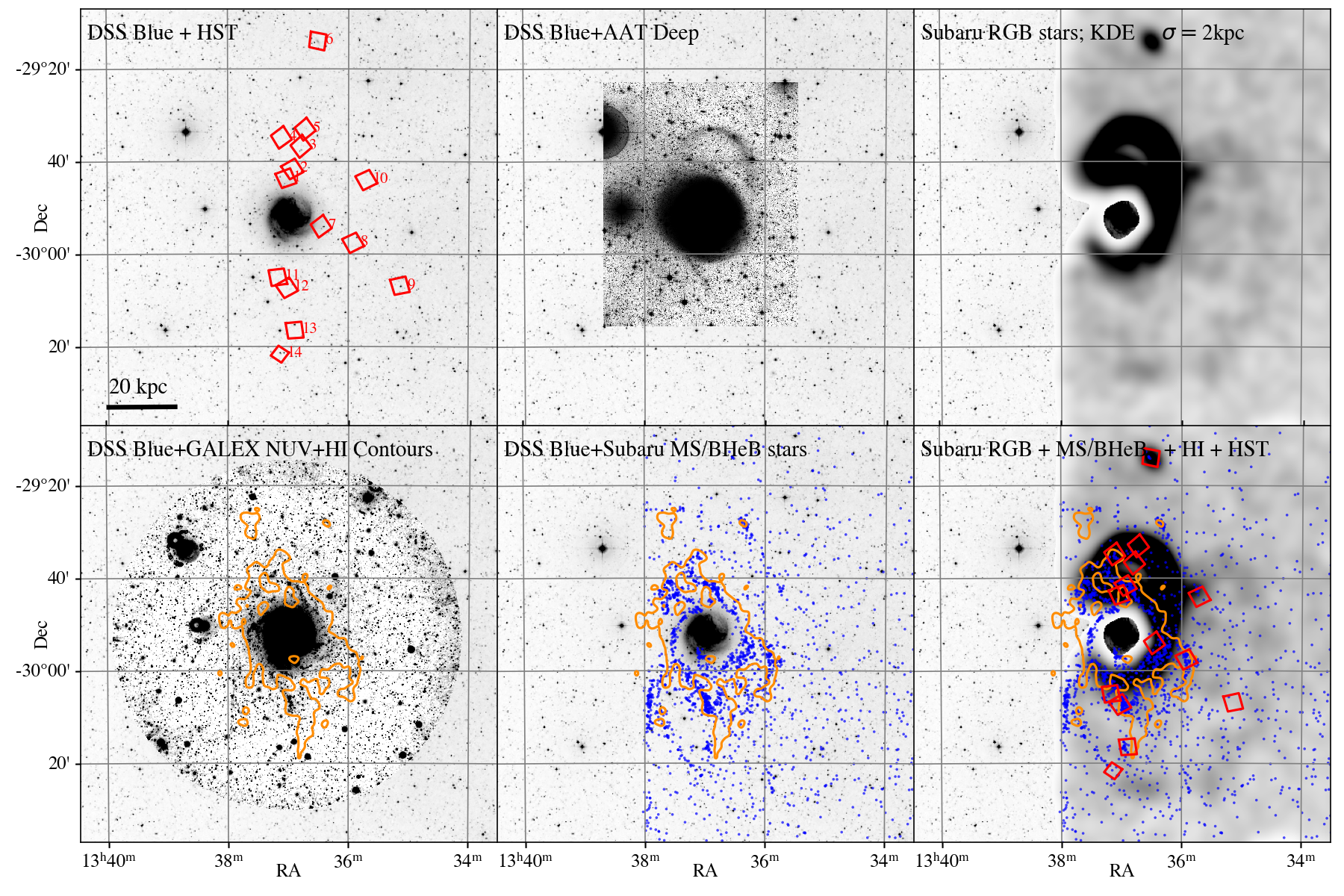}
    \caption{An overview of M83 stellar populations. In all panels, the blue Digitized Sky Survey image is shown as a backdrop, with North up and East to the left. Top left: the HST fields are overlaid in red, with the field numbers given to the right of each field. Top center: The deep AAT image \protect\citep{MalinHadley1997} is overlaid, showing M83's stellar stream. Top right: on the same scale, a Gaussian KDE with $\sigma = 2$\,kpc is applied to the RGB star map, bringing M83's stream and satellites into sharp focus as well as a new very low surface brightness stream to the South. Bottom left: Near-UV GALEX image with {\sc Hi} contours overlaid \protect\citep{Eibensteiner2023}. Bottom center: Subaru's blue stars highlight M83's extended star forming disk, covering a wider extent to the GALEX data with similar sensitivity. Bottom right: the distribution of M83's resolved stars (blue stars and RGB map) with the HST fields highlighted, to illustrate which halo components the HST fields probe. }
    \label{fig:M83_pano}
\end{figure*}

\subsection{A panoramic view of M83's halo reveals signs of hierarchical assembly}

Fig.\ \ref{fig:M83_2040} shows the color-magnitude diagrams of Subaru stars with $20<r_{proj}/kpc<40$ from M83, well outside M83's optical disk ($R_{25} \sim 7.7' \sim 10.7$\,kpc, containing $>90$\% of the B-band light; \citealt{ESOLV}). Recall that our nearest-neighbor star--galaxy separation selects `Subaru stars' that are similar in 10-dimensional space to the known stars in M83's halo, stream and satellites (Fig.\ \ref{fig:SGCMD}). This is important in understanding the CMD of Subaru stars in Fig.\ \ref{fig:M83_2040}. The left-hand panel shows a much stronger concentration of such stars, especially with $i>24.8$ and $1<g-i<2.2$ where M83's RGB is expected to be. The center panel shows a lower concentration of sources in our fiducial background region (projected radii between 85 and 100\,kpc) with properties consistent with M83 stars; these sources are all foreground MW stars or background unresolved galaxies. We note that none of the results in this paper are affected by the choice of the particular background region, as long as it is more than $r_{projected} > 70$\,kpc from M83. The right-hand panel shows an area-scaled subtraction of the background population from the $20<r_{proj}/kpc<40$ Subaru stars, showing a prominent relatively metal-poor RGB and a smattering of younger stars with $i>24.5$ and $g-i<0$. Isochrones from the Padova stellar population synthesis models\footnote{\url{http://stev.oapd.inaf.it/cmd} for a range of metallicities and 10\,Gyr age are overlaid, using the PARSEC evolutionary tracks version 1.2S \citep{Bressan2012}.} The Subaru data clearly resolves a population of luminous stars in the far outskirts of M83, dominated by a metal-poor RGB.  

These resolved stars reveal substantial new low surface brightness detail in M83's outskirts. M83 is unremarkable at typical depths (Digitized Sky Survey image; top left of Fig.\ \ref{fig:M83_pano}). Deep unsharp-masked imaging from the AAT (\citealt{MalinHadley1997}; top center) reveals a stellar stream to the North (e.g., \citealt{vandenBergh1980,MalinHadley1997,Barnes2014}, also called KK 208, \citealt{KK1998}). M83's RGB map, smoothed with a Gaussian kernel of 2kpc width to highlight low surface brightness features, brings into focus a variety of features. M83's stellar stream is extremely prominent in RGB stars. Two already-discovered satellite galaxies are clearly visible (see Fig.\ \ref{fig:SGCMD} for locations): UGCA 365 is prominent in the far north of the fields (overlapping with GHOSTS' Field 6), and a fainter satellite dw1335-29 \citep{Muller2015,Carrillo2017} just off to the west of M83's stream. The most substantial new feature in this map is a long, diffuse low surface brightness stream towards its south (hereafter the Southern Stream); we will explore its nature in \S \ref{sec:stream}. 

Conspicuous by its absence is the central part of M83, with no stellar detections owing to crowding that was not resolved by the Subaru pipeline. In addition, M83's outer disk is not readily apparent in these maps, primarily because the redder RGB stars characteristic of M83's stellar disk (see \S \ref{sec:HST_CMDs}) are too red to be well-detected in the $g$-band data (see e.g., \citealt{Smercina2020} for a discussion of this issue in the context of M81's metal-rich disk RGB stars).

The Subaru data also reveal bright young stellar populations, appearing largely unassociated with M83's stellar stream. M83 is well known for its very extended {\sc Hi} disk, with UV-bright regions of star formation (e.g., \citealt{HuchtmeierBohnenstengel1981}, \citealt{Thilker2005}; bottom left of Fig.\ \ref{fig:M83_pano}). There is a smooth distribution of misclassified background objects (more prominent towards the South of the mosaic, because of the slightly worse PSF in its southern parts).  Yet, there are also blue resolved stars that prominently trace out the same structures as the UV emission, including hints of star formation well outside the GALEX field of view. The lack of correspondence between the stellar stream and {\sc Hi}/young stellar disk is well-known  \citep{Barnes2014}. UGCA 365 does have a few stars visible in the blue stars map; in addition, there are some blue young stars that are projected close to dw1335-29 on the sky and may or may not be associated with it (see, e.g., \citealt{Carrillo2017} for more discussion of this point). Acknowledging this widespread star formation, we will focus for the remainder of this paper on the older stars that trace the accretion and merger history of the M83 system.

In the top right panel, there is little visual sign of a diffuse, phase-mixed stellar halo of the type that is so prominent in RGB star maps of M31 \citep{Ibata2014} or Cen A \citep{Crnojevic2016}. That such a halo is not prominent in such a map is already informative --- it tells us that M83's stellar halo has a relatively low mass, rather  similar to e.g., M101 ($\sim 0.8-3 \times 10^8 M_{\odot}$, where the halo is difficult to characterize with deep integrated light or resolved star observations; \citealt{vD14,Jang2020}), M94 ($\sim 3\times10^8 M_{\odot}$; \citealt{Gozman2023}), or M81 ($\sim 10^9 M_{\odot}$, \citealt{Harmsen2017,Smercina2020}). 

\subsection{HST fields resolve the stellar populations of halo structures} \label{sec:HST_CMDs}

\begin{figure*}
    \centering
    \includegraphics[width=180mm]{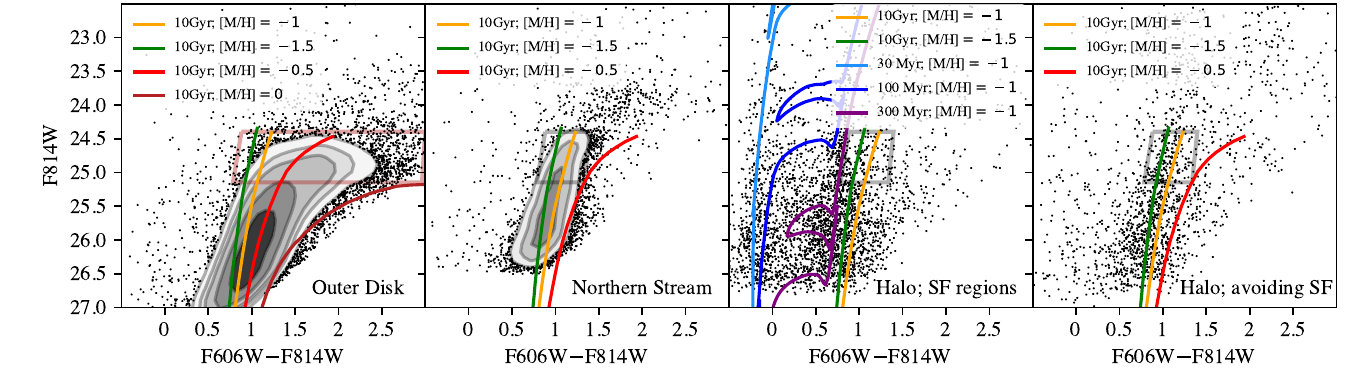}    
    \caption{HST CMDs of the outer disk, the Northern Stream, the parts of the halo with prominent young star populations, and the parts of the halo without prominent young populations. 10\,Gyr old isochrones are shown in green through dark red for $[M/H] =-1.5$ to $0$, and younger isochrones are shown on the halo SF regions panel at $[M/H]=-1$ for ages of 30, 100 and 300\,Myr in shades of blue and purple. The metal-poor RGB selection region is outlined in gray. Contours are placed in dense regions of the CMD at 20, 35, 50, 80 and 120 stars per (0.0875\,mag$\times$0.1125\,mag) bin.}
    \label{cmds3p}
\end{figure*}

With its superior depth and outstanding star--galaxy separation, HST data give a high-fidelity view of the stellar populations in select regions in M83's outskirts (shown in red in Fig.\ \ref{fig:M83_pano}). The HST data complements and enriches the panoramic view available from Subaru. 

We start with the M83's outer disk (left-hand panel Fig.\ \ref{cmds3p}; corresponding to HST Field 1). M83's outer disk shows a very wide and prominent RGB, corresponding to complex stellar populations with a wide range of metallicities, extending to high degrees of metal-enrichment, reaching towards solar metallicity, characteristic of M83's outer disk older populations. Because dynamical effects such as radial migration or tidal disruption of a disk that move disk stars outwards will affect {\it all} disk stars, this outer disk CMD gives an approximate expectation for how outer disk star contamination should appear in even more remote halo fields.  

In contrast, fields outside of galactocentric radii $>15\,$kpc show very different stellar populations from M83's outer disk. There are three main CMD morphologies that are seen at radii $>15\,$kpc. 

The second panel shows the CMD of the Northern Stream (Fields 4 and 5), again with old, metal-poor isochrones overplotted. This region shows a prominent metal-poor RGB with metallicity $[M/H]\sim -1$. In addition, there is a prominent AGB at $F606W-F814W\sim1.5$ and $F814W\sim24$ that we will later use to place age constraints on the epoch of last major star formation in the stream. 

The third panel shows a stacked CMD of areas of our fields that show young stellar populations (selected areas from fields 8, 11, 12 and 13), along with some younger isochrones (30, 100, 500\,Myr at [M/H]$\sim-1$). Large populations of young blue main sequence stars with $F606W-F814W \lesssim 0$, and massive helium burning stars with $F606W-F814W \sim 0$ and $F606W-F814W \sim 0.8$, are apparent. This reflects clumpy recent star formation in M83's extended H{\sc i} gas disk, as studied by \citet{Thilker08} and \citet{Bruzzese19}. These areas also contain a few RGB stars --- these RGB stars are unclustered, and reflect M83's underlying stellar halo.

The stacked CMD in the right-hand panel of Fig.\ \ref{cmds3p} shows the CMD of the unclumped parts of the sparse fields (the rest of fields 8 and 11--13, most of field 3 [avoiding a corner which clips the Northern Stream], most of field 10 [avoiding the part of the field containing dw1335-29] and including fields 9 and 14). In stark contrast to the outer disk CMD which shows a prominent and broad TRGB, a much narrower and bluer RGB is visible on top of a scattered population of foreground stars and unresolved background galaxies (see also e.g., Figs. A1 and A2 of \citealt{Jang2020}), with only a small number of younger, bluer stars. Unlike the Northern Stream CMD, there is no sign of a prominent AGB star population; instead, there is a modest population of foreground MW stars (F814W$<24$; $1<$F606W-F814W$<2$; see also \citealt{Robin2007}) whose density does not vary within Poisson uncertainties from field to field. By comparison with the isochrones, one can see that the RGB star population is substantially metal-poorer than those in M83's outer disk and are not drawn from the disk population; these metal-poor RGB stars are M83's stellar halo. 

In order to study the properties and spatial distribution of these metal-poor halo RGB stars, we have to be careful to restrict our attention to regions dominated by RGB stars. M83's halo RGB is metal-poor and has a distribution of colors that is very close to the reddest of the helium-burning stars in M83's extended {\sc Hi} disk (as seen as the nearly-vertical sequence with F606W$-$F814W$\sim 0.8$ and F814W$<25$ in the third panel of Fig.\ \ref{cmds3p}). For the purposes of our study of M83's stellar halo, we therefore select a relatively narrow window in CMD space (outlined by the gray selection region in Fig.\ \ref{cmds3p}) which captures the tip of the RGB while avoiding contamination from helium-burning stars. We use this selection for the construction of HST density and metallicity profiles. We note that for the outer disk (left-hand panel), we show instead a wider color selection (dark orange outline) which we use to study the full RGB star population profile.

\section{Quantifying and mapping M83's large stellar streams (KK208)} \label{sec:stream}

The (relatively) higher surface brightness part of M83's large stellar stream (the Northern Stream) was discovered already nearly 30 years ago in deep, specially-processed photographic imaging plates \citep{MalinHadley1997}, and is visible in deep IRAC imaging \citep{Barnes2014}; it is very prominent in our resolved star maps also. With the combination of Subaru and HST data, we can add to our knowledge of the stellar stream by placing constraints on its metallicity, stellar mass, time at which star formation stopped in the stream progenitor, and by tracing out a possible very low surface brightness extension towards the south.

While it is possible that the Southern Stream is a separate stream, its width and visual morphology appears to continue the arc of the eastern part of the Northern Stream towards the South. Because this tail is at larger projected radius than the higher surface brightness parts of the stream, and given the morphology of Northern part of stream, we tentatively suggest that this may be a trailing arm of the Northern Stream (we will discuss this possibility in depth in \S \ref{sec:dynmodel}).

\subsection{Stellar population constraints on the streams}

The Northern and Southern Streams (outlined in Fig.\ \ref{fig:Stream_Subaru_CMDs}) both show overdensities of RGB stars $g-i\sim1.5$ and $i \sim 25$. In the case of the Northern Stream, a faint echo of the AGB stellar population that is so prominent in Fig.\ \ref{cmds3p} is visible, at $i\sim 24.5$ and $g-i \sim 2.5$. The Southern Stream is considerably less dense, but shows a clear RGB. The RGB appears slightly bluer (mean $g-i\sim1.75\pm0.05$ for $i<25.5$ RGB stars) than the Northern Stream (mean $g-i\sim1.91\pm0.05$ for $i<25.5$ RGB stars). This impression is supported by HST CMDs (right panels of Fig.\ \ref{fig:Stream_Subaru_CMDs}), showing that the Northern Stream fields are considerably more dense than the Southern fields. The Southern fields, albeit not background subtracted, show a weak CMD with a slightly bluer color than the Northern fields. 

\subsubsection{The metallicity of the Northern and Southern Streams}

We constrain the metallicities of both streams using the Subaru data, cognizant of the inevitable limitations imposed by incompleteness and contamination (Fig.\ \ref{fig:compcont}). In order to translate position in color--magnitude space to metallicity, we follow \citet{Gozman2024} \citet{Ogami2024} by using radial basis function interpolation between 10\,Gyr old Padova isochrones to translate RGB star color into a metallicity estimate. We choose only $i<25.5$ RGB stars above the 50\% completeness limit. It is important to correct for the contribution of background objects, which bias the metallicity. We choose background regions far separated from M83 (with the same areas as the stream selection areas), and calculate equivalent `metallicities' for these background objects. In order to correct for background, we subtract the cumulative distribution of the inferred `metallicities' of objects in the background region from the cumulative distribution of metallicities in the stream region. The remaining distribution is the cumulative metallicity distribution of objects in the stream region, statistically corrected for background (this method is similar to that adopted by \citealt{Crnojevic2019} and \citealt{Fielder2025}). We adopt the median and percentiles of this distribution to be the stream metallicity and its spread. We estimate uncertainties by comparing different regions of the streams, quantifying their scatter; this estimate also marginalizes over background uncertainty. We also estimate $\sim 0.2$\,dex systematic error from isochrone color calibration and from completeness. 

The metallicity of the Northern Stream is thus found to be $[M/H] = -0.95\pm0.05$ (random) $\pm0.2$ (systematic) with a `spread' $\sigma_{Northern\,stream}$ (half of the 16\% -- 84\% interval) of $0.40\pm0.05$\,dex (random uncertainty only); the Southern Stream has $[M/H] = -1.1\pm0.1$ (random) $\pm0.2$ (systematic) with a `spread' of $\sigma_{Southern\,stream}$ of $0.45\pm0.11$\,dex (random uncertainty only). 

Independent insight into M83's Northern Stream can be gained by using the HST data for Fields 4 and 5 (Figs.\ \ref{cmds3p} and  \ref{fig:Stream_Subaru_CMDs}). 
We choose to follow \citet{Streich2014} in using the color of the RGB at a fiducial absolute magnitude to estimate the [M/H] and [Fe/H]. We account for the tilt of the RGB using a [M/H]$=-1$ 10\,Gyr old isochrone to scale the color of RGB stars to a fiducial absolute magnitude of $M_{\rm F814W} = -3$ ($m_{\rm F814W} = 25.41$). This yields (F606W-F814W)$_{-3} = 1.01\pm0.05$ (dominated by PSF photometry calibration uncertainty; \citealt{Radburn-Smith2011}), corresponding to [M/H]$=-0.9\pm0.3$, in agreement with the Subaru estimate [M/H]$_{Subaru}=-0.95\pm0.05\pm0.2$. Different methods for estimating metallicity (e.g., different choices in fiducial magnitude or from isochrones directly) give estimates that vary by considerably less than those expected from calibration error alone. 

\subsubsection{Constraints on Northern Stream SFH from AGB stars} \label{sec:AGBage}

The HST data allow us to show that the stream experienced star formation until relatively recently by measuring its AGB star population relative to its RGB population. Such a stellar populations probe is helpful for datasets like these that are too shallow to measure a full star formation history (which requires reaching the horizontal branch in depth; e.g., \citealt{Weisz2011}). The ratio of bright AGB stars to bright RGB stars correlates with $t_{90}$ --- the time before which 90\% of star formation in a system has happened (\citealt{Harmsen2023}; see also \citealt{LeeWiesz2024} for SFH fits from AGB stars in M31's halo). In stellar streams, $t_{90}$ places an upper limit on the disruption time: the satellite cannot continue to form stars after it is disrupted. Fig.\ \ref{cmds3p} shows a prominent and well-populated AGB, indicating relative stellar population youth. Following \citet{Harmsen2023}, and correcting for foreground/background following that work, we measure an AGB/RGB ratio of $\log_{10} N_{AGB}/N_{RGB} = -0.78\pm0.04$, corresponding to a $t_{90} \sim 2.1\pm1.5$\,Gyr following Equation 8 of \citet{Harmsen2023}, where the error in the inferred $t_{90}$ is dominated by the intrinsic scatter in the observational sample that calibrates the relationship between $\log_{10} N_{AGB}/N_{RGB}$ and $t_{90}$. We conclude that the main body of M83's prominent tidal stream was bound and able to form stars as recently as $\sim 2$\,Gyr ago. This stellar population age is consistent with completely independent estimates of $\sim 1$\,Gyr old derived by \citet{Barnes2014} from the relatively narrow width of the stellar stream and using the framework of \cite{Johnston2001}. 

\subsection{An estimate of the stream's mass}

\begin{figure*}
    \centering
    \includegraphics[width=170mm]{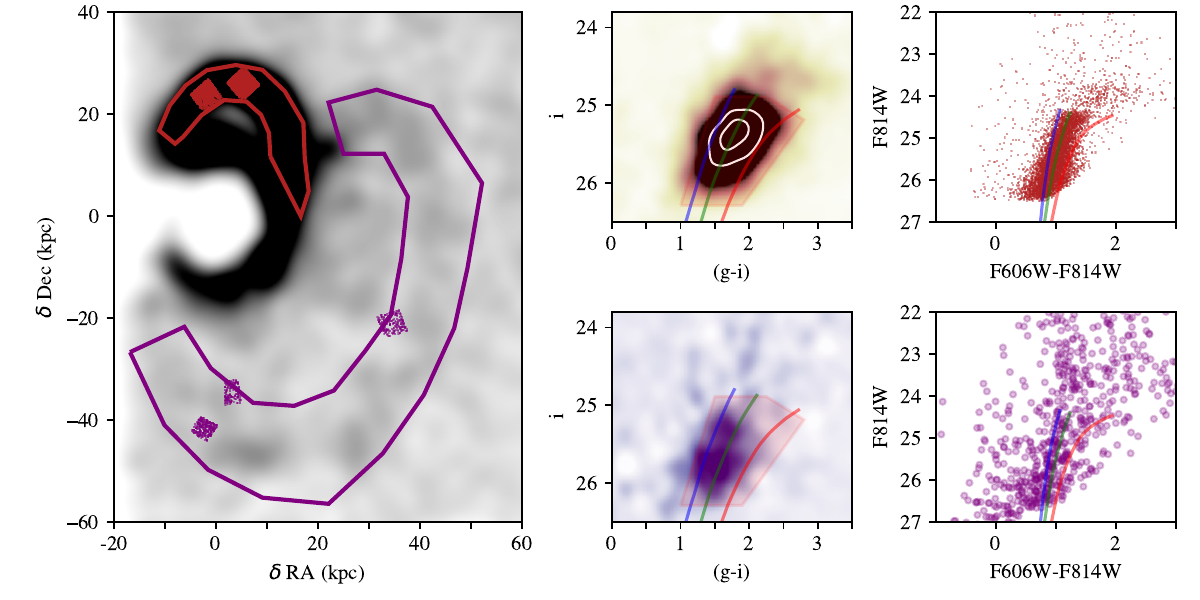}   
    \caption{Left: The distribution of Subaru stars consistent with being on the RGB, with the outlines of the Northern Stream (dark red) and the Southern extension of the stream (purple). The stars in the HST fields that overlap with each stream footprint are overlaid. Top panels: Sky subtracted Hess diagram (center) of Subaru data within the Northern Stream with 10\,Gyr $[M/H] = -1.5,-1,-0.5$ isochrones in blue, green and red, and the RGB selection overlaid. The AGB and RGB have such different densities that we have used contours to delineate the densest bins containing the bright RGB stars. The right panel shows the HST CMD of the stars in the Northern Stream footprint, with the same isochrones overlaid. Bottom panels: same as above, but for the Southern extension of the stream. }
    \label{fig:Stream_Subaru_CMDs}
\end{figure*}

An estimate of the Stream's mass requires both the wide-area coverage of Subaru, accounting for completeness and stellar populations in a way that is only possible using the deeper and more complete HST data. The Northern and Southern parts of the stream show very significant overdensities of RGB stars $g-i\sim1.5$ and $i \sim 25$ when corrected for background sources (Fig.\ \ref{fig:Stream_Subaru_CMDs}). Correcting for fore/back ground stars and galaxies, there are 2570$\pm50$
Northern RGB stars with $i<25.5$, and 500 $\pm60$ Southern RGB stars (the error bars include sky subtraction uncertainties). The Southern extension to the stream contains about $\sim 20$\% of the stellar mass contained in the denser and better-defined Northern part of the Stream. 

There are two independent ways to estimate the luminosity and mass of the streams; they agree to within their errors. One is by scaling Subaru star counts to dw1335-29 (41 $i<25.5$ Subaru RGB stars; $M_V = -10.1\pm0.4$; \citealt{Carrillo2017}). On this basis, the Southern and Northern Streams have $M_V \sim -12.8$ and $-14.5$ respectively, also $\pm0.4$\,mag. The corresponding masses, assuming a stellar $M/L_V = 1.5$ is $\log_{10}(M_*/M_\odot) \sim 7.2\pm0.2$ and $7.9\pm0.2$ respectively.  The second approach scales the number of detected RGB stars to the total stellar mass of a population using theoretical isochrones. Using a $[M/H]=-1$ 10\,Gyr old isochrone, a present-day stellar mass of 19030$M_{\odot}$ is needed to produce one RGB star in the CMD selection region, with a 20\% uncertainty from stellar population variations. 
Combined with the completeness in the TRGB region $24.88<i<25.5$ of 68\%$\pm$5\%, we recover $\log_{10}(M_*/M_\odot) \sim 7.16\pm0.10$ and $7.85\pm0.09$ for the Southern and Northern Streams respectively, in agreement with the purely empirical scaling. 

If the Northern and Southern Streams are associated, as discussed in \S \ref{sec:dynmodel}, their combined stellar mass would be $\log_{10}(M_*/M_\odot) \sim 7.93\pm0.1$.

\section{The properties of M83's diffuse stellar halo} \label{sec:halo}

\begin{figure*}
	\includegraphics[width=0.95\linewidth]{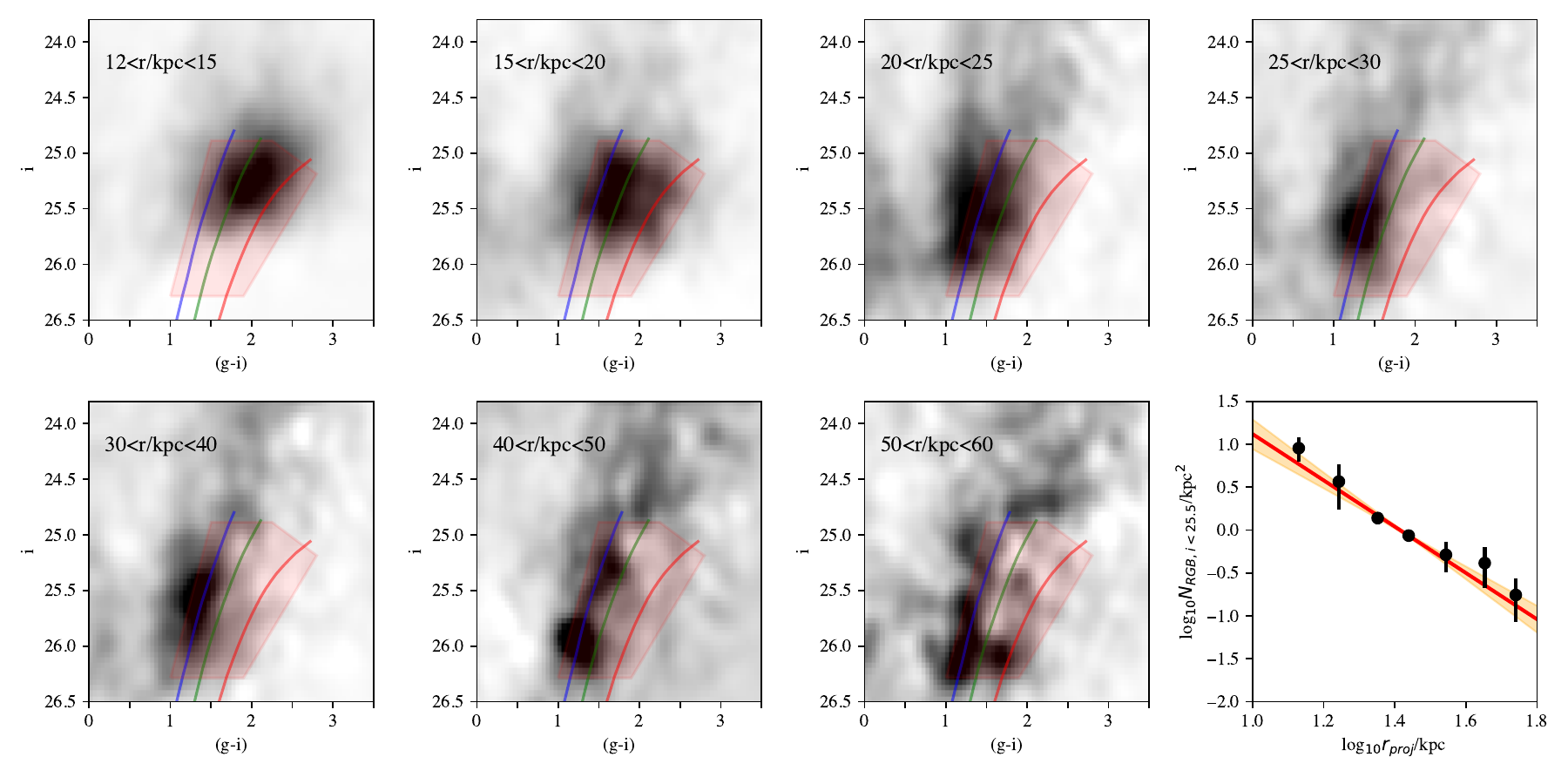}   
    \caption{First seven panels: Sky subtracted Hess diagrams of Subaru `stars' (classified with our star--galaxy separation) in various radial ranges, not in substructure or satellites, with 10\,Gyr $[M/H] = -1.5,-1,-0.5$ isochrones overplotted in blue, green and red, and the RGB selection region highlighted. The outermost radii are noisy owing to substantial uncertainties from background subtraction. Bottom right: The resulting number of detected $i<25.5$ RGB stars per square kpc with errors calculated from the azimuthal variation in each bin, with the best fit power law overplotted in red and 68\% confidence interval in orange.  }
    \label{fig:Subaru_haloprof}
\end{figure*}

\begin{figure}
	\includegraphics[width=0.95\columnwidth]{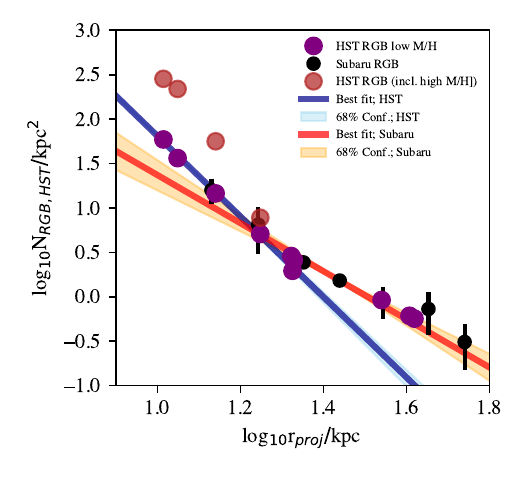}   
    \caption{The radial density profile of metal-poor RGB stars from the HST fields (purple data points with error bars that are often smaller than the points), omitting from consideration the fields dominated by streams or satellites. The three purple HST data points at $\log_{10} r_{proj}/{\rm kpc} \sim 1.55$ may have contributions from the Southern Stream. In dark blue is the best-fit power law profile for the HST metal-poor RGB star numbers. In translucent red are the total RGB counts for the inner four fields that include the redder metal-richer RGBs in M83's disk; counts of red stars are negligible outside that radius. The black datapoints with error bars and red line show the density profile from Subaru star counts translated into equivalent HST star counts. Both HST and Subaru paint a consistent picture of a steep power-law density profile for M83's metal-poor stellar halo population.  }
    \label{fig:HST_radial_mass}
\end{figure}

\begin{figure}\centering
	\includegraphics[width=0.95\columnwidth]{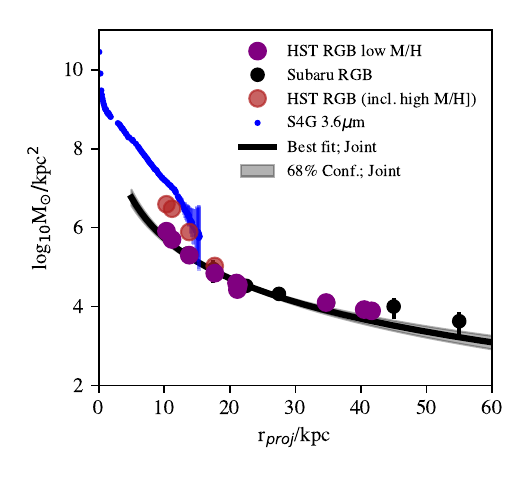}   
    \caption{The radial density profile as determined using 3.6$\mu$m data from Spitzer (blue), from Subaru halo star counts (black), from HST metal-poor RGB star counts (purple; the three data points at 30--40\,kpc may have contributions from the Southern Stream), including high-metallicity stars in the innermost 4 HST fields where they are important (dark red).  The best power law fit to the joint density profile (adding 0.1\,dex in quadrature to the data points' uncertainties) is shown in black. M83's disk profile drops steeply at radii $>12$\,kpc, matching smoothly onto HST total RGB star counts. Beyond $r\sim17$\,kpc, M83's metal-poor stellar halo dominates, with a roughly power-law profile. }
    \label{fig:Subaru_radial_mass}
\end{figure}

\begin{figure}\centering
  \includegraphics[width=0.95\columnwidth]{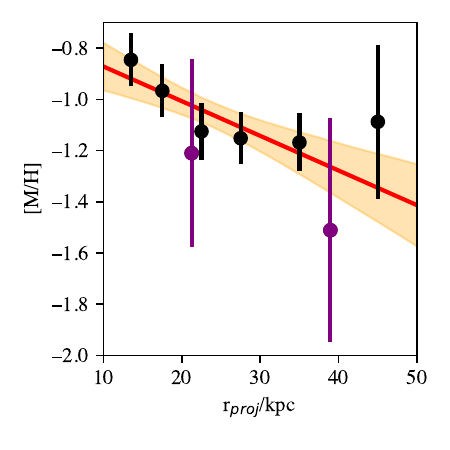}
      \vspace{-0.3cm}

    \caption{The radial metallicity profile, determined from RGB colors assuming 10\,Gyr old halo stellar populations. Measurements from the Subaru data are in black; measurements made by aggregating HST fields are in purple (the outermost HST data points may have contributions from the Southern Stream). As was apparent in the Hess diagrams in Fig.\ \ref{fig:Subaru_haloprof}, the color gradient in the halo stellar populations implies a metallicity gradient, dropping  by 0.3 dex from 12--40\,kpc in galactocentric radius.  }
    \label{fig:Subaru_radial_mh}
\end{figure}

In this section, we seek to quantify and understand the widespread, low-density `aggregate' stellar halo. That such a component exists (and is somewhat lower metallicity than M83's stream) is clear from the right-most panel of Fig. \ref{cmds3p}. Since M83's disk appears to be close to face-on (with an axis ratio of 0.89 from \citealp{RC3}), and since in our analyses we find no statistically-significant sign of asymmetry in halo star counts as a function of azimuth (beyond the tidal streams), we assume circular symmetry in what follows.

As in Section \ref{sec:stream}, we quantify the halo using the Subaru and HST data largely separately, using the Subaru data to quantify the wide-scale distribution of mass, and the HST data to quantify a few representative lines of sight as a check on the systematic errors inherent in the less sensitive but wider-area Subaru analyses. 

\subsection{Halo Stellar Density Profile}

\subsubsection{Subaru stellar halo density profile}

We first quantify the stellar halo RGB star density profile using Subaru data. In common with our analysis of M83's stream(s), we choose to focus our attention on RGB stars, restricting our attention to $i<25.5$ where the Subaru star completeness exceeds 50\%. We choose a background region spanning galactocentric radii of 85\,kpc to 100\,kpc; we note that none of our estimates vary significantly even with generous changes to the background region as long as it exceeds $r_{projected} > 70$\,kpc in distance, where there is no detectable sign of M83's stellar halo. Area calculations, for calculating the density of stars (number per unit area) and to properly scale the background counts by area, account for field edges and regions with no detections (e.g., around bright stars), and excluded the stream footprints, and exclusion regions of 3kpc around UGCA 365 and 1.1kpc (2 half light radii) around d1335-29. 

The Hess diagrams as a function of radius, and the star counts of $i<25.5$ RGB stars, are shown in Fig.\ \ref{fig:Subaru_haloprof}. M83's diffuse stellar halo --- unassociated with any recognized substructure or satellites --- is clearly detected for radii $r_{proj}<50\,$kpc. Its profile appears to follow a power law, including the most uncertain bins at $40<r_{proj}/{\rm kpc}<60$. 

Owing to the huge areas averaged over for these profiles, the formal and background uncertainties are small (typically 1-10\%). Cognizant of the possibility of undetected variation from region to region in what we expect {\it a priori} to be a structured stellar halo accretion-dominated stellar halo \citep{BullockJohnston2005,Harmsen2017}, we choose to estimate uncertainties by quantifying variations along each radial bin as a function of azimuth in 15 degree bins. We ignore azimuthal ranges with no star counts (as they lie outside the field boundaries, or are dominated by stream stars). Our estimate of the `error in the mean' is to take the median absolute deviation between different azimuthal regions in each radial bin scaled by a factor of 1.4826 to approximate a standard deviation, divided by the square root of the (number of values$-1$). These error bars are included in Figs.\ \ref{fig:Subaru_haloprof}. 

Given how closely these points seem to follow a power law, and given that stellar halos are typically quantified in terms of their power law profiles \citep[e.g.,][]{Bell2008,Merritt2016,Harmsen2017,Jang2020,Gozman2023}, we quantify the stellar halo RGB number $N_{RGB}$ profile as a power law with intrinsic log-normal scatter $N(0,\sigma_{RGB})$ using maximum-likelihood fitting with Markov Chain Monte Carlo minimization (using the package {\it emcee}):
\begin{equation}
N_{RGB} = N_{RGB,25\,kpc} (r_{proj}/25\,kpc)^{\alpha} + N(0,\sigma_{RGB}),  \label{eq:powerlaw}
\end{equation}
with best-fit values of power law slope $\alpha = -2.7\pm0.5$, $N_{RGB,25\,kpc} = 1.11\pm0.06$ stars/kpc$^2$ (number of stars per kpc$^2$ at 25\,kpc radius), and an 84\% upper limit of scatter $\sigma_{RGB}$ of $0.03$ stars/kpc$^2$. This integrates to a total number of stars in between 15 and 40\,kpc of 4500$\pm$300 stars. In order to convert this estimate into a stellar mass, we must assume a stellar population: we assume a 10\,Gyr old, $[M/H] \sim -1.1\pm0.2$ population, consistent with the range of metallicities seen in the halo (\S \ref{subsec:metprof}). Given typical 68\% completeness, we estimate a mass between 15 and 40\,kpc of $log_{10} M_{halo,15-40\,kpc} \sim 8.04\pm0.03$ formal uncertainty; we estimate $8.04\pm0.09$ to account for uncertainties in the conversion of star counts into stellar mass. 

\subsubsection{HST Stellar halo density profile}

While the Subaru data cover a wide area, one could always be concerned about the fidelity of star--galaxy separation. We therefore analyze the HST data --- with its outstanding star--galaxy separation --- as a totally independent constraint on M83's stellar halo. In this analysis, we follow closely the analyses of \citet{Harmsen2017} and \citet{Jang2020}.

As a first step, we attempt to characterize M83's diffuse stellar halo. For this purpose, we select RGB stars in the narrow selection region discussed in \S \ref{sec:HST_CMDs} and shown in Fig.\ \ref{cmds3p}. This restrictive cut avoids red Helium-burning stars on the blue side and focuses on brighter RGB stars to avoid numerous fainter Helium-burning stars with large photometric errors from scattering into our selection. In addition, this selection has a red cut that encompasses all of the metal-poor RGB stars at $r_{proj}>20\,kpc$, while avoiding the numerous metal-rich RGB stars that are so prominent in M83's disk (we will come back to these later). 

We use fields 1--3, 7--9, and 11--14 for this analysis (see Table \ref{tab:obs} and Figs.\ \ref{fig:M83_pano} and \ref{cmds3p}). Fields 4 and 5 are dominated by the Northern Stream. Fields 6 and 10 have dominant contributions from UGCA 365 and dw1335-29 respectively. Fields 9, 13 and 14 have some overlap with the footprint of the Southern Stream (Fig.\ \ref{fig:Stream_Subaru_CMDs}). Omitting these three outermost fields from consideration would give us extremely limited radial leverage to fit a power law profile, so will include them in this analysis here, cognizant that they place an upper limit on the stellar halo density at that radius. We correct our measurements for a (very modest) contribution from background objects in our RGB selection using blank sky fields (following \citealt{Radburn-Smith2011}, \citealt{Monachesi2016_GHOSTS}, \citealt{Harmsen2017}, \citealt{Jang2020}, \citealt{Harmsen2023}). The measurements account for completeness using artificial stars. The error bars incorporate counting and background uncertainties.   

The density profile of the metal-poor halo stars is shown in Fig.\ \ref{fig:HST_radial_mass} in purple. For comparison, in black datapoints with errors, we show the Subaru stellar mass density profile translated into HST equivalent metal-poor RGB star counts by applying HST's RGB star to stellar mass ratio; the star counts are consistent to within their uncertainties. We note that the HST star counts (from Fields 9, 13 and 14; the three purple points in Fig.\ \ref{fig:HST_radial_mass} at $\log_{10}r_{proj}/{\rm kpc} \sim 1.55$, and at 30--40\,kpc in Fig.\ \ref{fig:Subaru_radial_mass}) may include debris from the Southern tidal stream, and so may be biased high. 
Using Eqn.\ \ref{eq:powerlaw}, the HST metal-poor datapoints favor a steep power law finding $\alpha = -4.5\pm0.2$, $N_{RGB,25\,kpc} = 1.02\pm0.14$ stars/kpc$^2$ (number of stars per kpc$^2$ at 25\,kpc radius), with an inferred scatter $\sigma_{RGB}$ of $0.5\pm0.2$ stars/kpc$^2$ (blue line and confidence region). This integrates to a total number of stars in between 15 and 40\,kpc of 5300$\pm$500 stars. This power law is $\sim 3\sigma$ steeper than the Subaru power law profile, which reflects primarily the influence of $r_{proj}<15$\,kpc RGB stars. The outer parts deviate from this power law fit, but have larger uncertainties, so do not significantly affect the fit (except by driving a significant inferred scatter in the relation). Despite this difference in power-law parameters, the HST and Subaru density profiles are remarkably similar, and lend confidence in both detections of M83's metal-poor stellar halo. 

In order to estimate the surface brightnesses and stellar mass densities of M83's outskirts, it is necessary to use stellar population models. In common with our analysis of the halo using Subaru data, we assume a 10\,Gyr old population and $[M/H] = -1.1$, estimating 13800 solar masses per expected HST-detected RGB star (at the present day; the values are different from Subaru because the color--magnitude ranges for selection are different). We assign a systematic uncertainty of 20\%, accounting for a wide range of star formation history and metallicity variations that would affect the CMD enough to cause a significant mismatch with the observed sparse halo CMD (this value is consistent with those assumed by e.g., \citealp{Harmsen2017}). This yields a mass between 15 and 40\,kpc of $log_{10} M_{halo,15-40\,kpc} \sim 7.86\pm0.04$ formal uncertainty, with $\pm0.1$\,dex systematic uncertainty from converting RGB counts into stellar masses. 

\subsubsection{M83's overall surface density profile}

In order to place M83's stellar halo density profile in context, we must use integrated light to quantify the crowded inner parts of M83. We choose to use deep 3.6$\micron$ surface photometry from the S4G survey \citep{Sheth2010,Munoz-Mateos2015}, converting the surface brightness profile into a stellar mass profile   assuming  $M/L \sim 0.6$ and a 3.6$\micron$ magnitude of the Sun of 5.91 in AB magnitudes \citep[following][]{Meidt2012,Cluver2014,Querejeta2015}. 

The resulting profile is shown in Fig.\ \ref{fig:Subaru_radial_mass}, constraining M83's stellar mass density profile to a galactocentric radius of 60\,kpc, and probing a dynamic range of more than a million in surface density. It is clear that the 3.6$\micron$ imaging from S4G, despite its depth, is around 1--2 orders of magnitude too insensitive to pick up M83's low surface brightness, extremely diffuse halo. The transition from a relatively steep exponential profile to a much shallower profile at $r \sim 17$\,kpc is striking. This corresponds to a dramatic change in CMD morphology and therefore stellar populations (compare the left-most panel of Fig.\ \ref{cmds3p} for the outer disk vs.\ the right-most CMD for the composite sparse halo fields). This is clear from the total RGB star counts from HST --- when more metal-rich disk stars (orange region in the left-most panel of Fig.\ \ref{cmds3p}) are included in star counts (red points in Fig.\ \ref{fig:HST_radial_mass} and \ref{fig:Subaru_radial_mass}), one can see that metal-rich stars dominate this steep exponential drop-off, becoming undetectable at $r_{proj} \sim 17$\,kpc. This difference in metallicity between the metal-rich disk of M83 and its metal-poor halo allows a relatively clean separation between the two components.

In an effort to quantify M83's stellar halo with both the Subaru and HST star counts, we fit both datasets jointly, choosing to impose a minimum uncertainty on each datapoint of 0.1\,dex. Fitting it with a power law with intrinsic scatter (Eqn.\ \ref{eq:powerlaw}), we recover $\alpha = -3.4\pm0.3$, $\Sigma_{halo,25\,kpc} = 2.4\pm0.3 \times 10^4 M_{\odot}$/kpc$^2$, with an 84\% upper limit on the inferred scatter $\sigma_{RGB}$ of $500 M_{\odot}$/kpc$^2$. This yields a mass between 15 and 40\,kpc of $log_{10} M_{halo,15-40\,kpc} \sim 8.02\pm0.04$ formal uncertainty; we estimate $8.02\pm0.10$ to incorporate uncertainties from converting RGB counts into stellar masses. 

\subsection{Stellar halo metallicity profile} \label{subsec:metprof}

Given our detection of the stellar halo to $r_{proj}\sim40$\,kpc, we can quantify the stellar halo metallicity profile. Following our analysis of stream metallicities in Section \ref{sec:stream}, we can estimate the run of RGB color-inferred metallicities as a function of radius. These metallicities are much less certain than the overall densities; in particular, they are very weakly constrained for $r_{proj}>40$\,kpc. Uncertainties are calculated as for the density profiles by quantifying the variation between different azimuthal regions in a given radial bin. The results are shown in Fig.\ \ref{fig:Subaru_radial_mh}. The metallicity profile shows a broad decline as a function of radius, although a flattening outside of 20\,kpc is also quite consistent with the Subaru data.

The halo HST fields have few enough stars to make the metallicity measurements very uncertain. We combine halo fields 3 and 8 for the point at $r\sim 20$\,kpc, and fields 9, 13 and 14 for the point at $r\sim 40$\,kpc (all three of these outermost fields should contain some contribution from the Southern Stream). Using the same method we did for the stream HST-derived metallicity \citep{Streich2014,Monachesi2016_GHOSTS}, we find metallicities between [M/H]$\sim -1.2$ and $-1.5$, which are somewhat lower than the Subaru-derived metallicities but consistent within their joint uncertainties (purple data points in Fig.\ \ref{fig:Subaru_radial_mh}. Together, both datasets paint a consistent, if relatively noisy, picture of the metallicity of M83's stellar halo.

As a simple and possibly inadequate description of the broad run of the metallicity, we fit a linear function with log radius, with an intrinsic scatter term, of the form $[M/H] = m (r_{proj}-30\,kpc) + [M/H]_{30\,{\rm kpc}} + {N}(0,\sigma_{[M/H]})$ to all metallicity values (Subaru and HST), where the best fit values  are: $m = -0.014\pm0.006$ dex\,kpc$^{-1}$, the metallicity at 30\,kpc $[M/H]_{30\,{\rm kpc}}=-1.15\pm0.10$, and the intrinsic scatter is $<0.03$\,dex at 84\% confidence.

\subsection{Summary of Observational Results}

We summarize some basic quantities and our measurements of M83's stellar halo and streams in Table \ref{tab:haloprops}. For the diffuse halo properties, we adopt the measurements from the fit to the combined HST and Subaru datasets for both densities and metallicities. For the stream measurements, we adopt masses from isochrone scaling of Subaru star counts to a stellar mass, Subaru [M/H] values for both streams (as an HST estimate is not available in the South), and AGB/RGB star measurements and $t_{90}$ inferences from the HST data. 

Most studies (observational and simulation) do not differentiate between material in diffuse halos and recognizable streams (e.g., for M31, or NGC 891, both of which have prominent streams). Accordingly, for comparison, it is most meaningful to estimate the mass and metallicity of the `aggregate' stellar halo, containing both the diffuse component and the parts of the halo that are in recognizable streams. Because the Northern and Southern Streams are both broadly contained in the footprint 15--40\,kpc, we choose to sum the masses of all three components $M_{halo+stream,15-40kpc} = 8.28\pm0.13$, and compute a weighted average metallicity of all three components [M/H]$_{halo+stream,30kpc} = -1.05\pm0.20$. We report these summed halo masses and averaged metallicities in Table \ref{tab:haloprops}.

Inferred total stellar halo properties are presented also, but are derived in what follows.

\begingroup
\begin{table*}
\begin{center}
\caption {\label{tab:haloprops} M83 (NGC 5236) overall and stellar halo properties.} 
\begin{tabular}{llr}
 \hline\hline
 Property & Value & Source/Reference\\    \hline
 $M_{*,gal}$   & 
 $5.2\pm1.5 \times 10^{10} M_{\odot}$  & \protect\citet{Barnes2014}\\
 $M_{gas}$  &
 $6.05\times10^9$ $M_{\odot}$  &
 \protect\citet{Heald2016}\\ 
 $v_{max}$   & 
 $180\pm20 km/s$  &
 \protect\citet{Barnes2014} \\
 \hline
 $\log_{10} M_{halo, 15-40 kpc}/M_{\odot}$   &  
 $8.02\pm0.10$  &
 This paper; Subaru$+$HST\\ 
 Halo Power Law Slope  &  
 $-3.4\pm0.3$  &  
 This paper; Subaru$+$HST \\ 
 $[$M/H$]$ at 30 kpc   &  
 $-1.15\pm0.1$ &
 This paper; Subaru$+$HST\\
 $d$[M/H]$/dr$   &  
 $-0.014\pm0.006$ dex/kpc  &
 This paper; Subaru$+$HST\\
 \hline
 $\log_{10} M_{Northern\,Stream}/M_{\odot}$  &
 $7.85\pm0.09$  &
 This paper; Subaru\\
 $[$M/H$]_{Northern\,Stream}$  &
 $-0.9\pm0.2$  &
 This paper; Subaru\\
 $\sigma_{[{\rm M/H}],Northern\,Stream}$  &
 $0.40\pm0.05$  &
 This paper; Subaru \\
 $\log_{10} (N_{AGB}/N_{RGB})_{Northern\,stream}$  &
 $-0.78\pm0.04$  &
 This paper; HST\\
 $t_{90,Northern\,stream}$  &
 $2.1\pm1.5$\,Gyr  &
 This paper; HST\\
 $\log_{10} M_{Southern\,Stream}/M_{\odot}$  &
 $7.16\pm0.10$  &
 This paper; Subaru\\
 $[$M/H$]_{Southern\,Stream}$  &
 $-1.1\pm0.2$  &
 This paper; Subaru\\
 $\sigma_{[{\rm M/H}],Southern\,Stream}$  &
 $0.45\pm0.11$  &
 This paper; Subaru \\
 $\log_{10} M_{Combined\,Streams}/M_{\odot}$  &
 $7.93\pm0.10$  &
 This paper; Subaru\\
 \hline
 $M_{halo+stream,15-40kpc}$ & $8.28\pm0.13$ & This paper; Subaru/HST summed \\ 
 $[$M/H$]_{halo+stream,30kpc}$ & $ -1.05\pm0.20$ & This paper; Subaru/HST weighted average \\
 $\log_{10} M_{*,10-40,min}$ & $  8.28^{+0.16}_{-0.24}$ & This paper; Subaru/HST/TNG-50$^*$ \\ 
 $[$M/H$]_{30,min}$ & $-1.25^{+0.30}_{-0.45}$ & This paper; Subaru/HST/TNG-50$^*$ \\ 
\hline
 $\log_{10} M_{*,accreted}$ & $  8.78^{+0.22}_{-0.28}$ & This paper; Subaru/HST$+$models$^\dagger$ \\ 
 $\log_{10} M_{*,dom}$ & $  8.5\pm0.3$ & This paper; Subaru/HST$+$models$^\dagger$ \\ 
\hline
\end{tabular}
\label{haloprops}
\\[1ex]
$^*$ These values and uncertainties are inferred by using the observed halo masses and metallicities of M83 to select suitable analogs in TNG-50; the range of minor-axis equivalent halo masses and metallicities of those analogs are given here. See \S \ref{sec:faceedge}. \\
$^\dagger$ These values and uncertainties are inferred by combining $\log_{10} M_{*,10-40,min}$ with results from a number of stellar halo formation models in a cosmological context. See \S \ref{sec:mergerhist} for details. 
\end{center}
\end{table*}
\endgroup

\section{Discussion}
\label{imp}

In this paper, we have found that M83's diffuse outskirts are resolved into stars using both wide-field ground-based data and more accurate sparse pencil-beam data from HST. Both datasets clearly show a diffuse, metal-poor stellar halo with metallicity [M/H]$= -1.1\pm0.1$ and a density profile $\Sigma_{projected} \propto r_{projected}^{-3.4\pm0.3}$ very different from M83's metal-richer, exponential disk. This envelope has an integrated stellar mass between 15 and 40\,kpc in projected radius of $\log_{10} M_{halo,15-40\,kpc}/M_{\odot} = 8.02\pm0.10$. We also characterized M83's prominent Northern stellar stream ($\log_{10} M_*/M_{\odot} = 7.85\pm0.09$; [M/H]$=-0.95\pm0.20$), finding that it stopped forming stars around 2\,Gyr ago. The Subaru data reveal a new low surface brightness stream to the South of M83, containing about 20\% of the mass of the known Northern Stream, with $\sim 0.15$\,dex lower metallicity; morphologically, this stream appears to be a continuation of the Northern Stream. The `aggregate' stellar halo mass between 15 and 40 kpc is $M_{halo+stream,15-40kpc} = 8.28\pm0.13$, and its weighted average metallicity is [M/H]$_{halo+stream,30kpc} = -1.05\pm0.20$. 

In this section, we will first reflect on the origin of the Southern Stream, both by comparison with possible analogs (e.g., the Sagittarius stream) and a preliminary set of stream models. We will then place M83's halo and stream properties in context with both simulated and observed stellar halos, and will attempt to constrain M83's merger history. With these constraints in hand, we will revisit the possibility of merging playing a role in M83's extended {\sc Hi} disk, its active star formation, and its enigmatic double nucleus. 

\subsection{Constraining the origin and implications of the Southern Stream}\label{sec:dynmodel}

In Section \ref{sec:stream} 
we raised the possibility that the newly discovered Southern Stream  (Fig. \ref{fig:Stream_Subaru_CMDs}, purple) can be associated with the previously detected Northern Stream (see Fig \ref{fig:Stream_Subaru_CMDs}, red). We quantified its mass, metallicity and placed constraints on its star formation history. With these estimates in hand, we can now revisit our tentative attribution of the Southern Stream with an extension of M83's known Northern stellar stream to the South. 

The Southern Stream has only 20\% of the mass of the Northern Stream, and is 0.2\,dex more metal poor. In these properties, we see parallels with the Sagittarius stream, where the trailing tail is significantly (factors of several--five) less dense than the leading tail, and extends to galactocentric radii of $\sim 100$\,kpc \citep{Hernitschek_2017_Sgr}. In addition, if the streams were associated, the 0.2\,dex difference in metallicity between the Northern and Southern Streams would suggest a modest metallicity gradient along M83's stellar stream, in accord with the kind of gradients seen in the Sagittarius tidal stream \citep{Bellazzini2006} and M31's giant stellar stream \citep{Escala2021}. 

It might also be possible that this stream is unassociated with the Northern Stream --- such an interpretation would be suggested if wider coverage showed other streams as possible continuations of the Northern Stream, or if large stellar population differences became apparent. One possibility worth considering is an association of the Southern Stream with the $M_V = -10.1$ satellite dw1335-29 \citep{Muller2015,Carrillo2017}, motivated by an overlap between dw1335-29 and the Southern Stream on the plane of the sky (see Fig.\ \ref{fig:SGCMD}). We tentatively argue against such an interpretation for three reasons. The most compelling reason is that the upper RGB of dw1335-29 is significantly bluer $g-i=1.61\pm0.05$ than the stream $1.78\pm0.05$, implying that dw1335-29 has a {\it lower} metallicity than the stream; this runs counter to other known streams which have more metal-rich regions close to the progenitor \citep{Bellazzini2006,Escala2021}. The stream has 12$\times$ the mass of dw1335-29, suggesting a very advanced stage of tidal disruption. If this were the case, one would expect a closer match between the stream's stellar populations and dw1335-29. Finally, dw1335 does not appear to be elongated along the stream direction, instead showing elongation perpendicular to the stream \citep{Carrillo2017}. While we do not conclusively rule out an association with dw1335-29, such an association appears disfavored by the evidence.

\subsubsection{Preliminary dynamical models for the Streams}\label{sec:model}

\begin{figure}[t]
    \centering
\includegraphics[width=0.95\columnwidth]{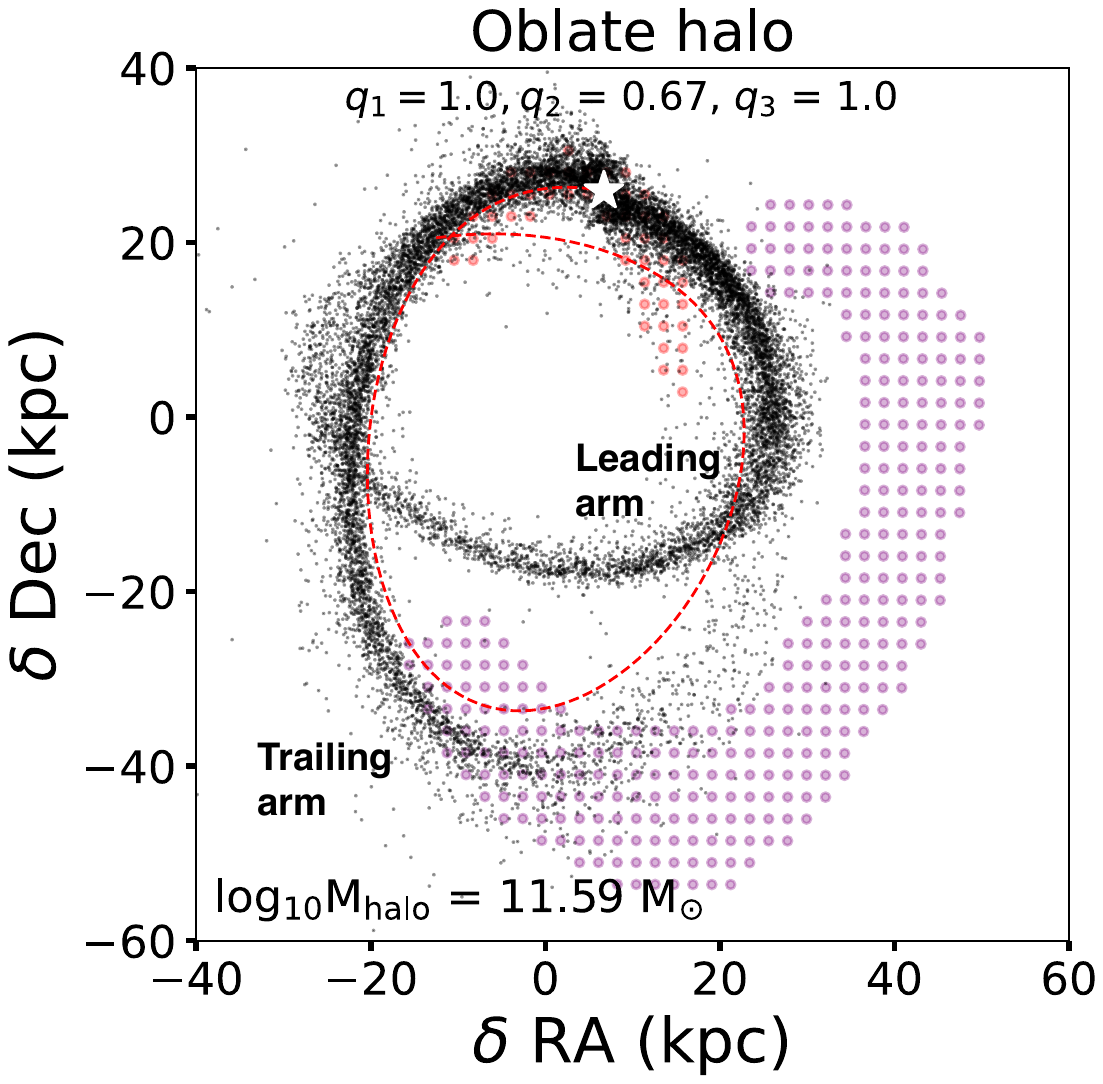}
    \vspace{-1cm}\caption{
    Stream model (black points) evolved in an static oblate dark matter halo with flattening along the line-of-sight (y)-direction. 
    The white star indicates the present day location of the progenitor in the plane of the sky. The red points and purple points shows the northern and southern parts of the stream from Figure \ref{fig:Stream_Subaru_CMDs}. The red dashed lines show the past 1 Gyr of evolution of the progenitor orbit. The stream model is not a perfect match to the data but 
    populates the northern and southern regions of the stream simultaneously. }
    \label{fig:streamfit}
\end{figure}

By attempting to approximately reproduce the Northern and Southern streams simultaneously witn one progenitor, dynamical modeling can also constrain the possible origin of M83's Southern Stream. 
To search for progenitor orbits and dark matter halo configurations that reproduce the two stream segments, we use the custom extragalactic stream sampler from \citet{NibPear2025}. 
We attempt to fit the stream in an NFW dark matter halo \citep{navarro1997}: 
\begin{equation}
\Phi_{\rm NFW} = -\frac{GM_{\rm halo}}{r_s} \frac{{\rm ln} (1+a)}{a}
\end{equation}
where
\begin{equation}
a = \frac{\sqrt{(x/q_1)^2 + (y/q_2)^2 + (z/q_3)^2}}{r_s}
\end{equation}
and where $r_s$ is the scale radius of the halo, 
$q_1$, $q_2$, $q_3$ is the flattening in the $x,y,z$ directions, and $M_{\rm halo}$ is the dark matter halo mass. Here, ($x, z$) is the sky plane (equivalent to ($\delta$RA, $\delta$Dec) in Fig. \ref{fig:Stream_Subaru_CMDs}) and $y$ the line-of-sight direction.
In this work we fix $q_1 =1$ and $q_3=1$ and only explore flattening along $q_2$, the direction along the line of sight, perpendicular to M83's disk. 
We also include a \citet{Miyamoto1975} disk representing the stellar disk of M83, where we fix $M_{\rm disk} = 5.2\times 10^{10}$ M$_{\odot}$, the disk scale length = 2.5 kpc and scale height = 0.2 kpc.

In this proof-of-concept modeling, we fix $r_s = 22$, $c = 15$ and limit our parameter search and explore progenitors with log$_{10}$m$_{\rm *,prog}$ = $7.5-8.5$ M$_{\odot}$ and a wide range of dark matter halos with log$_{10}$m$_{\rm DM,halo}$ = $11.25-12$ M$_{\odot}$ allowed by the baryon fraction. We fix the on-sky position ($x,z$) of the progenitor (following \citealt{Mueller2025}, their Table 1), leaving other parameters free.  

In Figure \ref{fig:streamfit}, we show an example orbit which generates a stream that reproduces parts of the Northern and Southern streams simultaneously. The white star shows the progenitor location at present day, and the red dashed line shows the past 1 Gyr of the 5 Gyr evolution for the progenitor orbit. We overplot the model stream (black) to the red and purple regions from Fig. \ref{fig:Stream_Subaru_CMDs}. 
The Northern Stream consists of leading arm debris which was stripped more recently, and the outer parts of the trailing stream line up with the Southern Stream. An oblate halo with $q_2=0.67$, flattened in the disk plane (see e.g., \citealt{Chua2019}), is necessary to broadly reproduce the two arms simultaneously.  
The present-day phase space coordinates of the progenitor in M83's galactocentric frame are: ($x,y,z$) = ($6.7,-13.6,25.9$) kpc, and ($v_x,v_y,v_z$) =  ($ 121.4, -101.2,  -24.4$) km s$^{-1}$, and log$_{\rm 10}$m$_{\rm prog} = 8.2$ M$_{\odot}$. 

It is promising that this preliminary investigation produces a stream model which fits the broad morphological features of the Northern and Southern Streams simultaneously. Because the stream probes more than one full wrap, the streams are sensitive to the potential at a wide range of galactocentric radii, promising to constrain the dark matter halo parameters \citep{Johnston2001,belokurov2014,hendel2015,walder2024,NibPear2025}. Further modeling (S.\ Pearson et al., in prep.), involving triaxiality \citep[e.g.,][]{Law2009}, dynamical friction \citep[see e.g.,][]{Fardal2019} and a time-dependent growing potential \citep{buist2015,Lilleengen2023,Brooks2025}, is warranted to more faithfully reproduce the stream morphologies.

\begin{figure*}\centering
	\includegraphics[width=0.95\linewidth]{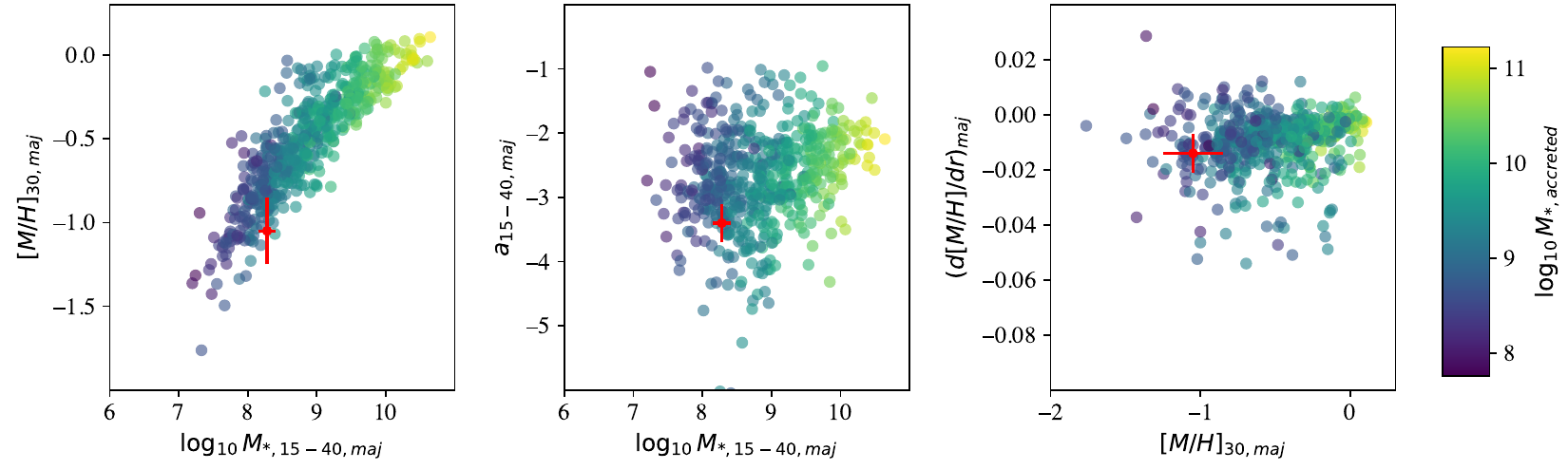}   
    \caption{ Left: The run of stellar halo metallicity at 30\,kpc projected face-on radius as a function of stellar mass in accreted stars between 15 and 40\,kpc in face-on projected radius for stellar halos in the TNG-50 simulation suite. Center: Power law slope of the density profile against 15--40\,kpc halo mass. Right: metallicity gradient as a function of metallicity at 30\,kpc. Colors denote the total accreted mass of the TNG-50 halo.  Overplotted in all panels are the measurements for M83, with uncertainties. }
    \label{fig:M83_TNG50}
\end{figure*}

\subsection{Comparison to the TNG-50 galaxy simulation suite} \label{subsec:m83-TNG}

Because M83's halo is face-on, and most other measured halos are characterized around edge-on galaxies, we choose to take a somewhat unconventional, but principled, approach to this comparison. We choose to first explore stellar halos around massive galaxies in the TNG-50 simulation, comparing their properties as observed from the face-on and edge-on orientations. In this way, we not only place M83 in context with predicted halos, but are able to account for the predicted differences in halo measurements that are due entirely to viewing angle when comparing with the wider halo population. This enables a more robust and ideally unbiased comparison with previously-observed stellar halos. 

\subsubsection{Measuring the stellar halos and growth histories of TNG-50 galaxies}

The TNG-50 simulation is well-suited for this work, combining a sizeable cosmological volume (51.7 Mpc)$^3$ \citep{Pillepich2019} containing hundreds of hosts with similar stellar masses to the Milky Way \citep{Pillepich2024}, but with resolutions approaching those achieved by zoom-in simulations (baryonic mass resolution of $\sim 8.5 \times 10^4 M_{\odot}$ and spatial resolutions $\sim$300pc). With this baryonic resolution, even low-mass stellar halos with masses $\sim 10^8$ to $10^9$ solar masses have thousands of star particles, ensuring the ability to measure stellar halo properties such as masses or metallicities spatially resolved into regions, or measure the gradients of these quantities. 

We select a wide range of galaxies for consideration, spanning stellar masses (within a 30\,kpc aperture) of  $10^{10}M_\odot \le M_{\ast, \text{30kpc aperture}} \le 2\times10^{11}M_\odot$. This stellar mass interval contains most of the observed systems with measurements of stellar halo properties. We selected halo dark matter (DM) masses ($M_{halo}$)  $M_{200c}\le10^{13}M_\odot$. The intention of this upper bound was to prevent contamination of the sample by MW-mass DM subhalos from larger groups and clusters. This cut was largely successful, but in practice some such subhalos entered the sample and are omitted at a later step. No further constraints are imposed on $M_{200c}$ for observationally motivated reasons: it is difficult to determine and constrain the halo dark matter mass of MW-like galaxies in observational data.

For the purposes of characterizing stellar halos, we choose to compare with {\it accreted} particles only. Our primary reason for doing this is observational: with resolved stars, stellar populations information can be used to effectively  distinguish between {\it in-situ} and accreted stars, particularly at large projected galactic radii $r_{proj}>10-15$\,kpc \citep[e.g.,][]{Harmsen2017,Jang2020,Gozman2023}. In addition, many current hydrodynamical models, including the Illustris TNG suite, produce an excess of relatively metal-rich stars at 10--40\,kpc radius compared to observations \citep[e.g.,][]{Monachesi_2016_Auriga,Harmsen2017,Monachesi_2019,Merritt2020,Wright2024}; in all cases, this excess is attributed to metal-rich {\it in-situ} stars formed in the potential well of the main galaxy. This overproduction is particularly sensitive to model details \citep{Zolotov2009,Pillepich2015}, and therefore varies dramatically from model to model. In contrast, predictions for accreted halos are much more robust, depending only on the well-constrained halo occupation of satellites (which is constrained by the luminosity function of the whole simulation, and is therefore close to correct for many simulations) and halo mergers, which are a robust prediction across simulations. Indeed, the predictions for accreted halos are very stable across simulations, including even particle tagging techniques using halo occupation or semi-analytic techniques \citep{BullockJohnston2005,Cooper2010,Pillepich2015,Deason2016,Monachesi_2016_Auriga,DSouza2018_MNRAS,Monachesi_2019}, and these predictions agree very well with observations \citep{DSouza2018_MNRAS, Monachesi_2019}. 

We adopt accreted masses from \citet{Rodriguez-Gomez2016}, and tag individual particles as accreted following \citet{DSouza2018_MNRAS}. In brief, particles formed in the main progenitor branch of the galaxy (using the SUBLINK merger trees; \citealt{Rodriguez-Gomez2015}) are deemed to be {\it in-situ} stars. The other stellar particles found in the subhalo of the main galaxy are tagged as ‘accreted’; this definition is consistent with \citet{Rodriguez-Gomez2016}. 
 
Using the tagged particle data, we create a wide range of metrics that are directly comparable with resolved-star studies. In particular, we construct mock accreted masses in the 10--40\,kpc radial range $M_{*,10-40,min}$ for edge-on galaxies following \citep{Harmsen2017}, in integrating the density profile at minor axis radii of 10--40\,kpc, accounting for the projected ellipticity of the edge-on halo $b/a$. For face-on galaxies, we integrate the accreted mass between 15--40\,kpc for the face-on projection, denoted $M_{*,15-40,maj}$. Metallicities at 30\,kpc projected distance were calculated, as well as the metallicity gradient in the 10--40\,kpc range. Power laws were fit to the (either minor axis or face-on) density profile, giving the power law density profile slope $a$. We calculate $t_{90}$, the time before which 90\% of the accreted stars in that aperture were formed, as an approximate measure of when star formation in the disrupting satellite galaxies ceased, for comparison with approximate comparison with estimates of $t_{90}$ from the ratio of AGB/RGB stars \citep{Harmsen2023}. 
For the purposes of constraining the properties of all accreted stars, or the most massive merger event, we measure the mass and time of the most massive merger event $M_{*,dom}$ and $t_{merg,dom}$, as well as adopting the stellar mass of accreted star particles $M_{*,acc}$, adopted from \citet{Rodriguez-Gomez2016}. 

One limitation that is worth noting immediately with TNG-50 galaxies is that, for $M_* <10^{10} M_{\odot}$, they tend to be more metal-rich for a given stellar mass than the observed galaxy population. For example, \citet{Sarmiento2023} perform extremely careful comparisons of mock MANGA datacubes with MANGA, finding that the inner $R_e$ (containing 50\% of the stars of a galaxy) of $M_* < 10^{10} M_{\odot}$ galaxies appear 0.2--0.4\,dex too metal rich for their stellar mass. Such offsets are to be expected in such complex modeling efforts, as they represent the complex combination of star formation histories, metal production (including yields), outflows and mixing of gas --- indeed, that TNG-50's metallicities within a factor of 2 or 3 of the observed galaxy population is a significant model achievement. Nonetheless, for our purposes it is important to remember this offset, as we will see it reflected in an offset between TNG-50 accreted star metallicities and observed metallicities from resolved star observations of halos.  

\subsubsection{Comparing M83's halo to TNG-50}

We show the comparison of M83's major axis halo mass, metallicity, and density/metallicity gradients (red point with error bar) with the accreted halos of TNG-50 galaxies (circles, color-coded with total accreted mass) in Fig.\ \ref{fig:M83_TNG50}. Broadly, M83's stellar halo properties are within the envelope of halos expected for $M*>10^{10} M_{\odot}$ galaxies. As expected, M83's halo metallicity lies a little low compared to TNG-50's, reflecting the overall offset of TNG-50 towards high metallicity at a given stellar mass, although its metallicity gradient is reasonably typical for TNG-50 accreted halos. Both the mass and density profile of M83's halo are low compared to most TNG-50 accreted halos in this broad host galaxy mass range $M*>10^{10} M_{\odot}$, with only 20\% of accreted halos having lower mass than M83's, and 15\% of accreted halos having steeper density gradients than observed in M83.

\subsection{Comparison of M83's stellar halo with other observed stellar halos} \label{sub:m83_others}

\subsubsection{Accounting for expected differences between face-on and edge-on halos} \label{sec:faceedge}

\begin{figure}\centering
	\includegraphics[width=0.95\columnwidth]{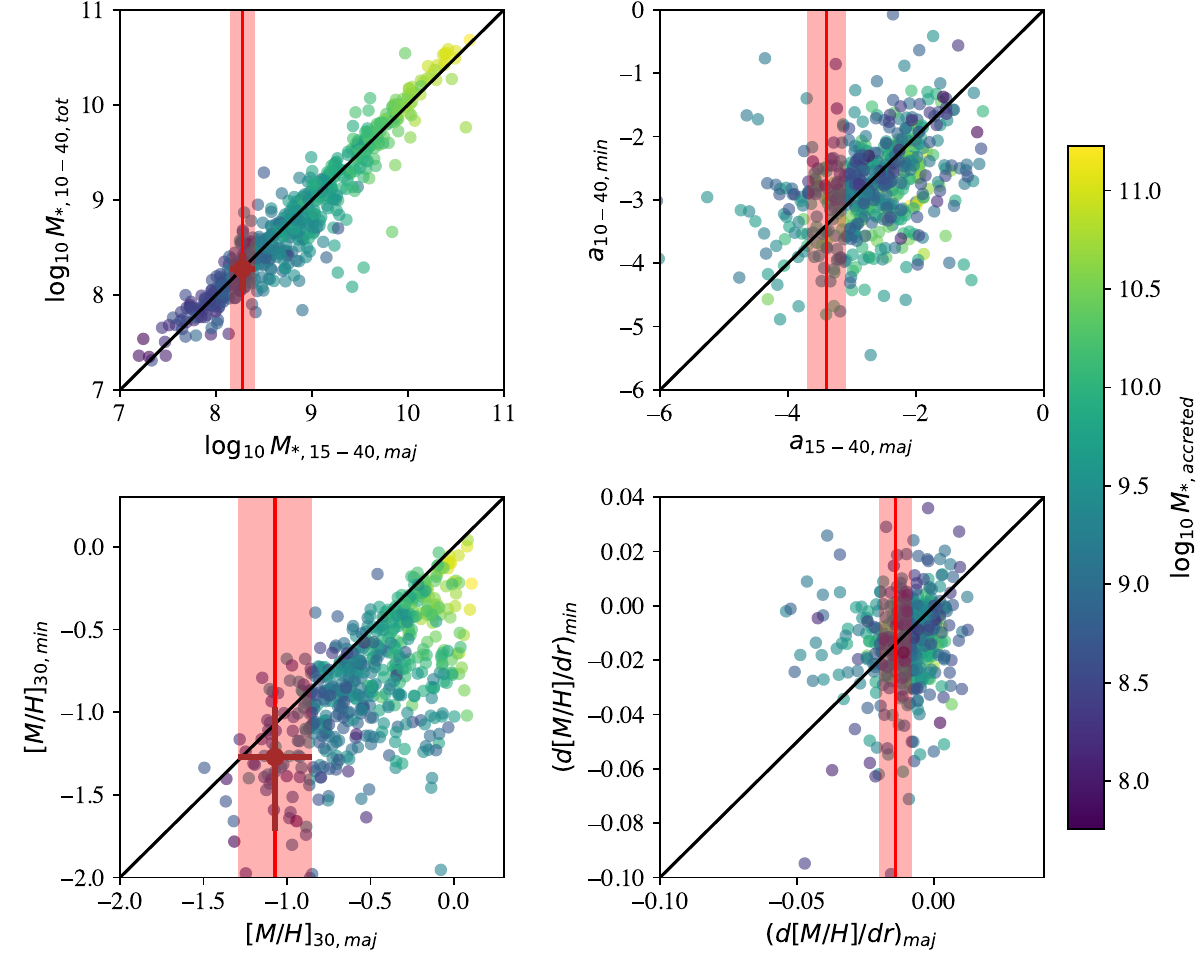}   
    \caption{Predicted relationship between face-on and edge-on accreted quantities, from TNG-50. Top left: stellar halo aperture masses. Top right: stellar halo density power law slopes. Bottom left: stellar halo metallicities at 30\,kpc. Bottom right: stellar halo metallicity gradients. In all panels, measurements from M83 are shown as red lines with shaded regions denoting uncertainties. In the left-hand panels, we show the estimates for edge-on stellar halo measurements of `aggregate' stellar halo mass and metallicity with the brown data points and error bars. The line of equality is shown in black. Gradients between face-on and edge-on are similar; masses are similar and metallicites tend to be lower in the edge-on orientation. The TNG-50 symbols are color-coded by the log of total accreted mass.  }
    \label{fig:face_edge}
\end{figure}

Most other stellar halo measurements have been performed for edge-on galaxies. While we could attempt to directly compare the measurements for face-on and edge-on halos, we choose instead to apply corrections from the TNG-50 simulation suite to give the likely edge-on halo measurements for halos with measurements consistent with M83's face-on halo. 

Fig.\ \ref{fig:face_edge} shows the relationship between face-on and edge-on quantities for the sample of TNG-50 accreted halos. Gradients are very broadly similar (density and metallicty) between face-on and edge-on, but with scatter that approaches the dynamic range of the measurements. The minor-axis derived masses are generally quite similar to the face-on derived masses: $\log_{10} M_{*,10-40,min} \sim \log_{10} M_{*,15-40,maj} +0.0^{+0.1}_{-0.2}$. The metallicities for edge-on halos are lower than the face-on case, $[M/H]_{30,min} \sim [M/H]_{30,maj} -0.2^{+0.2}_{-0.4}$. On this basis, and adding the observational errors in quadature, we estimate an equivalent edge-on halo mass for M83 (accounting for both the diffuse halo and streams) of $\log_{10} M_{*,10-40,min} \sim  8.28^{+0.16}_{-0.24}$ and $[M/H]_{30,min} \sim -1.25^{+0.30}_{-0.45}$; we report these derived values in Table \ref{tab:haloprops} and show then in Fig.\ \ref{fig:face_edge} with brown error bars. Given that the scatter between face-on and edge-on gradients is so substantial, we refrain from estimating the edge-on density and metallicity gradients.

\subsubsection{Comparison with resolved star measurements of other stellar halos} \label{sec:comphalo}

\begin{figure}\centering
	\includegraphics[width=0.98\columnwidth]{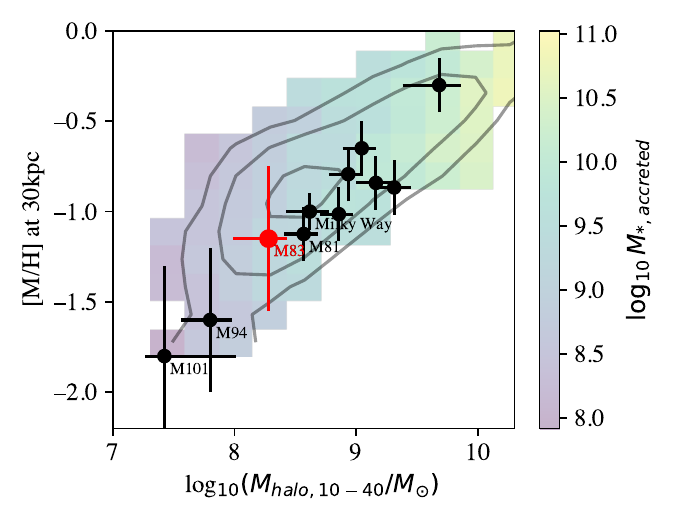}    
    \caption{We compare M83's stellar halo with the Milky Way, M31 and eight other MW-mass galaxies with resolved stellar populations \protect{\citep{Harmsen2017,Jang2020,Gozman2023}}. M83's halo stellar mass and metallicity fit in with the stellar halo properties of other MW-mass galaxies, but filling out the low-mass and low-metallicity end of the stellar halo population, similar to M81, the Milky Way, M101 and M94. Contours show the distribution of TNG-50 stellar halos, with the median total accreted mass color-coded in the background. }
    \label{galaxycomp}
\end{figure}

We compare M83's stellar halo with those of ten other galaxies --- the Milky Way, M31 and eight other galaxies with resolved star measurements. In what follows, we will follow \citet{Smercina2020}, \citet{Jang2020} and \citet{Gozman2023} in focusing on the stellar halo masses and metallicities, and their correlation with each other --- there are no other correlations of note between halo or galaxy parameters (e.g., \citealt{Harmsen2017}). Because most of these systems are highly inclined, we will compare values of $\log_{10}M_{*,10-40,min}$ and [M/H]$_{30,min}$, which are observed for most systems and inferred for the face-on systems M83, M94 and M101. Recall that M83's inferred values account for contributions both from the diffuse stellar halo and its stream. 

Values for the Milky Way, M31, M81, NGC 253, NGC 891, NGC 4565, NGC 4945 and NGC 7814 are taken from \citet[their section 7.1]{Harmsen2017}, with some updates. The MW’s stellar halo mass and metallicity are taken from \citet{Mackereth2020} and \citet{Conroy2019}, respectively. We adjust the metallicity of M31's stellar halo from \citet{Harmsen2017} to [M/H] $\sim -0.3$ following \citet{DSouza2018_M31}.

M94's stellar halo measurements are also from the face-on orientation, so we attempt to translate M94's face-on estimates of $M_{*,10-40,maj} = 9.3\times10^7 M_{\odot}$ (we adopt $\pm0.1$\,dex uncertainties) and $[M/H]_{30,maj}=-1.4\pm0.1$ \citep{Gozman2023} to edge-on values using TNG-50 following \S \ref{sec:faceedge}. We find that the relationship between $M_{*,10-40,maj}$ and $M_{*,10-40,min}$ depends mildly on the major axis power law profile slope. Accordingly, restricting our attention to low-mass TNG-50 halos with $log_{10} M_{*,10-40,maj}/M_{odot}<8.5$ and $\alpha_{maj}=-2.61\pm0.17$, we find $\log_{10} M_{*,10-40,min}/M_{*,10-40,maj} = -0.2\pm0.1$.  On that basis, we infer an equivalent edge-on accreted mass in 10-40\,kpc radius range of $\log_{10} M_{*,10-40,min}/M_{\odot} = 7.8\pm0.2$. The metallicity is evaluated at 30\,kpc projected radius, therefore, $[M/H]_{min} = -1.6_{-0.45}^{+0.30}$. 

M101's radial coverage is very different from M83 or M94. Accordingly, we adopt their total stellar halo mass and scale down to a 10--40\,kpc aperture stellar halo mass. We adopt a total stellar halo mass for M101 of $8.3\times 10^7M_\odot$ following the exponential disk + power-law halo fit of \citet{Jang2020}. We adopt their upper limit on halo mass of $M_{halo,M101} \le 3.2 \times 10^8$ $M_{\odot}$ as the error bar in the positive direction, making the final value and uncertainties $M_{halo,M101}=8.3_{-2.6}^{+24}\times 10^7M_\odot$. We plot these values multiplied by 1/3 to approximate the 10--40\,kpc aperture masses plotted in Fig. \ref{galaxycomp} following \citet{Harmsen2017}. We adopt a metallicity of M101's halo of [M/H]$_{30,maj}$ $\sim -1.6^{+0.3}_{-0.6}$, adjusting the [Fe/H] estimate of \citet{Jang2020} to [M/H] \citep{Streich2014}. To this value we apply the average offset of between face-on and edge-on metallicity of $-0.2$\,dex; thereby estimating [M/H]$_{30,min}$ $\sim -1.8^{+0.4}_{-0.6}$.  

Fig.\ \ref{galaxycomp} shows the result of this comparison. As hinted at in Fig.\ \ref{fig:M83_TNG50}, where M83's halo was in the range of but towards the low mass side of TNG-50 predicted accreted stellar halos, M83's halo lies within the range of properties seen in the stellar halos of other nearby, MW-mass galaxies. Yet, M83's halo is amongst the low mass and low metallicity subset of the stellar halos of other comparably-massed galaxies. Such relatively low mass halos include M101, M94, M83, M81 and the Milky Way --- M83 is in the middle of that range. 

Importantly, M83 lies on the the stellar halo mass-metallicity relation revealed by (\citealt{Harmsen2017}; Fig.\ \ref{galaxycomp}). While these galaxies carry larger uncertainties owing to their face-on orientation (and low mass stellar halos), M83, M94 \citep{Gozman2023} and M101 \citep{Jang2020} extend this relationship towards low halo masses and metallicities, confirming that it continues broadly as predicted \citep{Cooper2010,Deason2016,Harmsen2017,DSouza2018_MNRAS,Monachesi_2019} towards low metallicities for lower stellar halo mass. A tight correlation between stellar halo metallicity and mass is a robust expectation of models in which stellar halos are formed by the disruption of dwarf galaxies in a cosmological context \citep{BullockJohnston2005,Cooper2010,Deason2016,Harmsen2017,DSouza2018_MNRAS,Monachesi_2019}, reflecting the mass-metallicity relation of the progenitor dwarf galaxies and the tendency of most halos to have their masses and metallicities dominated by the one (or at most few) most massive accreted galaxies \citep{DSouza2018_MNRAS,Monachesi_2019}. This is reflected in the predictions from TNG-50. The trend in minor axis metallicity with aperture halo mass in TNG-50 appears broadly similar to the data, matching well at high mass with a rather shallower slope, resulting in a significant $\sim0.3$\,dex offset for low mass halos similar to M83.

\subsection{What does this tell us about M83's merger and accretion history?} \label{sec:mergerhist}

Our measurements of stellar masses and metallicities of both M83's overall stellar halo and stellar stream can provide considerable insight into M83's merger and accretion history --- M83 has experienced an rather quiet (for MW-mass galaxies) merger and accretion history. 

Previous work shows that measurements of outer stellar halo masses can strongly constrain the total accreted stellar mass \citep{Harmsen2017,Bell2017,DSouza2018_MNRAS,DSouza2018_M31,Jang2020}.  In particular, \citet{Harmsen2017} and \citet{Bell2017} find that a number of simulation approaches find that the total accreted mass is three times larger than the 10--40\,kpc summed aperture halo mass ($\log_{10} M_{*,10-40,min}/M_{\odot} = 8.28^{+0.16}_{-0.24}$), with a $\pm40$\% uncertainty. Consequently, we estimate M83's total accreted stellar mass to be $\log_{10} M_{*,accreted} = 8.78^{+0.22}_{-0.28}$.  

While stellar halos form from the disruption of many dwarf galaxies \citep{Bullock2001,BullockJohnston2005,Cooper2010}, their total stellar mass and metallicity is dominated by a small number of mergers/accretions \citep{Deason2016,DSouza2018_MNRAS,Monachesi_2019}, with many halos gaining more than half of their mass in a single event \citep{Deason2016,DSouza2018_MNRAS}. While this is most true for high-mass halos (like e.g., M31's; \citealp{DSouza2018_M31}), it broadly applies even for low-mass halos like M83's or the MW's (see, e.g., \citealp{Belukorov_2018,Helmi_2018} for arguments that the MW's halo is dominated by a single accretion event), and the total accreted mass constrains the expected mass of the most important past merger.  Following \citet{DSouza2018_MNRAS} and \citet{Gozman2023}, the stellar mass of M83's most important (dominant) merger is expected to be $-0.3^{+0.2}_{-0.1}$\,dex lower in mass than the total accreted mass. We thus infer that M83's dominant merger had a stellar mass of $\log_{10} M_{*,dom}/M_{\odot} = 8.5\pm0.3$, similar to the present-day stellar mass of the Small Magellanic Cloud ($\log_{10} M_{*,SMC}/M_{\odot} \sim 8.7$; \citealt{McConnachie2012})
\footnote{We note that these inferences are the same (to within the errorbars) as those that we would determine using simulation-based inference techniques on the TNG-50 simulation (E.\ Bell et al., in preparation; see e.g., \citealt{Sante2025} for an application of these techniques to the Milky Way's stellar halo). }. 

In the case of M83, we have more information about its merger history as we see a stellar stream from a recent large accretion. If we assume that M83's streams both stem from a common progenitor, the combined stellar mass would be $\log_{10}(M_*/M_\odot) \sim 7.93\pm0.1$ (or roughly $8.5\pm2 \times 10^7 M_{\odot}$). As aged by the ratio of AGB to RGB stars in the Northern Stream, star formation in this satellite stopped around $t_{90} \sim 2$\,Gyr ago, placing an upper limit on when star formation in the satellite stopped (as $t_{90}$ necessarily is earlier than the time at which star formation is quenched completely). In models of satellite accretion, star formation stops at the time that the main body of the satellite is disrupted \citep[e.g.][]{DSouza2018_M31}. Consequently, we tentatively suggest that this satellite was disrupted $\lesssim2$\,Gyr ago; broadly consistent with an independent estimate of $\sim 1$\,Gyr from the morphology of the stream from \citet{Barnes2014}.

\begin{figure*}\centering
	\includegraphics[width=0.95\linewidth]{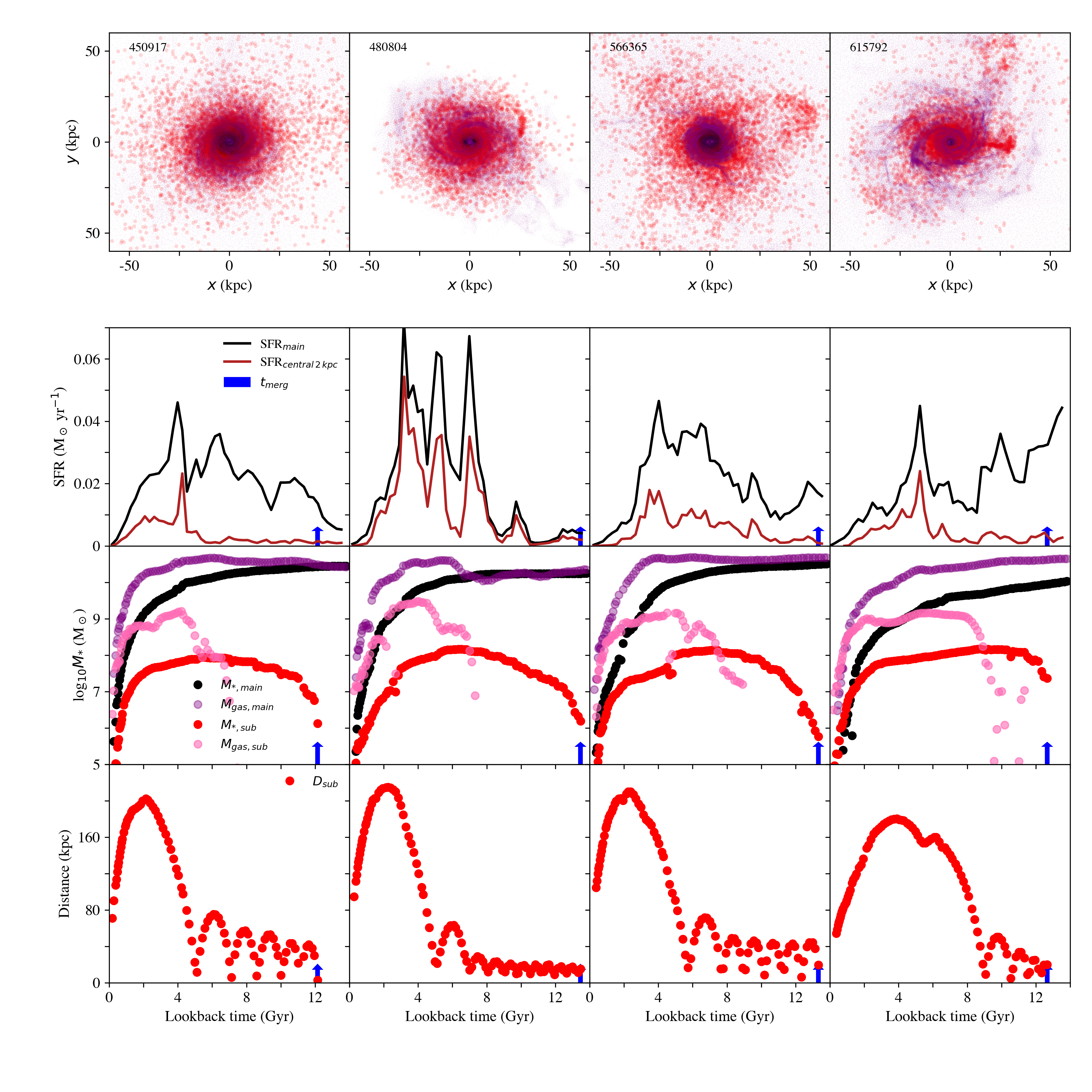}    
    \caption{Four plausible M83 analogs from TNG-50 that share many of the same halo and merger features. Top panels: Black points are in situ particles; these are confined to small radii. Red circles show accreted particles, which show both extended accreted halos and in 3/4 cases tidal features, including streams with many outward similarities to M83's large tidal stream. Purple shows gas; some analogs show extended gas disks reminiscent of M83. 2nd row: SFH of the main galaxy (black) and central 2\,kpc (dark red), with the merger time of the recent accretion shown as a blue arrow. 3rd row: the stellar and gas masses of the main galaxy (black and purple) and the recent merger partner (red and pink). 4th row: The distance of the recent merger partner from the main galaxy. }
    \label{fig:m83analogs}
\end{figure*}

With this information in hand, for illustrative purposes, we can attempt to isolate some plausible M83 analogs from TNG-50. In addition to our overall selection for TNG-50 analogs ($\log_{10}M_{200}/M_{\odot}<13$ and $10<\log_{10}M_{*}/M_{\odot}<11.3$), we require a face-on halo aperture mass close to M83's observed value, $|\log_{10}M_{*,15-40,halo+stream}/M_{\odot} - 8.28| < 0.3$, a merger time for the dominant merger $T_{dom}>4$\,Gyr, and a recent merger with mass $|\log_{10}M_{*,recent}/M_{\odot} - 7.93| < 0.3$ and $T_{recent}<2$\,Gyr. Four TNG-50 halos satisfied these criteria, shown in Fig.\ \ref{fig:m83analogs}. The top panels show that these analogs share many outward similarities with M83, showing an accreted halo that extends well outside the main disk of the galaxy, and some of them showing extended gas distributions. Three out of the four analogs show prominent tidal streams, including loops that appear very similar outwardly to M83's massive tidal stream. We will explore the properties of these analogs in the next section.

\subsection{What does this tell us about the origin of M83's starburst and {\sc Hi} disc?}

We return to one of our central questions: `Is there evidence for a massive accretion/merger origin for M83's unusually massive {\sc Hi} envelope and nuclear starburst, or must other drivers be sought?' Recall that M83's {\sc Hi } disk gives a sense of scale of the size of the relevant progenitor: with a mass in the outer disk of $6 \times 10^9 M_{\odot}$ \citep{Heald2016} and a metallicity of around 1/3 solar \citep{Bresolin2009}, a stellar mass at least  comparable to the LMC ($\sim 5 \times 10^9 M_{\odot}$) might have been expected. M83's stellar halo would be the repository for this debris; consequently, one might have expected a stellar halo mass $\sim 5 \times 10^9 M_{\odot}$ and metallicity [M/H]$\sim -0.5$.  

Instead, we have found that M83's halo has --- compared both to this expectation and to other nearby MW-mass galaxies --- a factor of 10 lower mass than this expectation, with only 1/3 of this expected metallicity. M83's most massive merger event should have a stellar mass of $\sim 3\times10^8 M_{\odot}$, and a metallicity similar to M83's halo metallicity $[M/H]\sim-1.15$, and by virtue of a lack of prominent AGB stars appears to have happened a long time ago. M83's most massive merger event was both (much) too low mass and too long ago to affect M83's present day {\sc Hi} disk and SF properties.  

In M83's stellar stream, we do know that a $M_* \sim 8.5\times10^7 M_{\odot}$ satellite with [M/H]$\sim -1$ was disrupted $<2$\,Gyr ago. While the time frame of this accretion makes it seem possible that this accretion could be associated with the acquisition of a large gas disk and triggering of a nuclear starburst, the stellar mass and stellar metallicity of this accreted dwarf appear far smaller than those required to be responsible for M83's massive, metal-rich extended gas disc. 

It is interesting to turn to the M83 analogs from TNG-50 to ask if large amounts of gas delivery or starburst triggering are expected for such low-mass accretions. These questions are explored in the 2nd, 3rd and 4th row of panels in Fig.\ \ref{fig:m83analogs}. 

The second row shows the star formation history of the four M83 analogs. The total SFH is shown in black; in dark red is the SFH of stars within 2 projected kpc of the center of the M83 analog. The merger time for the recent, $M_* \sim 10^8 M_{\odot}$ accretion is shown by the blue arrow. Strikingly, it is clear that the final disruption of the satellite within the last 2\,Gyr does not result in a significant enhancement in either the global (black) or central (dark red) SFR in TNG-50, with no more than a $<25$\% enhancement in SFR seen at the merger time for all analogs. Assuming that TNG-50 captures some of the relevant physics, this supports the idea that M83's starburst is unlikely to be driven by the same late-stage merger that made M83's stream (a similar conclusion was reached for Sagittarius-stream analogs selected from TNG-50; \citealt{Semczuk2025}). 

The bottom panels of Fig.\ \ref{fig:m83analogs} give some interesting insight into the role of the stream progenitor in sculpting M83's massive, extended disk. In all cases, the stream progenitor brings in $\sim 10^9 M_{\odot}$ of gas into the M83 analog (pink symbols in 3rd row of panels) --- a substantial amount, but a factor of several too little to create M83's $M_{HI} \sim 6\times10^9 M_{\odot}$ metal-rich disk. In at least one case, that delivery of gas might have shaped the main galaxy's SFH (i.e., the enhancement of star formation $\sim 5$\,Gyr ago in subhalo ID 615792). 

Yet, we conclude that this satellite accretion could not meaningfully contribute to M83's present-day metal-rich massive {\sc Hi} disk. In TNG-50, none of this gas remains bound to the satellite past the satellite's second or third pericenter passage $>$3\,Gyr ago --- well before the present day, and well before final disruption of the satellite $<2$\,Gyr ago. Most of this gas was in the satellite's circumgalactic medium and was stripped as the satellite approaches its first pericenter ($\sim90$\% of the gas). We recognize that M83's stream has $t_{90}\lesssim2$\,Gyr ago, and so kept at least a modest amount of gas for much longer than all of the TNG-50 analogs. The hydrodynamics of gas stripping is notoriously sensitive to physics on small scales that are not well resolved in even high-resolution simulations like TNG-50 \citep[see e.g.,][for a discussion of this issue focused on the modest recent star formation seen in the Fornax and Carina dwarf satellites of the MW]{Engler2023}. We suggest that observed systems with $\sim 10^8 M_{\odot}$ are a better guide (e.g., Fornax or Sgr), finding that they have a modest amount of star formation that continues to enrich the galaxy, particularly in $\alpha$ elements \citep{Hasselquist2021}. At any rate, the amount of gas that remains after first pericenter is $\lesssim10^8 M_{\sun}$, a factor of 50 or more insufficient to form M83's extended, metal-rich star forming {\sc Hi} disk.

Consequently, we conclude that M83's extended {\sc Hi} disk and nuclear starburst {\it did not} result from a large merger event. Instead, other non-merger mechanisms, such as secular evolution (e.g., \citealp{KormendyKennicutt2004}) or  accretion of gas from the IGM (e.g., \citealp{Putman_2012}) are likely to be responsible for M83's unique features. 

\section{Conclusions}

Motivated by a desire to understand whether or not mergers and accretions play an important role in driving M83's nuclear starburst and contributed to its massive $M_{HI}\sim6\times10^9M_{\odot}$, metal-rich extended {\sc Hi} disk, we carried out a comprehensive resolved-star study of M83's stellar halo. Stellar halos contain much of the debris from past mergers and accretions, and by studying this debris in concert with hydrodynamical model suites we can gain insight into M83's past merger history. 

Our combination of fourteen narrow-area HST fields with wide-area Subaru observations of $\sim2/3$ of M83's stellar halo was able to both provide accurate stellar population measures in the HST fields and inform $k$-nearest neighbor supervised star--galaxy separation that yielded modest contamination by background galaxies, but was $66$\% complete for bright, metal-poor RGB stars in M83's halo. 

The relatively uncrowded HST data is able to resolve the outskirts of M83's metal-rich outer disk, which drops as a steep exponential profile and becomes undetectable at $r\sim17$\,kpc. Outside this radius, both datasets reveal that M83 has an extended, very low density `smooth' aggregate stellar halo of old and metal-poor [M/H]$\sim -1.15$ RGB stars. Our HST data can follow the metal-poor population to smaller radii, and we estimate a mass in this aggregate stellar halo between 15 and 40\,kpc of $\log_{10} M_{*,15-40,maj}/M_{\odot} = 8.02\pm0.10$.  

Our Subaru and HST data strongly detect M83's well known tidal stream to the north, with a stellar mass of $\log_{10} M_{Northern\,stream}/M_{\odot} = 7.85\pm0.09$ and metallicity [M/H]$= -0.95\pm0.20$. Our HST data reveals a prominent AGB star population in the Northern part of the stream, implying a recent shut-off in star formation with a star formation shutoff time $\tau_{90} = 2.1\pm1.5$\,Gyr ago, suggesting that its progenitor was only recently accreted. Our ground-based Subaru imaging also detects a diffuse extended stream to M83's south, with mass $\log_{10} M_{Southern\,stream}/M_{\odot} = 7.16\pm0.10$ and metallicity [M/H]$= -1.1\pm0.2$. Preliminary stream modeling suggests that both streams could result from the tidal disruption of one progenitor, with the Northern Stream representing primarily its core and leading arm, and the Southern Stream being an extended trailing arm. In this scenario, the joint mass and metallicity are $\log_{10} M_{Northern\,stream}/M_{\odot} = 7.93\pm0.10$ and  [M/H]$= -1.0\pm0.2$. These properties are quite similar to the Sagittarius stream around the Milky Way. 

In order to compare M83 to resolved-star measurements for other stellar halos, we must use simulations of stellar halo formation to translate our face-on quantities to edge-on equivalents for comparison with other halos. We use a set of accreted halos from the TNG-50 simulation suite, finding that our measurements of halo mass and metallicity strongly constrain expected edge-on quantities but that density and metallicity gradients are much less well correlated. M83's stellar halo has a lower mass than the typical observed or TNG-50 stellar halo of Milky Way mass galaxies, and a low metallicity. M83's halo falls squarely on stellar halo mass--metallicity correlation seen for other Milky Way mass galaxies, as strongly predicted by more than a decade of theoretical work. 

We used published relationships between observed halo mass and total accreted mass determined from a range of different stellar halo modeling techniques to estimate M83's total accreted mass, $\log_{10} M_{*,accreted}/M_{\odot} = 8.78^{+0.22}_{-0.28}$. In these models, the accreted mass of a galaxy is usually dominated by a single progenitor, and we used published scalings to estimate that M83's largest past merger had a stellar mass of $\log_{10} M_{*,dom}/M_{\odot} = 8.5\pm0.3$.
 
We identified plausible M83 analogues in TNG-50 with a recent $10^8M_{\odot}$ accretion and a face-on stellar halo mass in the right range. The dominant mergers of these systems are all $>$8\,Gyr ago, and should be unassociated with present-day starburst and gas properties. We found that while a recent galaxy accretion can create a prominent stream, these accretions do not trigger starburst activity. Furthermore, while these accretions delivered $\sim10^9M_{\odot}$ of gas to the M83 analogues, most of this gas mass was delivered before or during its first pericenter, 4--8\,Gyr in the past, too long ago to contribute meaningfully to M83's extended {\sc Hi} disk, and with masses that are $<30$\% of its present-day mass. Consequently, we concluded that other non-merger mechanisms, such as secular evolution or accretion of gas from the IGM, are likely to be responsible for M83's remarkable nuclear starburst and extended {\sc Hi} disk.

\begin{acknowledgments}
We thank the referee for their careful reading of the manuscript and constructive suggestions. 
This work was partly supported by the National Science Foundation through grants NSF-AST 1514835 and 2007065, HST grants GO-13696 and GO-15230 provided by NASA through a grant from the Space Telescope Science Institute, which is operated by the Association of Universities for Research in Astronomy, Inc., under NASA contract NAS5-26555. We acknowledge partial support from the WFIRST Infrared Nearby
Galaxies Survey (WINGS) collaboration through
NASA grant NNG16PJ28C through subcontract from
the University of Washington. This work was supported in part by NASA under grant number 80NSSC24K0084 under solicitation NNH22ZDA001N-ROMAN: Nancy Grace Roman Space Telescope Research and Support Participation Opportunities.
AS is supported by NASA through the Hubble Fellowship grant HST-HF2-51567 awarded by STScI.
AM acknowledges support from the ANID FONDECYT Regular 1251882, from the ANID BASAL project FB210003, and funding from the HORIZON-MSCA-2021-SE-01 Research and Innovation Programme under the Marie Sklodowska-Curie grant agreement number 101086388.
JB acknowledges support from NASA grant 80NSSC24K1227. 
This work was partially supported by a research grant (VIL53081) from VILLUM FONDEN. 
This work was co-funded by the European Union (ERC, BeyondSTREAMS, 101115754). Views and opinions expressed are, however, those of the author(s) only and do not necessarily reflect those of the European Union or the European Research Council. Neither the European Union nor the granting authority can be held responsible for them. The Tycho supercomputer hosted at the SCIENCE HPC center at the University of Copenhagen was used in support of this work.

Based on observations utilizing Pan-STARRS1 Survey. The Pan-STARRS1 Surveys (PS1) and the PS1 public science archive have been made possible through contributions by the Institute for Astronomy, the University of Hawaii, the Pan-STARRS Project Office, the Max Planck Society and its participating institutes, the Max Planck Institute for Astronomy, Heidelberg and the Max Planck Institute for Extraterrestrial Physics, Garching, The Johns Hopkins University, Durham University, the University of Edinburgh, the Queen's University Belfast, the Harvard-Smithsonian Center for Astrophysics, the Las Cumbres Observatory Global Telescope Network Incorporated, the National Central University of Taiwan, the Space Telescope Science Institute, the National Aeronautics and Space Administration under grant No. NNX08AR22G issued through the Planetary Science Division of the NASA Science Mission Directorate, the National Science Foundation grant No. AST-1238877, the University of Maryland, Eotvos Lorand University (ELTE), the Los Alamos National Laboratory, and the Gordon and Betty Moore Foundation.

Based on observations obtained at the Subaru Observatory, which is operated by the National Astronomical Observatory of Japan, via the Gemini/Subaru Time Exchange Program. We thank the Subaru support staff --- particularly Akito Tajitsu, Tsuyoshi Terai, Dan Birchall, and Fumiaki Nakata --- for invaluable help in preparing the observations and navigating the Subaru Queue Mode.

The authors wish to recognize and acknowledge the very significant cultural role and reverence that the summit of Maunakea has always had within the indigenous Hawaiian community. We are most fortunate to have the opportunity to conduct observations from this mountain.

\end{acknowledgments}

\begin{contribution}

EFB was responsible for project design and leadership, analysis and writing. BH and MC led the HST density analysis. PAP is responsible for reducing and managing the Subaru data. SP and JN are responsible for the stream modeling and writing of that subsection. RSdJ led the HST/GHOSTS survey. All other authors contributed towards project execution, writing, review and editing.

\end{contribution}

%
\facilities{HST(ACS/WFC3), Subaru(HSC)}

\software{\texttt{HSC Pipeline} \citep{Bosch2018}, \texttt{Matplotlib} \citep{matplotlib}, \texttt{NumPy} \citep{numpy-guide,numpy}, \texttt{Astropy} \citep{Astropy13,astropy}, \texttt{SciPy} \citep{scipy},
\texttt{emcee} \citep{Foreman-Mackey_2013},\texttt{JAX} \citep{jax2018github} }

\bibliography{M83Ref}{}

@ARTICLE{dustmaps,
       author = {{Green}, {Gregory M.}},
        title = "{dustmaps: A Python interface for maps of interstellar dust}",
      journal = {The Journal of Open Source Software},
         year = "2018",
        month = "Jun",
       volume = {3},
       number = {26},
        pages = {695},
          doi = {10.21105/joss.00695},
       adsurl = {https://ui.adsabs.harvard.edu/abs/2018JOSS....3..695G}
}

@ARTICLE{Planck,
       author = {{Planck Collaboration} and {Aghanim}, N. and {Ashdown}, M. and {Aumont}, J. and {Baccigalupi}, C. and {Ballardini}, M. and {Banday}, A.~J. and {Barreiro}, R.~B. and {Bartolo}, N. and {Basak}, S. and {Benabed}, K. and {Bernard}, J. -P. and {Bersanelli}, M. and {Bielewicz}, P. and {Bonavera}, L. and {Bond}, J.~R. and {Borrill}, J. and {Bouchet}, F.~R. and {Boulanger}, F. and {Burigana}, C. and {Calabrese}, E. and {Cardoso}, J. -F. and {Carron}, J. and {Chiang}, H.~C. and {Colombo}, L.~P.~L. and {Comis}, B. and {Couchot}, F. and {Coulais}, A. and {Crill}, B.~P. and {Curto}, A. and {Cuttaia}, F. and {de Bernardis}, P. and {de Zotti}, G. and {Delabrouille}, J. and {Di Valentino}, E. and {Dickinson}, C. and {Diego}, J.~M. and {Dor{\'e}}, O. and {Douspis}, M. and {Ducout}, A. and {Dupac}, X. and {Dusini}, S. and {Elsner}, F. and {En{\ss}lin}, T.~A. and {Eriksen}, H.~K. and {Falgarone}, E. and {Fantaye}, Y. and {Finelli}, F. and {Forastieri}, F. and {Frailis}, M. and {Fraisse}, A.~A. and {Franceschi}, E. and {Frolov}, A. and {Galeotta}, S. and {Galli}, S. and {Ganga}, K. and {G{\'e}nova-Santos}, R.~T. and {Gerbino}, M. and {Ghosh}, T. and {Giraud-H{\'e}raud}, Y. and {Gonz{\'a}lez-Nuevo}, J. and {G{\'o}rski}, K.~M. and {Gruppuso}, A. and {Gudmundsson}, J.~E. and {Hansen}, F.~K. and {Helou}, G. and {Henrot-Versill{\'e}}, S. and {Herranz}, D. and {Hivon}, E. and {Huang}, Z. and {Jaffe}, A.~H. and {Jones}, W.~C. and {Keih{\"a}nen}, E. and {Keskitalo}, R. and {Kiiveri}, K. and {Kisner}, T.~S. and {Krachmalnicoff}, N. and {Kunz}, M. and {Kurki-Suonio}, H. and {Lamarre}, J. -M. and {Langer}, M. and {Lasenby}, A. and {Lattanzi}, M. and {Lawrence}, C.~R. and {Le Jeune}, M. and {Levrier}, F. and {Lilje}, P.~B. and {Lilley}, M. and {Lindholm}, V. and {L{\'o}pez-Caniego}, M. and {Ma}, Y. -Z. and {Mac{\'\i}as-P{\'e}rez}, J.~F. and {Maggio}, G. and {Maino}, D. and {Mandolesi}, N. and {Mangilli}, A. and {Maris}, M. and {Martin}, P.~G. and {Mart{\'\i}nez-Gonz{\'a}lez}, E. and {Matarrese}, S. and {Mauri}, N. and {McEwen}, J.~D. and {Melchiorri}, A. and {Mennella}, A. and {Migliaccio}, M. and {Miville-Desch{\^e}nes}, M. -A. and {Molinari}, D. and {Moneti}, A. and {Montier}, L. and {Morgante}, G. and {Moss}, A. and {Natoli}, P. and {Oxborrow}, C.~A. and {Pagano}, L. and {Paoletti}, D. and {Patanchon}, G. and {Perdereau}, O. and {Perotto}, L. and {Pettorino}, V. and {Piacentini}, F. and {Plaszczynski}, S. and {Polastri}, L. and {Polenta}, G. and {Puget}, J. -L. and {Rachen}, J.~P. and {Racine}, B. and {Reinecke}, M. and {Remazeilles}, M. and {Renzi}, A. and {Rocha}, G. and {Rosset}, C. and {Rossetti}, M. and {Roudier}, G. and {Rubi{\~n}o-Mart{\'\i}n}, J.~A. and {Ruiz-Granados}, B. and {Salvati}, L. and {Sandri}, M. and {Savelainen}, M. and {Scott}, D. and {Sirignano}, C. and {Sirri}, G. and {Soler}, J.~D. and {Spencer}, L.~D. and {Suur-Uski}, A. -S. and {Tauber}, J.~A. and {Tavagnacco}, D. and {Tenti}, M. and {Toffolatti}, L. and {Tomasi}, M. and {Tristram}, M. and {Trombetti}, T. and {Valiviita}, J. and {Van Tent}, F. and {Vielva}, P. and {Villa}, F. and {Vittorio}, N. and {Wandelt}, B.~D. and {Wehus}, I.~K. and {Zacchei}, A. and {Zonca}, A.},
        title = "{Planck intermediate results. XLVIII. Disentangling Galactic dust emission and cosmic infrared background anisotropies}",
      journal = {\aap},
         year = 2016,
        month = dec,
       volume = {596},
          eid = {A109},
        pages = {A109},
          doi = {10.1051/0004-6361/201629022},
archivePrefix = {arXiv},
       eprint = {1605.09387},
 primaryClass = {astro-ph.CO},
       adsurl = {https://ui.adsabs.harvard.edu/abs/2016A&A...596A.109P}
}

@ARTICLE{Smercina2020,
       author = {{Smercina}, Adam and {Bell}, Eric F. and {Price}, Paul A. and {Slater}, Colin T. and {D'Souza}, Richard and {Bailin}, Jeremy and {de Jong}, Roelof S. and {Jang}, In Sung and {Monachesi}, Antonela and {Nidever}, David},
        title = "{The Saga of M81: Global View of a Massive Stellar Halo in Formation}",
      journal = {\apj},
         year = 2020,
        month = dec,
       volume = {905},
       number = {1},
          eid = {60},
        pages = {60},
          doi = {10.3847/1538-4357/abc485},
archivePrefix = {arXiv},
       eprint = {1910.14672},
 primaryClass = {astro-ph.GA},
       adsurl = {https://ui.adsabs.harvard.edu/abs/2020ApJ...905...60S}
}

@ARTICLE{Okamoto2015,
       author = {{Okamoto}, Sakurako and {Arimoto}, Nobuo and {Ferguson}, Annette M.~N. and {Bernard}, Edouard J. and {Irwin}, Mike J. and {Yamada}, Yoshihiko and {Utsumi}, Yousuke},
        title = "{A Hyper Suprime-Cam View of the Interacting Galaxies of the M81 Group}",
      journal = {\apjl},
         year = 2015,
        month = aug,
       volume = {809},
       number = {1},
          eid = {L1},
        pages = {L1},
          doi = {10.1088/2041-8205/809/1/L1},
archivePrefix = {arXiv},
       eprint = {1507.04889},
 primaryClass = {astro-ph.GA},
       adsurl = {https://ui.adsabs.harvard.edu/abs/2015ApJ...809L...1O}
}

@ARTICLE{LeeWiesz2024,
       author = {{Lee}, Abigail J. and {Weisz}, Daniel R. and {Ren}, Yi and {Savino}, Alessandro and {Dolphin}, Andrew E.},
        title = "{Measuring Star Formation Histories from Asymptotic Giant Branch Stars: A Demonstration in M31}",
      journal = {arXiv e-prints},
         year = 2024,
        month = oct,
          eid = {arXiv:2410.09256},
        pages = {arXiv:2410.09256},
          doi = {10.48550/arXiv.2410.09256},
archivePrefix = {arXiv},
       eprint = {2410.09256},
 primaryClass = {astro-ph.GA},
       adsurl = {https://ui.adsabs.harvard.edu/abs/2024arXiv241009256L}
}

@ARTICLE{Zhu2022a,
       author = {{Zhu}, Ling and {Pillepich}, Annalisa and {van de Ven}, Glenn and {Leaman}, Ryan and {Hernquist}, Lars and {Nelson}, Dylan and {Pakmor}, Ruediger and {Vogelsberger}, Mark and {Zhang}, Le},
        title = "{Mass of the dynamically hot inner stellar halo predicts the ancient accreted stellar mass}",
      journal = {\aap},
         year = 2022,
        month = apr,
       volume = {660},
          eid = {A20},
        pages = {A20},
          doi = {10.1051/0004-6361/202142496},
archivePrefix = {arXiv},
       eprint = {2110.13172},
 primaryClass = {astro-ph.GA},
       adsurl = {https://ui.adsabs.harvard.edu/abs/2022A&A...660A..20Z}
}

@ARTICLE{Hernandez23,
       author = {{Hernandez}, Svea and {Jones}, Logan and {Smith}, Linda J. and {Togi}, Aditya and {Aloisi}, Alessandra and {Blair}, William P. and {Hirschauer}, Alec S. and {Hunt}, Leslie K. and {James}, Bethan L. and {Kumari}, Nimisha and {Lebouteiller}, Vianney and {Mingozzi}, Matilde and {Ramambason}, Lise},
        title = "{Dissecting the Mid-infrared Heart of M83 with JWST}",
      journal = {\apj},
         year = 2023,
        month = may,
       volume = {948},
       number = {2},
          eid = {124},
        pages = {124},
          doi = {10.3847/1538-4357/acc837},
archivePrefix = {arXiv},
       eprint = {2301.09582},
 primaryClass = {astro-ph.GA},
       adsurl = {https://ui.adsabs.harvard.edu/abs/2023ApJ...948..124H}
}

@software{DOLPHOT16,
       author = {{Dolphin}, Andrew},
        title = "{DOLPHOT: Stellar photometry}",
 howpublished = {Astrophysics Source Code Library, record ascl:1608.013},
         year = 2016,
        month = aug,
          eid = {ascl:1608.013},
       adsurl = {https://ui.adsabs.harvard.edu/abs/2016ascl.soft08013D}
}

@ARTICLE{Soumagnac2015,
       author = {{Soumagnac}, M.~T. and {Abdalla}, F.~B. and {Lahav}, O. and {Kirk}, D. and {Sevilla}, I. and {Bertin}, E. and {Rowe}, B.~T.~P. and {Annis}, J. and {Busha}, M.~T. and {Da Costa}, L.~N. and {Frieman}, J.~A. and {Gaztanaga}, E. and {Jarvis}, M. and {Lin}, H. and {Percival}, W.~J. and {Santiago}, B.~X. and {Sabiu}, C.~G. and {Wechsler}, R.~H. and {Wolz}, L. and {Yanny}, B.},
        title = "{Star/galaxy separation at faint magnitudes: application to a simulated Dark Energy Survey}",
      journal = {\mnras},
         year = 2015,
        month = jun,
       volume = {450},
       number = {1},
        pages = {666-680},
          doi = {10.1093/mnras/stu1410},
archivePrefix = {arXiv},
       eprint = {1306.5236},
 primaryClass = {astro-ph.CO},
       adsurl = {https://ui.adsabs.harvard.edu/abs/2015MNRAS.450..666S}
}

@ARTICLE{Chua2019,
       author = {{Chua}, Kun Ting Eddie and {Pillepich}, Annalisa and {Vogelsberger}, Mark and {Hernquist}, Lars},
        title = "{Shape of dark matter haloes in the Illustris simulation: effects of baryons}",
      journal = {\mnras},
         year = 2019,
        month = mar,
       volume = {484},
       number = {1},
        pages = {476-493},
          doi = {10.1093/mnras/sty3531},
archivePrefix = {arXiv},
       eprint = {1809.07255},
 primaryClass = {astro-ph.GA},
       adsurl = {https://ui.adsabs.harvard.edu/abs/2019MNRAS.484..476C}
}

@ARTICLE{Baldry2010,
       author = {{Baldry}, I.~K. and {Robotham}, A.~S.~G. and {Hill}, D.~T. and {Driver}, S.~P. and {Liske}, J. and {Norberg}, P. and {Bamford}, S.~P. and {Hopkins}, A.~M. and {Loveday}, J. and {Peacock}, J.~A. and {Cameron}, E. and {Croom}, S.~M. and {Cross}, N.~J.~G. and {Doyle}, I.~F. and {Dye}, S. and {Frenk}, C.~S. and {Jones}, D.~H. and {van Kampen}, E. and {Kelvin}, L.~S. and {Nichol}, R.~C. and {Parkinson}, H.~R. and {Popescu}, C.~C. and {Prescott}, M. and {Sharp}, R.~G. and {Sutherland}, W.~J. and {Thomas}, D. and {Tuffs}, R.~J.},
        title = "{Galaxy And Mass Assembly (GAMA): the input catalogue and star-galaxy separation}",
      journal = {\mnras},
         year = 2010,
        month = may,
       volume = {404},
       number = {1},
        pages = {86-100},
          doi = {10.1111/j.1365-2966.2010.16282.x},
archivePrefix = {arXiv},
       eprint = {0910.5120},
 primaryClass = {astro-ph.CO},
       adsurl = {https://ui.adsabs.harvard.edu/abs/2010MNRAS.404...86B}
}

@ARTICLE{Desai12,
       author = {{Desai}, S. and {Armstrong}, R. and {Mohr}, J.~J. and {Semler}, D.~R. and {Liu}, J. and {Bertin}, E. and {Allam}, S.~S. and {Barkhouse}, W.~A. and {Bazin}, G. and {Buckley-Geer}, E.~J. and {Cooper}, M.~C. and {Hansen}, S.~M. and {High}, F.~W. and {Lin}, H. and {Lin}, Y. -T. and {Ngeow}, C. -C. and {Rest}, A. and {Song}, J. and {Tucker}, D. and {Zenteno}, A.},
        title = "{The Blanco Cosmology Survey: Data Acquisition, Processing, Calibration, Quality Diagnostics, and Data Release}",
      journal = {\apj},
         year = 2012,
        month = sep,
       volume = {757},
       number = {1},
          eid = {83},
        pages = {83},
          doi = {10.1088/0004-637X/757/1/83},
archivePrefix = {arXiv},
       eprint = {1204.1210},
 primaryClass = {astro-ph.CO},
       adsurl = {https://ui.adsabs.harvard.edu/abs/2012ApJ...757...83D}
}

@ARTICLE{Slater20,
       author = {{Slater}, Colin T. and {Ivezi{\'c}}, {\v{Z}}eljko and {Lupton}, Robert H.},
        title = "{Morphological Star-Galaxy Separation}",
      journal = {\aj},
         year = 2020,
        month = feb,
       volume = {159},
       number = {2},
          eid = {65},
        pages = {65},
          doi = {10.3847/1538-3881/ab6166},
archivePrefix = {arXiv},
       eprint = {1912.08210},
 primaryClass = {astro-ph.IM},
       adsurl = {https://ui.adsabs.harvard.edu/abs/2020AJ....159...65S}
}

@ARTICLE{Jones24,
       author = {{Jones}, Logan H. and {Hernandez}, Svea and {Smith}, Linda J. and {Togi}, Aditya and {Diaz-Santos}, Tanio and {Aloisi}, Alessandra and {Blair}, William and {Hirschauer}, Alec S. and {Hunt}, Leslie K. and {James}, Bethan L. and {Kumari}, Nimisha and {Lebouteiller}, Vianney and {Mingozzi}, Matilde and {Ramambason}, Lise},
        title = "{A JWST/MIRI View of the ISM in M83: I. Resolved Molecular Hydrogen Properties, Star Formation, and Feedback}",
      journal = {arXiv e-prints},
         year = 2024,
        month = oct,
          eid = {arXiv:2410.09020},
        pages = {arXiv:2410.09020},
          doi = {10.48550/arXiv.2410.09020},
archivePrefix = {arXiv},
       eprint = {2410.09020},
 primaryClass = {astro-ph.GA},
       adsurl = {https://ui.adsabs.harvard.edu/abs/2024arXiv241009020J}
}

@ARTICLE{Hernandez25,
       author = {{Hernandez}, Svea and {Smith}, Linda J. and {Jones}, Logan H. and {Togi}, Aditya and {Melendez}, Marcio B. and {Abril-Melgarejo}, Valentina and {Adamo}, Angela and {Alonso Herrero}, Almudena and {Diaz-Santos}, Tanio and {Fischer}, Travis C. and {Garcia-Burillo}, Santiago and {Hirschauer}, Alec S. and {Hunt}, Leslie K. and {James}, Bethan and {Lebouteiller}, Vianney and {Long}, Knox S. and {Mingozzi}, Matilde and {Ramambason}, Lise and {Ramos Almeida}, Cristina},
        title = "{JWST/MIRI detection of [Ne V] and [Ne VI] in M83: Evidence for the long sought-after AGN?}",
      journal = {arXiv e-prints},
         year = 2025,
        month = feb,
          eid = {arXiv:2502.17621},
        pages = {arXiv:2502.17621},
          doi = {10.48550/arXiv.2502.17621},
archivePrefix = {arXiv},
       eprint = {2502.17621},
 primaryClass = {astro-ph.GA},
       adsurl = {https://ui.adsabs.harvard.edu/abs/2025arXiv250217621H}
}

@ARTICLE{DellaBruna2022,
       author = {{Della Bruna}, Lorenza and {Adamo}, Angela and {Amram}, Philippe and {Rosolowsky}, Erik and {Usher}, Christopher and {Sirressi}, Mattia and {Schruba}, Andreas and {Emsellem}, Eric and {Leroy}, Adam and {Bik}, Arjan and {Blair}, William P. and {McLeod}, Anna F. and {{\"O}stlin}, G{\"o}ran and {Renaud}, Florent and {Robert}, Carmelle and {Rousseau-Nepton}, Laurie and {Smith}, Linda J.},
        title = "{Stellar feedback in M83 as observed with MUSE. I. Overview, an unprecedented view of the stellar and gas kinematics and evidence of outflowing gas}",
      journal = {\aap},
         year = 2022,
        month = apr,
       volume = {660},
          eid = {A77},
        pages = {A77},
          doi = {10.1051/0004-6361/202142315},
archivePrefix = {arXiv},
       eprint = {2202.01738},
 primaryClass = {astro-ph.GA},
       adsurl = {https://ui.adsabs.harvard.edu/abs/2022A&A...660A..77D}
}

@ARTICLE{Cai2025,
       author = {{Cai}, Runsheng and {Zhu}, Ling and {Shen}, Shiyin and {Wang}, Wenting and {Pillepich}, Annalisa and {Falc{\'o}n-Barroso}, Jes{\'u}s},
        title = "{Robust machine learning model of inferring the ex-situ stellar fraction of galaxies from photometric data}",
      journal = {arXiv e-prints},
         year = 2025,
        month = feb,
          eid = {arXiv:2502.13216},
        pages = {arXiv:2502.13216},
          doi = {10.48550/arXiv.2502.13216},
archivePrefix = {arXiv},
       eprint = {2502.13216},
 primaryClass = {astro-ph.GA},
       adsurl = {https://ui.adsabs.harvard.edu/abs/2025arXiv250213216C}
}

@ARTICLE{Sarmiento2023,
       author = {{Sarmiento}, Regina and {Huertas-Company}, Marc and {Knapen}, Johan H. and {Ibarra-Medel}, H{\'e}ctor and {Pillepich}, Annalisa and {S{\'a}nchez}, Sebasti{\'a}n F. and {Boecker}, Alina},
        title = "{MaNGIA: 10 000 mock galaxies for stellar population analysis}",
      journal = {\aap},
         year = 2023,
        month = may,
       volume = {673},
          eid = {A23},
        pages = {A23},
          doi = {10.1051/0004-6361/202245509},
archivePrefix = {arXiv},
       eprint = {2211.11790},
 primaryClass = {astro-ph.GA},
       adsurl = {https://ui.adsabs.harvard.edu/abs/2023A&A...673A..23S}
}

@ARTICLE{Rodriguez-Gomez2015,
       author = {{Rodriguez-Gomez}, Vicente and {Genel}, Shy and {Vogelsberger}, Mark and {Sijacki}, Debora and {Pillepich}, Annalisa and {Sales}, Laura V. and {Torrey}, Paul and {Snyder}, Greg and {Nelson}, Dylan and {Springel}, Volker and {Ma}, Chung-Pei and {Hernquist}, Lars},
        title = "{The merger rate of galaxies in the Illustris simulation: a comparison with observations and semi-empirical models}",
      journal = {\mnras},
         year = 2015,
        month = may,
       volume = {449},
       number = {1},
        pages = {49-64},
          doi = {10.1093/mnras/stv264},
archivePrefix = {arXiv},
       eprint = {1502.01339},
 primaryClass = {astro-ph.GA},
       adsurl = {https://ui.adsabs.harvard.edu/abs/2015MNRAS.449...49R}
}

@ARTICLE{Rodriguez-Gomez2016,
       author = {{Rodriguez-Gomez}, Vicente and {Pillepich}, Annalisa and {Sales}, Laura V. and {Genel}, Shy and {Vogelsberger}, Mark and {Zhu}, Qirong and {Wellons}, Sarah and {Nelson}, Dylan and {Torrey}, Paul and {Springel}, Volker and {Ma}, Chung-Pei and {Hernquist}, Lars},
        title = "{The stellar mass assembly of galaxies in the Illustris simulation: growth by mergers and the spatial distribution of accreted stars}",
      journal = {\mnras},
         year = 2016,
        month = may,
       volume = {458},
       number = {3},
        pages = {2371-2390},
          doi = {10.1093/mnras/stw456},
archivePrefix = {arXiv},
       eprint = {1511.08804},
 primaryClass = {astro-ph.GA},
       adsurl = {https://ui.adsabs.harvard.edu/abs/2016MNRAS.458.2371R}
}

@ARTICLE{Wright2024,
       author = {{Wright}, Anna C. and {Tumlinson}, Jason and {Peeples}, Molly S. and {O'Shea}, Brian W. and {Lochhaas}, Cassandra and {Corlies}, Lauren and {Smith}, Britton D. and {Binh}, Nguyen and {Augustin}, Ramona and {Simons}, Raymond C.},
        title = "{Figuring Out Gas and Galaxies in Enzo (FOGGIE). VII. The (Dis)assembly of Stellar Halos}",
      journal = {\apj},
         year = 2024,
        month = jul,
       volume = {970},
       number = {1},
          eid = {70},
        pages = {70},
          doi = {10.3847/1538-4357/ad49a3},
archivePrefix = {arXiv},
       eprint = {2309.10039},
 primaryClass = {astro-ph.GA},
       adsurl = {https://ui.adsabs.harvard.edu/abs/2024ApJ...970...70W}
}

@ARTICLE{Merritt2020,
       author = {{Merritt}, Allison and {Pillepich}, Annalisa and {van Dokkum}, Pieter and {Nelson}, Dylan and {Hernquist}, Lars and {Marinacci}, Federico and {Vogelsberger}, Mark},
        title = "{A missing outskirts problem? Comparisons between stellar haloes in the Dragonfly Nearby Galaxies Survey and the TNG100 simulation}",
      journal = {\mnras},
         year = 2020,
        month = jul,
       volume = {495},
       number = {4},
        pages = {4570-4604},
          doi = {10.1093/mnras/staa1164},
archivePrefix = {arXiv},
       eprint = {2004.11402},
 primaryClass = {astro-ph.GA},
       adsurl = {https://ui.adsabs.harvard.edu/abs/2020MNRAS.495.4570M}
}

@ARTICLE{Pillepich2019,
       author = {{Pillepich}, Annalisa and {Nelson}, Dylan and {Springel}, Volker and {Pakmor}, R{\"u}diger and {Torrey}, Paul and {Weinberger}, Rainer and {Vogelsberger}, Mark and {Marinacci}, Federico and {Genel}, Shy and {van der Wel}, Arjen and {Hernquist}, Lars},
        title = "{First results from the TNG50 simulation: the evolution of stellar and gaseous discs across cosmic time}",
      journal = {\mnras},
         year = 2019,
        month = dec,
       volume = {490},
       number = {3},
        pages = {3196-3233},
          doi = {10.1093/mnras/stz2338},
archivePrefix = {arXiv},
       eprint = {1902.05553},
 primaryClass = {astro-ph.GA},
       adsurl = {https://ui.adsabs.harvard.edu/abs/2019MNRAS.490.3196P}
}

@ARTICLE{Sante2025,
       author = {{Sante}, Andrea and {Kawata}, Daisuke and {Font}, Andreea S. and {Grand}, Robert J.~J.},
        title = "{GalactiKit: reconstructing mergers from $z=0$ debris using simulation-based inference in Auriga}",
      journal = {arXiv e-prints},
         year = 2025,
        month = feb,
          eid = {arXiv:2502.14972},
        pages = {arXiv:2502.14972},
          doi = {10.48550/arXiv.2502.14972},
archivePrefix = {arXiv},
       eprint = {2502.14972},
 primaryClass = {astro-ph.GA},
       adsurl = {https://ui.adsabs.harvard.edu/abs/2025arXiv250214972S}
}

@ARTICLE{Hasselquist2021,
       author = {{Hasselquist}, Sten and {Hayes}, Christian R. and {Lian}, Jianhui and {Weinberg}, David H. and {Zasowski}, Gail and {Horta}, Danny and {Beaton}, Rachael and {Feuillet}, Diane K. and {Garro}, Elisa R. and {Gallart}, Carme and {Smith}, Verne V. and {Holtzman}, Jon A. and {Minniti}, Dante and {Lacerna}, Ivan and {Shetrone}, Matthew and {J{\"o}nsson}, Henrik and {Cioni}, Maria-Rosa L. and {Fillingham}, Sean P. and {Cunha}, Katia and {O'Connell}, Robert and {Fern{\'a}ndez-Trincado}, Jos{\'e} G. and {Mu{\~n}oz}, Ricardo R. and {Schiavon}, Ricardo and {Almeida}, Andres and {Anguiano}, Borja and {Beers}, Timothy C. and {Bizyaev}, Dmitry and {Brownstein}, Joel R. and {Cohen}, Roger E. and {Frinchaboy}, Peter and {Garc{\'\i}a-Hern{\'a}ndez}, D.~A. and {Geisler}, Doug and {Lane}, Richard R. and {Majewski}, Steven R. and {Nidever}, David L. and {Nitschelm}, Christian and {Povick}, Joshua and {Price-Whelan}, Adrian and {Roman-Lopes}, Alexandre and {Rosado}, Margarita and {Sobeck}, Jennifer and {Stringfellow}, Guy and {Valenzuela}, Octavio and {Villanova}, Sandro and {Vincenzo}, Fiorenzo},
        title = "{APOGEE Chemical Abundance Patterns of the Massive Milky Way Satellites}",
      journal = {\apj},
         year = 2021,
        month = dec,
       volume = {923},
       number = {2},
          eid = {172},
        pages = {172},
          doi = {10.3847/1538-4357/ac25f9},
archivePrefix = {arXiv},
       eprint = {2109.05130},
 primaryClass = {astro-ph.GA},
       adsurl = {https://ui.adsabs.harvard.edu/abs/2021ApJ...923..172H}
}

@ARTICLE{matplotlib,
       author = {{Hunter}, John D.},
        title = "{Matplotlib: A 2D Graphics Environment}",
      journal = {Computing in Science and Engineering},
         year = 2007,
        month = may,
       volume = {9},
       number = {3},
        pages = {90-95},
          doi = {10.1109/MCSE.2007.55},
       adsurl = {https://ui.adsabs.harvard.edu/abs/2007CSE.....9...90H}
}

@book{numpy-guide,
 title={A guide to NumPy},
 author={Oliphant, Travis E},
 volume={1},
 year={2006},
 publisher={Trelgol Publishing USA}
}

@article{numpy,
  title={The NumPy array: a structure for efficient numerical computation},
  author={Van Der Walt, Stefan and Colbert, S Chris and Varoquaux, Gael},
  journal={Computing in Science \& Engineering},
  volume={13},
  number={2},
  pages={22},
  year={2011},
  publisher={IEEE Computer Society}
}

@ARTICLE{scipy,
       author = {{Virtanen}, Pauli and {Gommers}, Ralf and {Oliphant}, Travis E. and
         {Haberland}, Matt and {Reddy}, Tyler and {Cournapeau}, David and
         {Burovski}, Evgeni and {Peterson}, Pearu and {Weckesser}, Warren and
         {Bright}, Jonathan and {van der Walt}, St{\'e}fan J. and
         {Brett}, Matthew and {Wilson}, Joshua and {Millman}, K. Jarrod and
         {Mayorov}, Nikolay and {Nelson}, Andrew R.~J. and {Jones}, Eric and
         {Kern}, Robert and {Larson}, Eric and {Carey}, C.~J. and
         {Polat}, {\.I}lhan and {Feng}, Yu and {Moore}, Eric W. and {Vand
        erPlas}, Jake and {Laxalde}, Denis and {Perktold}, Josef and
         {Cimrman}, Robert and {Henriksen}, Ian and {Quintero}, E.~A. and
         {Harris}, Charles R. and {Archibald}, Anne M. and
         {Ribeiro}, Ant{\^o}nio H. and {Pedregosa}, Fabian and
         {van Mulbregt}, Paul and {SciPy 1. 0 Contributors}},
        title = "{SciPy 1.0: fundamental algorithms for scientific computing in Python}",
      journal = {Nature Methods},
         year = 2020,
        month = feb,
       volume = {17},
        pages = {261-272},
          doi = {10.1038/s41592-019-0686-2},
archivePrefix = {arXiv},
       eprint = {1907.10121},
 primaryClass = {cs.MS},
       adsurl = {https://ui.adsabs.harvard.edu/abs/2020NatMe..17..261V}
}

@BOOK{ESOLV,
       author = {{Lauberts}, Andris and {Valentijn}, Edwin A.},
        title = "{The surface photometry catalogue of the ESO-Uppsala galaxies}",
         year = 1989,
       adsurl = {https://ui.adsabs.harvard.edu/abs/1989spce.book.....L}
}

@ARTICLE{Bressan2012,
       author = {{Bressan}, Alessandro and {Marigo}, Paola and {Girardi}, L{\'e}o. and {Salasnich}, Bernardo and {Dal Cero}, Claudia and {Rubele}, Stefano and {Nanni}, Ambra},
        title = "{PARSEC: stellar tracks and isochrones with the PAdova and TRieste Stellar Evolution Code}",
      journal = {\mnras},
         year = 2012,
        month = nov,
       volume = {427},
       number = {1},
        pages = {127-145},
          doi = {10.1111/j.1365-2966.2012.21948.x},
archivePrefix = {arXiv},
       eprint = {1208.4498},
 primaryClass = {astro-ph.SR},
       adsurl = {https://ui.adsabs.harvard.edu/abs/2012MNRAS.427..127B}
}

@ARTICLE{Ogami2024,
       author = {{Ogami}, Itsuki and {Tanaka}, Mikito and {Komiyama}, Yutaka and {Chiba}, Masashi and {Guhathakurta}, Puragra and {Kirby}, Evan N. and {Wyse}, Rosemary F.~G. and {Filion}, Carrie and {Gilbert}, Karoline M. and {Escala}, Ivanna and {Mori}, Masao and {Kirihara}, Takanobu and {Tanaka}, Masayuki and {Ishigaki}, Miho N. and {Hayashi}, Kohei and {Lee}, Myun Gyoon and {Sharma}, Sanjib and {Kalirai}, Jason S. and {Lupton}, Robert H.},
        title = "{The structure of the stellar halo of the Andromeda galaxy explored with the NB515 for Subaru/HSC - I. New insights on the stellar halo up to 120 kpc}",
      journal = {\mnras},
         year = 2025,
        month = jan,
       volume = {536},
       number = {1},
        pages = {530-553},
          doi = {10.1093/mnras/stae2527},
archivePrefix = {arXiv},
       eprint = {2401.00668},
 primaryClass = {astro-ph.GA},
       adsurl = {https://ui.adsabs.harvard.edu/abs/2025MNRAS.536..530O}
}

@ARTICLE{Semczuk2025,
       author = {{Semczuk}, Marcin and {Antoja}, Teresa and {Gir{\'o}n-Soto}, Alexandra and {Laporte}, Chervin F.~P.},
        title = "{Analogues of the Milky Way-Sagittarius interaction in the TNG50: effect on the Milky Way}",
      journal = {arXiv e-prints},
         year = 2025,
        month = aug,
          eid = {arXiv:2508.00690},
        pages = {arXiv:2508.00690},
          doi = {10.48550/arXiv.2508.00690},
archivePrefix = {arXiv},
       eprint = {2508.00690},
 primaryClass = {astro-ph.GA},
       adsurl = {https://ui.adsabs.harvard.edu/abs/2025arXiv250800690S}
}

@ARTICLE{Crnojevic2019,
       author = {{Crnojevi{\'c}}, D. and {Sand}, D.~J. and {Bennet}, P. and {Pasetto}, S. and {Spekkens}, K. and {Caldwell}, N. and {Guhathakurta}, P. and {McLeod}, B. and {Seth}, A. and {Simon}, J.~D. and {Strader}, J. and {Toloba}, E.},
        title = "{The Faint End of the Centaurus A Satellite Luminosity Function}",
      journal = {\apj},
         year = 2019,
        month = feb,
       volume = {872},
       number = {1},
          eid = {80},
        pages = {80},
          doi = {10.3847/1538-4357/aafbe7},
archivePrefix = {arXiv},
       eprint = {1809.05103},
 primaryClass = {astro-ph.GA},
       adsurl = {https://ui.adsabs.harvard.edu/abs/2019ApJ...872...80C}
}

@ARTICLE{Fielder2025,
       author = {{Fielder}, Catherine E. and {Sand}, David J. and {Jones}, Michael G. and {Crnojevi{\'c}}, Denija and {Drlica-Wagner}, Alex and {Bennet}, Paul and {Carlin}, Jeffrey L. and {Cerny}, William and {Doliva-Dolinsky}, Amandine and {Hunter}, Laura C. and {Karunakaran}, Ananthan and {Limberg}, Guilherme and {Mutlu-Pakdil}, Bur{\c{c}}in and {Pace}, Andrew B. and {Pearson}, Sarah and {Smercina}, Adam and {Spekkens}, Kristine and {Starkenburg}, Tjitske and {Strader}, Jay and {Stringfellow}, Guy S. and {Tollerud}, Erik and {Bom}, Clecio R. and {Carballo-Bello}, Julio A. and {Chaturvedi}, Astha and {Choi}, Yumi and {James}, David J. and {Mart{\'\i}nez-V{\'a}zquez}, Clara E. and {Riley}, Alexander H. and {Sakowska}, Joanna and {Vivas}, Kathy},
        title = "{Streams, Shells, and Substructures in the Accretion-built Stellar Halo of NGC 300}",
      journal = {\apjl},
         year = 2025,
        month = apr,
       volume = {982},
       number = {2},
          eid = {L41},
        pages = {L41},
          doi = {10.3847/2041-8213/adbf17},
archivePrefix = {arXiv},
       eprint = {2501.04089},
 primaryClass = {astro-ph.GA},
       adsurl = {https://ui.adsabs.harvard.edu/abs/2025ApJ...982L..41F}
}

@ARTICLE{Mueller2025,
       author = {{M{\"u}ller}, Oliver and {Rejkuba}, Marina and {Fahrion}, Katja and {Pawlowski}, Marcel S. and {Famaey}, Benoit and {Libeskind}, Noam and {Heesters}, Nick and {Lelli}, Federico and {Hilker}, Michael and {Taibi}, Salvatore and {Pearson}, Sarah},
        title = "{MUSE observations of dwarf galaxies and a stellar stream in the M83 group}",
      journal = {arXiv e-prints},
         year = 2025,
        month = apr,
          eid = {arXiv:2504.20765},
        pages = {arXiv:2504.20765},
          doi = {10.48550/arXiv.2504.20765},
archivePrefix = {arXiv},
       eprint = {2504.20765},
 primaryClass = {astro-ph.GA},
       adsurl = {https://ui.adsabs.harvard.edu/abs/2025arXiv250420765M}
}

@article{Foreman-Mackey_2013,
doi = {10.1086/670067},
url = {https://dx.doi.org/10.1086/670067},
year = {2013},
month = {feb},
publisher = {University of Chicago Press},
volume = {125},
number = {925},
pages = {306},
author = {Foreman-Mackey, Daniel and Hogg, David W. and Lang, Dustin and Goodman, Jonathan},
title = {emcee: The MCMC Hammer},
journal = {Publications of the Astronomical Society of the Pacific}
}

@ARTICLE{Gozman2024,
       author = {{Gozman}, Katya and {Bell}, Eric F. and {Jang}, In Sung and {Arias}, Jose Marco and {Bailin}, Jeremy and {de Jong}, Roelof S. and {D'Souza}, Richard and {Gnedin}, Oleg Y. and {Monachesi}, Antonela and {Price}, Paul A. and {Rao}, Vaishnav V. and {Smercina}, Adam},
        title = "{Exploring the Diversity of Faint Satellites in the M81 Group}",
      journal = {\apj},
         year = 2024,
        month = dec,
       volume = {977},
       number = {2},
          eid = {179},
        pages = {179},
          doi = {10.3847/1538-4357/ad8c3a},
archivePrefix = {arXiv},
       eprint = {2412.02697},
 primaryClass = {astro-ph.GA},
       adsurl = {https://ui.adsabs.harvard.edu/abs/2024ApJ...977..179G}
}

@software{jax2018github,
  author = {James Bradbury and Roy Frostig and Peter Hawkins and Matthew James Johnson and Chris Leary and Dougal Maclaurin and George Necula and Adam Paszke and Jake Vander{P}las and Skye Wanderman-{M}ilne and Qiao Zhang},
  title = {{JAX}: composable transformations of {P}ython+{N}um{P}y programs},
  url = {http://github.com/jax-ml/jax},
  version = {0.3.13},
  year = {2018},
}

@ARTICLE{Astropy13,
       author = {{Astropy Collaboration} and {Robitaille}, Thomas P. and {Tollerud}, Erik J. and {Greenfield}, Perry and {Droettboom}, Michael and {Bray}, Erik and {Aldcroft}, Tom and {Davis}, Matt and {Ginsburg}, Adam and {Price-Whelan}, Adrian M. and {Kerzendorf}, Wolfgang E. and {Conley}, Alexander and {Crighton}, Neil and {Barbary}, Kyle and {Muna}, Demitri and {Ferguson}, Henry and {Grollier}, Fr{\'e}d{\'e}ric and {Parikh}, Madhura M. and {Nair}, Prasanth H. and {Unther}, Hans M. and {Deil}, Christoph and {Woillez}, Julien and {Conseil}, Simon and {Kramer}, Roban and {Turner}, James E.~H. and {Singer}, Leo and {Fox}, Ryan and {Weaver}, Benjamin A. and {Zabalza}, Victor and {Edwards}, Zachary I. and {Azalee Bostroem}, K. and {Burke}, D.~J. and {Casey}, Andrew R. and {Crawford}, Steven M. and {Dencheva}, Nadia and {Ely}, Justin and {Jenness}, Tim and {Labrie}, Kathleen and {Lim}, Pey Lian and {Pierfederici}, Francesco and {Pontzen}, Andrew and {Ptak}, Andy and {Refsdal}, Brian and {Servillat}, Mathieu and {Streicher}, Ole},
        title = "{Astropy: A community Python package for astronomy}",
      journal = {\aap},
         year = 2013,
        month = oct,
       volume = {558},
          eid = {A33},
        pages = {A33},
          doi = {10.1051/0004-6361/201322068},
archivePrefix = {arXiv},
       eprint = {1307.6212},
 primaryClass = {astro-ph.IM},
       adsurl = {https://ui.adsabs.harvard.edu/abs/2013A&A...558A..33A}
}

@ARTICLE{Engler2023,
       author = {{Engler}, Christoph and {Pillepich}, Annalisa and {Joshi}, Gandhali D. and {Pasquali}, Anna and {Nelson}, Dylan and {Grebel}, Eva K.},
        title = "{Satellites of Milky Way- and M31-like galaxies with TNG50: quenched fractions, gas content, and star formation histories}",
      journal = {\mnras},
         year = 2023,
        month = jul,
       volume = {522},
       number = {4},
        pages = {5946-5972},
          doi = {10.1093/mnras/stad1357},
archivePrefix = {arXiv},
       eprint = {2211.00010},
 primaryClass = {astro-ph.GA},
       adsurl = {https://ui.adsabs.harvard.edu/abs/2023MNRAS.522.5946E}
}

@ARTICLE{McConnachie2012,
       author = {{McConnachie}, Alan W.},
        title = "{The Observed Properties of Dwarf Galaxies in and around the Local Group}",
      journal = {\aj},
         year = 2012,
        month = jul,
       volume = {144},
       number = {1},
          eid = {4},
        pages = {4},
          doi = {10.1088/0004-6256/144/1/4},
archivePrefix = {arXiv},
       eprint = {1204.1562},
 primaryClass = {astro-ph.CO},
       adsurl = {https://ui.adsabs.harvard.edu/abs/2012AJ....144....4M}
}

@ARTICLE{Conroy2019,
       author = {{Conroy}, Charlie and {Naidu}, Rohan P. and {Zaritsky}, Dennis and {Bonaca}, Ana and {Cargile}, Phillip and {Johnson}, Benjamin D. and {Caldwell}, Nelson},
        title = "{Resolving the Metallicity Distribution of the Stellar Halo with the H3 Survey}",
      journal = {\apj},
         year = 2019,
        month = dec,
       volume = {887},
       number = {2},
          eid = {237},
        pages = {237},
          doi = {10.3847/1538-4357/ab5710},
archivePrefix = {arXiv},
       eprint = {1909.02007},
 primaryClass = {astro-ph.GA},
       adsurl = {https://ui.adsabs.harvard.edu/abs/2019ApJ...887..237C}
}

@ARTICLE{Mackereth2020,
       author = {{Mackereth}, J. Ted and {Bovy}, Jo},
        title = "{Weighing the stellar constituents of the galactic halo with APOGEE red giant stars}",
      journal = {\mnras},
         year = 2020,
        month = mar,
       volume = {492},
       number = {3},
        pages = {3631-3646},
          doi = {10.1093/mnras/staa047},
archivePrefix = {arXiv},
       eprint = {1910.03590},
 primaryClass = {astro-ph.GA},
       adsurl = {https://ui.adsabs.harvard.edu/abs/2020MNRAS.492.3631M}
}

@ARTICLE{Pillepich2024,
       author = {{Pillepich}, Annalisa and {Sotillo-Ramos}, Diego and {Ramesh}, Rahul and {Nelson}, Dylan and {Engler}, Christoph and {Rodriguez-Gomez}, Vicente and {Fournier}, Martin and {Donnari}, Martina and {Springel}, Volker and {Hernquist}, Lars},
        title = "{Milky Way and Andromeda analogues from the TNG50 simulation}",
      journal = {\mnras},
         year = 2024,
        month = dec,
       volume = {535},
       number = {2},
        pages = {1721-1762},
          doi = {10.1093/mnras/stae2165},
archivePrefix = {arXiv},
       eprint = {2303.16217},
 primaryClass = {astro-ph.GA},
       adsurl = {https://ui.adsabs.harvard.edu/abs/2024MNRAS.535.1721P}
}

@ARTICLE{Munoz-Mateos2015,
       author = {{Mu{\~n}oz-Mateos}, Juan Carlos and {Sheth}, Kartik and {Regan}, Michael and {Kim}, Taehyun and {Laine}, Jarkko and {Erroz-Ferrer}, Santiago and {Gil de Paz}, Armando and {Comeron}, Sebastien and {Hinz}, Joannah and {Laurikainen}, Eija and {Salo}, Heikki and {Athanassoula}, E. and {Bosma}, Albert and {Bouquin}, Alexandre Y.~K. and {Schinnerer}, Eva and {Ho}, Luis and {Zaritsky}, Dennis and {Gadotti}, Dimitri A. and {Madore}, Barry and {Holwerda}, Benne and {Men{\'e}ndez-Delmestre}, Kar{\'\i}n and {Knapen}, Johan H. and {Meidt}, Sharon and {Querejeta}, Miguel and {Mizusawa}, Trisha and {Seibert}, Mark and {Laine}, Seppo and {Courtois}, Helene},
        title = "{The Spitzer Survey of Stellar Structure in Galaxies (S$^{4}$G): Stellar Masses, Sizes, and Radial Profiles for 2352 Nearby Galaxies}",
      journal = {\apjs},
         year = 2015,
        month = jul,
       volume = {219},
       number = {1},
          eid = {3},
        pages = {3},
          doi = {10.1088/0067-0049/219/1/3},
archivePrefix = {arXiv},
       eprint = {1505.03534},
 primaryClass = {astro-ph.GA},
       adsurl = {https://ui.adsabs.harvard.edu/abs/2015ApJS..219....3M}
}

@ARTICLE{Cluver2014,
       author = {{Cluver}, M.~E. and {Jarrett}, T.~H. and {Hopkins}, A.~M. and {Driver}, S.~P. and {Liske}, J. and {Gunawardhana}, M.~L.~P. and {Taylor}, E.~N. and {Robotham}, A.~S.~G. and {Alpaslan}, M. and {Baldry}, I. and {Brown}, M.~J.~I. and {Peacock}, J.~A. and {Popescu}, C.~C. and {Tuffs}, R.~J. and {Bauer}, A.~E. and {Bland-Hawthorn}, J. and {Colless}, M. and {Holwerda}, B.~W. and {Lara-L{\'o}pez}, M.~A. and {Leschinski}, K. and {L{\'o}pez-S{\'a}nchez}, A.~R. and {Norberg}, P. and {Owers}, M.~S. and {Wang}, L. and {Wilkins}, S.~M.},
        title = "{Galaxy and Mass Assembly (GAMA): Mid-infrared Properties and Empirical Relations from WISE}",
      journal = {\apj},
         year = 2014,
        month = feb,
       volume = {782},
       number = {2},
          eid = {90},
        pages = {90},
          doi = {10.1088/0004-637X/782/2/90},
archivePrefix = {arXiv},
       eprint = {1401.0837},
 primaryClass = {astro-ph.GA},
       adsurl = {https://ui.adsabs.harvard.edu/abs/2014ApJ...782...90C}
}

@ARTICLE{Querejeta2015,
       author = {{Querejeta}, Miguel and {Meidt}, Sharon E. and {Schinnerer}, Eva and {Cisternas}, Mauricio and {Mu{\~n}oz-Mateos}, Juan Carlos and {Sheth}, Kartik and {Knapen}, Johan and {van de Ven}, Glenn and {Norris}, Mark A. and {Peletier}, Reynier and {Laurikainen}, Eija and {Salo}, Heikki and {Holwerda}, Benne W. and {Athanassoula}, E. and {Bosma}, Albert and {Groves}, Brent and {Ho}, Luis C. and {Gadotti}, Dimitri A. and {Zaritsky}, Dennis and {Regan}, Michael and {Hinz}, Joannah and {Gil de Paz}, Armando and {Menendez-Delmestre}, Karin and {Seibert}, Mark and {Mizusawa}, Trisha and {Kim}, Taehyun and {Erroz-Ferrer}, Santiago and {Laine}, Jarkko and {Comer{\'o}n}, S{\'e}bastien},
        title = "{The Spitzer Survey of Stellar Structure in Galaxies (S$^{4}$G): Precise Stellar Mass Distributions from Automated Dust Correction at 3.6 {\ensuremath{\mu}}m}",
      journal = {\apjs},
         year = 2015,
        month = jul,
       volume = {219},
       number = {1},
          eid = {5},
        pages = {5},
          doi = {10.1088/0067-0049/219/1/5},
archivePrefix = {arXiv},
       eprint = {1410.0009},
 primaryClass = {astro-ph.GA},
       adsurl = {https://ui.adsabs.harvard.edu/abs/2015ApJS..219....5Q}
}

@ARTICLE{Meidt2012,
       author = {{Meidt}, Sharon E. and {Schinnerer}, Eva and {Knapen}, Johan H. and {Bosma}, Albert and {Athanassoula}, E. and {Sheth}, Kartik and {Buta}, Ronald J. and {Zaritsky}, Dennis and {Laurikainen}, Eija and {Elmegreen}, Debra and {Elmegreen}, Bruce G. and {Gadotti}, Dimitri A. and {Salo}, Heikki and {Regan}, Michael and {Ho}, Luis C. and {Madore}, Barry F. and {Hinz}, Joannah L. and {Skibba}, Ramin A. and {Gil de Paz}, Armando and {Mu{\~n}oz-Mateos}, Juan-Carlos and {Men{\'e}ndez-Delmestre}, Kar{\'\i}n and {Seibert}, Mark and {Kim}, Taehyun and {Mizusawa}, Trisha and {Laine}, Jarkko and {Comer{\'o}n}, S{\'e}bastien},
        title = "{Reconstructing the Stellar Mass Distributions of Galaxies Using S$^{4}$G IRAC 3.6 and 4.5 {\ensuremath{\mu}}m Images. I. Correcting for Contamination by Polycyclic Aromatic Hydrocarbons, Hot Dust, and Intermediate-age Stars}",
      journal = {\apj},
         year = 2012,
        month = jan,
       volume = {744},
       number = {1},
          eid = {17},
        pages = {17},
          doi = {10.1088/0004-637X/744/1/17},
archivePrefix = {arXiv},
       eprint = {1110.2683},
 primaryClass = {astro-ph.CO},
       adsurl = {https://ui.adsabs.harvard.edu/abs/2012ApJ...744...17M}
}

@ARTICLE{Eibensteiner2023,
       author = {{Eibensteiner}, C. and {Bigiel}, F. and {Leroy}, A.~K. and {Koch}, E.~W. and {Rosolowsky}, E. and {Schinnerer}, E. and {Sardone}, A. and {Meidt}, S. and {de Blok}, W.~J.~G. and {Thilker}, D. and {Pisano}, D.~J. and {Ott}, J. and {Barnes}, A. and {Querejeta}, M. and {Emsellem}, E. and {Puschnig}, J. and {Utomo}, D. and {Be{\v{s}}li{\'c}}, I. and {den Brok}, J. and {Faridani}, S. and {Glover}, S.~C.~O. and {Grasha}, K. and {Hassani}, H. and {Henshaw}, J.~D. and {Jim{\'e}nez-Donaire}, M.~J. and {Kerp}, J. and {Dale}, D.~A. and {Kruijssen}, J.~M.~D. and {Laudage}, S. and {Sanchez-Blazquez}, P. and {Smith}, R. and {Stuber}, S. and {Pessa}, I. and {Watkins}, E.~J. and {Williams}, T.~G. and {Winkel}, B.},
        title = "{Kinematic analysis of the super-extended H I disk of the nearby spiral galaxy M 83}",
      journal = {\aap},
         year = 2023,
        month = jul,
       volume = {675},
          eid = {A37},
        pages = {A37},
          doi = {10.1051/0004-6361/202245290},
archivePrefix = {arXiv},
       eprint = {2304.02037},
 primaryClass = {astro-ph.GA},
       adsurl = {https://ui.adsabs.harvard.edu/abs/2023A&A...675A..37E}
}

@ARTICLE{Gozman2023,
       author = {{Gozman}, Katya and {Bell}, Eric F. and {Smercina}, Adam and {Price}, Paul and {Bailin}, Jeremy and {de Jong}, Roelof S. and {D'Souza}, Richard and {Jang}, In Sung and {Monachesi}, Antonela and {Slater}, Colin},
        title = "{Saying Hallo to M94's Stellar Halo: Investigating the Accretion History of the Largest Pseudobulge Host in the Local Universe}",
      journal = {\apj},
         year = 2023,
        month = apr,
       volume = {947},
       number = {1},
          eid = {21},
        pages = {21},
          doi = {10.3847/1538-4357/acbe3a},
archivePrefix = {arXiv},
       eprint = {2304.08436},
 primaryClass = {astro-ph.GA},
       adsurl = {https://ui.adsabs.harvard.edu/abs/2023ApJ...947...21G}
}

@ARTICLE{Escala2021,
       author = {{Escala}, Ivanna and {Gilbert}, Karoline M. and {Wojno}, Jennifer and {Kirby}, Evan N. and {Guhathakurta}, Puragra},
        title = "{Elemental Abundances in M31: Gradients in the Giant Stellar Stream}",
      journal = {\aj},
         year = 2021,
        month = aug,
       volume = {162},
       number = {2},
          eid = {45},
        pages = {45},
          doi = {10.3847/1538-3881/abfec4},
archivePrefix = {arXiv},
       eprint = {2105.02339},
 primaryClass = {astro-ph.GA},
       adsurl = {https://ui.adsabs.harvard.edu/abs/2021AJ....162...45E}
}

@ARTICLE{Bellazzini2006,
       author = {{Bellazzini}, M. and {Newberg}, H.~J. and {Correnti}, M. and {Ferraro}, F.~R. and {Monaco}, L.},
        title = "{Detection of a population gradient in the Sagittarius stream}",
      journal = {\aap},
         year = 2006,
        month = oct,
       volume = {457},
       number = {2},
        pages = {L21-L24},
          doi = {10.1051/0004-6361:20066002},
archivePrefix = {arXiv},
       eprint = {astro-ph/0608513},
 primaryClass = {astro-ph},
       adsurl = {https://ui.adsabs.harvard.edu/abs/2006A&A...457L..21B}
}

@ARTICLE{Weisz2011,
       author = {{Weisz}, Daniel R. and {Dalcanton}, Julianne J. and {Williams}, Benjamin F. and {Gilbert}, Karoline M. and {Skillman}, Evan D. and {Seth}, Anil C. and {Dolphin}, Andrew E. and {McQuinn}, Kristen B.~W. and {Gogarten}, Stephanie M. and {Holtzman}, Jon and {Rosema}, Keith and {Cole}, Andrew and {Karachentsev}, Igor D. and {Zaritsky}, Dennis},
        title = "{The ACS Nearby Galaxy Survey Treasury. VIII. The Global Star Formation Histories of 60 Dwarf Galaxies in the Local Volume}",
      journal = {\apj},
         year = 2011,
        month = sep,
       volume = {739},
       number = {1},
          eid = {5},
        pages = {5},
          doi = {10.1088/0004-637X/739/1/5},
archivePrefix = {arXiv},
       eprint = {1101.1093},
 primaryClass = {astro-ph.CO},
       adsurl = {https://ui.adsabs.harvard.edu/abs/2011ApJ...739....5W}
}

@ARTICLE{Harmsen2023,
       author = {{Harmsen}, Benjamin and {Bell}, Eric F. and {D'Souza}, Richard and {Monachesi}, Antonela and {de Jong}, Roelof S. and {Smercina}, Adam and {Jang}, In Sung and {Holwerda}, Benne W.},
        title = "{Constraining the assembly time of the stellar haloes of nearby Milky Way-mass galaxies through AGB populations}",
      journal = {\mnras},
         year = 2023,
        month = nov,
       volume = {525},
       number = {3},
        pages = {4497-4514},
          doi = {10.1093/mnras/stad2480},
archivePrefix = {arXiv},
       eprint = {2308.11499},
 primaryClass = {astro-ph.GA},
       adsurl = {https://ui.adsabs.harvard.edu/abs/2023MNRAS.525.4497H}
}

@ARTICLE{Hernitschek_2017_Sgr,
       author = {{Hernitschek}, Nina and {Sesar}, Branimir and {Rix}, Hans-Walter and {Belokurov}, Vasily and {Martinez-Delgado}, David and {Martin}, Nicolas F. and {Kaiser}, Nick and {Hodapp}, Klaus and {Chambers}, Kenneth C. and {Wainscoat}, Richard and {Magnier}, Eugene and {Kudritzki}, Rolf-Peter and {Metcalfe}, Nigel and {Draper}, Peter W.},
        title = "{The Geometry of the Sagittarius Stream from Pan-STARRS1 3{\ensuremath{\pi}} RR Lyrae}",
      journal = {\apj},
         year = 2017,
        month = nov,
       volume = {850},
       number = {1},
          eid = {96},
        pages = {96},
          doi = {10.3847/1538-4357/aa960c},
archivePrefix = {arXiv},
       eprint = {1710.09436},
 primaryClass = {astro-ph.GA},
       adsurl = {https://ui.adsabs.harvard.edu/abs/2017ApJ...850...96H}
}

@ARTICLE{vD14,
       author = {{van Dokkum}, Pieter G. and {Abraham}, Roberto and {Merritt}, Allison},
        title = "{First Results from the Dragonfly Telephoto Array: The Apparent Lack of a Stellar Halo in the Massive Spiral Galaxy M101}",
      journal = {\apjl},
         year = 2014,
        month = feb,
       volume = {782},
       number = {2},
          eid = {L24},
        pages = {L24},
          doi = {10.1088/2041-8205/782/2/L24},
archivePrefix = {arXiv},
       eprint = {1401.5467},
 primaryClass = {astro-ph.GA},
       adsurl = {https://ui.adsabs.harvard.edu/abs/2014ApJ...782L..24V}
}

@ARTICLE{Crnojevic2016,
       author = {{Crnojevi{\'c}}, D. and {Sand}, D.~J. and {Spekkens}, K. and {Caldwell}, N. and {Guhathakurta}, P. and {McLeod}, B. and {Seth}, A. and {Simon}, J.~D. and {Strader}, J. and {Toloba}, E.},
        title = "{The Extended Halo of Centaurus A: Uncovering Satellites, Streams, and Substructures}",
      journal = {\apj},
         year = 2016,
        month = may,
       volume = {823},
       number = {1},
          eid = {19},
        pages = {19},
          doi = {10.3847/0004-637X/823/1/19},
archivePrefix = {arXiv},
       eprint = {1512.05366},
 primaryClass = {astro-ph.GA},
       adsurl = {https://ui.adsabs.harvard.edu/abs/2016ApJ...823...19C}
}

@ARTICLE{Smercina2017,
       author = {{Smercina}, Adam and {Bell}, Eric F. and {Slater}, Colin T. and {Price}, Paul A. and {Bailin}, Jeremy and {Monachesi}, Antonela},
        title = "{D1005+68: A New Faint Dwarf Galaxy in the M81 Group}",
      journal = {\apjl},
         year = 2017,
        month = jul,
       volume = {843},
       number = {1},
          eid = {L6},
        pages = {L6},
          doi = {10.3847/2041-8213/aa78fa},
archivePrefix = {arXiv},
       eprint = {1706.07039},
 primaryClass = {astro-ph.GA},
       adsurl = {https://ui.adsabs.harvard.edu/abs/2017ApJ...843L...6S}
}

@ARTICLE{Bell2022,
       author = {{Bell}, Eric F. and {Smercina}, Adam and {Price}, Paul A. and {D'Souza}, Richard and {Bailin}, Jeremy and {de Jong}, Roelof S. and {Gozman}, Katya and {Jang}, In Sung and {Monachesi}, Antonela and {Gnedin}, Oleg Y. and {Slater}, Colin T.},
        title = "{Ultrafaint Dwarf Galaxy Candidates in the M81 Group: Signatures of Group Accretion}",
      journal = {\apjl},
         year = 2022,
        month = sep,
       volume = {937},
       number = {1},
          eid = {L3},
        pages = {L3},
          doi = {10.3847/2041-8213/ac8e5e},
archivePrefix = {arXiv},
       eprint = {2209.06009},
 primaryClass = {astro-ph.GA},
       adsurl = {https://ui.adsabs.harvard.edu/abs/2022ApJ...937L...3B}
}

@ARTICLE{Magnier2013,
       author = {{Magnier}, E.~A. and {Schlafly}, E. and {Finkbeiner}, D. and {Juric}, M. and {Tonry}, J.~L. and {Burgett}, W.~S. and {Chambers}, K.~C. and {Flewelling}, H.~A. and {Kaiser}, N. and {Kudritzki}, R. -P. and {Morgan}, J.~S. and {Price}, P.~A. and {Sweeney}, W.~E. and {Stubbs}, C.~W.},
        title = "{The Pan-STARRS 1 Photometric Reference Ladder, Release 12.01}",
      journal = {\apjs},
         year = 2013,
        month = apr,
       volume = {205},
       number = {2},
          eid = {20},
        pages = {20},
          doi = {10.1088/0067-0049/205/2/20},
archivePrefix = {arXiv},
       eprint = {1303.3634},
 primaryClass = {astro-ph.IM},
       adsurl = {https://ui.adsabs.harvard.edu/abs/2013ApJS..205...20M}
}

@ARTICLE{Bosch2018,
       author = {{Bosch}, James and {Armstrong}, Robert and {Bickerton}, Steven and {Furusawa}, Hisanori and {Ikeda}, Hiroyuki and {Koike}, Michitaro and {Lupton}, Robert and {Mineo}, Sogo and {Price}, Paul and {Takata}, Tadafumi and {Tanaka}, Masayuki and {Yasuda}, Naoki and {AlSayyad}, Yusra and {Becker}, Andrew C. and {Coulton}, William and {Coupon}, Jean and {Garmilla}, Jose and {Huang}, Song and {Krughoff}, K. Simon and {Lang}, Dustin and {Leauthaud}, Alexie and {Lim}, Kian-Tat and {Lust}, Nate B. and {MacArthur}, Lauren A. and {Mandelbaum}, Rachel and {Miyatake}, Hironao and {Miyazaki}, Satoshi and {Murata}, Ryoma and {More}, Surhud and {Okura}, Yuki and {Owen}, Russell and {Swinbank}, John D. and {Strauss}, Michael A. and {Yamada}, Yoshihiko and {Yamanoi}, Hitomi},
        title = "{The Hyper Suprime-Cam software pipeline}",
      journal = {\pasj},
         year = 2018,
        month = jan,
       volume = {70},
          eid = {S5},
        pages = {S5},
          doi = {10.1093/pasj/psx080},
archivePrefix = {arXiv},
       eprint = {1705.06766},
 primaryClass = {astro-ph.IM},
       adsurl = {https://ui.adsabs.harvard.edu/abs/2018PASJ...70S...5B}
}

@ARTICLE{Monachesi2013,
       author = {{Monachesi}, Antonela and {Bell}, Eric F. and {Radburn-Smith}, David J. and {Vlaji{\'c}}, Marija and {de Jong}, Roelof S. and {Bailin}, Jeremy and {Dalcanton}, Julianne J. and {Holwerda}, Benne W. and {Streich}, David},
        title = "{Testing Galaxy Formation Models with the GHOSTS Survey: The Color Profile of M81's Stellar Halo}",
      journal = {\apj},
         year = 2013,
        month = apr,
       volume = {766},
       number = {2},
          eid = {106},
        pages = {106},
          doi = {10.1088/0004-637X/766/2/106},
archivePrefix = {arXiv},
       eprint = {1302.2626},
 primaryClass = {astro-ph.CO},
       adsurl = {https://ui.adsabs.harvard.edu/abs/2013ApJ...766..106M}
}

@ARTICLE{Jang2020,
       author = {{Jang}, In Sung and {de Jong}, Roelof S. and {Holwerda}, Benne W. and
         {Monachesi}, Antonela and {Bell}, Eric F. and {Bailin}, Jeremy},
        title = "{Tracing the anemic stellar halo of M 101}",
      journal = {\aap},
         year = 2020,
        month = may,
       volume = {637},
          eid = {A8},
        pages = {A8},
          doi = {10.1051/0004-6361/201936994},
archivePrefix = {arXiv},
       eprint = {2001.12007},
 primaryClass = {astro-ph.GA},
       adsurl = {https://ui.adsabs.harvard.edu/abs/2020A&A...637A...8J}
}

@ARTICLE{Kalberla2009,
       author = {{Kalberla}, Peter M.~W. and {Kerp}, J{\"u}rgen},
        title = "{The Hi Distribution of the Milky Way}",
      journal = {\araa},
         year = 2009,
        month = sep,
       volume = {47},
       number = {1},
        pages = {27-61},
          doi = {10.1146/annurev-astro-082708-101823},
       adsurl = {https://ui.adsabs.harvard.edu/abs/2009ARA&A..47...27K}
}

@ARTICLE{Diaz2006,
       author = {{D{\'\i}az}, Rub{\'e}n J. and {Dottori}, Horacio and {Aguero}, Maria P. and
         {Mediavilla}, Evencio and {Rodrigues}, Irapuan and {Mast}, Damian},
        title = "{Hidden Trigger for the Giant Starburst Arc in M83?}",
      journal = {\apj},
         year = 2006,
        month = dec,
       volume = {652},
       number = {2},
        pages = {1122-1128},
          doi = {10.1086/507886},
archivePrefix = {arXiv},
       eprint = {astro-ph/0611771},
 primaryClass = {astro-ph},
       adsurl = {https://ui.adsabs.harvard.edu/abs/2006ApJ...652.1122D}
}

@ARTICLE{Knapen2010,
       author = {{Knapen}, J.~H. and {Sharp}, R.~G. and {Ryder}, S.~D. and
         {Falc{\'o}n-Barroso}, J. and {Fathi}, K. and {Guti{\'e}rrez}, L.},
        title = "{The central region of M83: massive star formation, kinematics, and the location and origin of the nucleus}",
      journal = {\mnras},
         year = 2010,
        month = oct,
       volume = {408},
       number = {2},
        pages = {797-811},
          doi = {10.1111/j.1365-2966.2010.17180.x},
archivePrefix = {arXiv},
       eprint = {1006.2698},
 primaryClass = {astro-ph.CO},
       adsurl = {https://ui.adsabs.harvard.edu/abs/2010MNRAS.408..797K}
}

@BOOK{RC3,
       author = {{de Vaucouleurs}, Gerard and {de Vaucouleurs}, Antoinette and
         {Corwin}, Herold G., Jr. and {Buta}, Ronald J. and {Paturel}, Georges and
         {Fouque}, Pascal},
        title = "{Third Reference Catalogue of Bright Galaxies}",
         year = "1991",
       adsurl = {https://ui.adsabs.harvard.edu/abs/1991rc3..book.....D}
}

@ARTICLE{Jang19,
       author = {{Jang}, I.~S. and {de Jong}, R.~S. and {Holwerda}, B.~W. and
         {Monachesi}, A. and {Bell}, E.~F. and {Bailin}, J.},
        title = "{Tracing the Anemic Stellar Halo of M101}",
      journal = {submitted to \aap},
         year = "2019",
        month = "",
        pages = {}
}

@ARTICLE{Bruzzese19,
       author = {{Bruzzese}, S.~M. and {Thilker}, David A. and {Meurer}, G.~R. and
         {Bianchi}, Luciana and {Watts}, A.~B. and {Ferguson}, A.~M.~N. and
         {Gil de Paz}, A. and {Madore}, B. and {Martin}, D. Christopher and
         {Rich}, R. Michael},
        title = "{The initial mass function in the extended ultraviolet disk of M83}",
      journal = {\mnras},
         year = "2019",
        month = "Nov",
        pages = {2843},
          doi = {10.1093/mnras/stz3151},
       adsurl = {https://ui.adsabs.harvard.edu/abs/2019MNRAS.tmp.2843B}
}

@ARTICLE{Sheth2010,
   author = {{Sheth}, K. and {Regan}, M. and {Hinz}, J.~L. and {Gil de Paz}, A. and 
	{Men{\'e}ndez-Delmestre}, K. and {Mu{\~n}oz-Mateos}, J.-C. and 
	{Seibert}, M. and {Kim}, T. and {Laurikainen}, E. and {Salo}, H. and 
	{Gadotti}, D.~A. and {Laine}, J. and {Mizusawa}, T. and {Armus}, L. and 
	{Athanassoula}, E. and {Bosma}, A. and {Buta}, R.~J. and {Capak}, P. and 
	{Jarrett}, T.~H. and {Elmegreen}, D.~M. and {Elmegreen}, B.~G. and 
	{Knapen}, J.~H. and {Koda}, J. and {Helou}, G. and {Ho}, L.~C. and 
	{Madore}, B.~F. and {Masters}, K.~L. and {Mobasher}, B. and 
	{Ogle}, P. and {Peng}, C.~Y. and {Schinnerer}, E. and {Surace}, J.~A. and 
	{Zaritsky}, D. and {Comer{\'o}n}, S. and {de Swardt}, B. and 
	{Meidt}, S.~E. and {Kasliwal}, M. and {Aravena}, M.},
    title = "{The Spitzer Survey of Stellar Structure in Galaxies (S4G)}",
  journal = {\pasp},
archivePrefix = "arXiv",
   eprint = {1010.1592},
     year = 2010,
    month = dec,
   volume = 122,
    pages = {1397},
      doi = {10.1086/657638},
   adsurl = {https://ui.adsabs.harvard.edu/abs/2010PASP..122.1397S}
}

@ARTICLE{Chabrier,
       author = {{Chabrier}, Gilles},
        title = "{Galactic Stellar and Substellar Initial Mass Function}",
      journal = {Publications of the Astronomical Society of the Pacific},
         year = "2003",
        month = "Jul",
       volume = {115},
       number = {809},
        pages = {763-795},
          doi = {10.1086/376392},
archivePrefix = {arXiv},
       eprint = {astro-ph/0304382},
 primaryClass = {astro-ph},
       adsurl = {https://ui.adsabs.harvard.edu/abs/2003PASP..115..763C}
}

@ARTICLE{Streich2014,
       author = {{Streich}, D. and {de Jong}, R.~S. and {Bailin}, J. and
         {Goudfrooij}, P. and {Radburn-Smith}, D. and {Vlajic}, M.},
        title = "{On the relation between metallicity and RGB color in HST/ACS data}",
      journal = {Astronomy and Astrophysics},
         year = "2014",
        month = "Mar",
       volume = {563},
          eid = {A5},
        pages = {A5},
          doi = {10.1051/0004-6361/201220956},
archivePrefix = {arXiv},
       eprint = {1401.2847},
 primaryClass = {astro-ph.GA},
       adsurl = {https://ui.adsabs.harvard.edu/abs/2014A&A...563A...5S}
}

@ARTICLE{KK1998,
       author = {{Karachentseva}, V.~E. and {Karachentsev}, I.~D.},
        title = "{A list of new nearby dwarf galaxy candidates}",
      journal = {Astronomy and Astrophysics Supplement Series},
         year = "1998",
        month = "Feb",
       volume = {127},
        pages = {409-419},
          doi = {10.1051/aas:1998109},
       adsurl = {https://ui.adsabs.harvard.edu/abs/1998A&AS..127..409K}
}

@ARTICLE{astropy,
   author = {{Astropy Collaboration} and {Robitaille}, T.~P. and {Tollerud}, E.~J. and 
	{Greenfield}, P. and {Droettboom}, M. and {Bray}, E. and {Aldcroft}, T. and 
	{Davis}, M. and {Ginsburg}, A. and {Price-Whelan}, A.~M. and 
	{Kerzendorf}, W.~E. and {Conley}, A. and {Crighton}, N. and 
	{Barbary}, K. and {Muna}, D. and {Ferguson}, H. and {Grollier}, F. and 
	{Parikh}, M.~M. and {Nair}, P.~H. and {Unther}, H.~M. and {Deil}, C. and 
	{Woillez}, J. and {Conseil}, S. and {Kramer}, R. and {Turner}, J.~E.~H. and 
	{Singer}, L. and {Fox}, R. and {Weaver}, B.~A. and {Zabalza}, V. and 
	{Edwards}, Z.~I. and {Azalee Bostroem}, K. and {Burke}, D.~J. and 
	{Casey}, A.~R. and {Crawford}, S.~M. and {Dencheva}, N. and 
	{Ely}, J. and {Jenness}, T. and {Labrie}, K. and {Lim}, P.~L. and 
	{Pierfederici}, F. and {Pontzen}, A. and {Ptak}, A. and {Refsdal}, B. and 
	{Servillat}, M. and {Streicher}, O.},
    title = "{Astropy: A community Python package for astronomy}",
  journal = {\aap},
archivePrefix = "arXiv",
   eprint = {1307.6212},
 primaryClass = "astro-ph.IM",
     year = 2013,
    month = oct,
   volume = 558,
      eid = {A33},
    pages = {A33},
      doi = {10.1051/0004-6361/201322068},
   adsurl = {http://adsabs.harvard.edu/abs/2013A%26A...558A..33A}
}

@ARTICLE{Putman_2012,
       author = {{Putman}, M.~E. and {Peek}, J.~E.~G. and {Joung}, M.~R.},
        title = "{Gaseous Galaxy Halos}",
      journal = {\araa},
         year = "2012",
        month = "Sep",
       volume = {50},
        pages = {491-529},
          doi = {10.1146/annurev-astro-081811-125612},
archivePrefix = {arXiv},
       eprint = {1207.4837},
 primaryClass = {astro-ph.GA},
       adsurl = {https://ui.adsabs.harvard.edu/abs/2012ARA&A..50..491P}
}

@ARTICLE{Helmi_2018,
       author = {{Helmi}, Amina and {Babusiaux}, Carine and {Koppelman}, Helmer H. and
         {Massari}, Davide and {Veljanoski}, Jovan and {Brown}, Anthony G.~A.},
        title = "{The merger that led to the formation of the Milky Way's inner stellar halo and thick disk}",
      journal = {\nat},
         year = "2018",
        month = "Nov",
       volume = {563},
       number = {7729},
        pages = {85-88},
          doi = {10.1038/s41586-018-0625-x},
archivePrefix = {arXiv},
       eprint = {1806.06038},
 primaryClass = {astro-ph.GA},
       adsurl = {https://ui.adsabs.harvard.edu/abs/2018Natur.563...85H}
}

@ARTICLE{Belukorov_2018,
       author = {{Belokurov}, V. and {Erkal}, D. and {Evans}, N.~W. and {Koposov}, S.~E. and
         {Deason}, A.~J.},
        title = "{Co-formation of the disc and the stellar halo}",
      journal = {\mnras},
         year = "2018",
        month = "Jul",
       volume = {478},
       number = {1},
        pages = {611-619},
          doi = {10.1093/mnras/sty982},
archivePrefix = {arXiv},
       eprint = {1802.03414},
 primaryClass = {astro-ph.GA},
       adsurl = {https://ui.adsabs.harvard.edu/abs/2018MNRAS.478..611B}
}

@ARTICLE{Monachesi_2019,
       author = {{Monachesi}, Antonela and {G{\'o}mez}, Facundo A. and {Grand
        }, Robert J.~J. and {Simpson}, Christine M. and {Kauffmann}, Guinevere and
         {Bustamante}, Sebasti{\'a}n and {Marinacci}, Federico and
         {Pakmor}, R{\"u}diger and {Springel}, Volker and {Frenk}, Carlos S. and
         {White}, Simon D.~M. and {Tissera}, Patricia B.},
        title = "{The Auriga stellar haloes: connecting stellar population properties with accretion and merging history}",
      journal = {\mnras},
         year = "2019",
        month = "May",
       volume = {485},
       number = {2},
        pages = {2589-2616},
          doi = {10.1093/mnras/stz538},
archivePrefix = {arXiv},
       eprint = {1804.07798},
 primaryClass = {astro-ph.GA},
       adsurl = {https://ui.adsabs.harvard.edu/abs/2019MNRAS.485.2589M}
}

@ARTICLE{Monachesi_2016_Auriga,
       author = {{Monachesi}, Antonela and {G{\'o}mez}, Facundo A. and {Grand
        }, Robert J.~J. and {Kauffmann}, Guinevere and {Marinacci}, Federico and
         {Pakmor}, R{\"u}diger and {Springel}, Volker and {Frenk}, Carlos S.},
        title = "{On the stellar halo metallicity profile of Milky Way-like galaxies in the Auriga simulations}",
      journal = {\mnras},
         year = "2016",
        month = "Jun",
       volume = {459},
       number = {1},
        pages = {L46-L50},
          doi = {10.1093/mnrasl/slw052},
archivePrefix = {arXiv},
       eprint = {1512.03064},
 primaryClass = {astro-ph.GA},
       adsurl = {https://ui.adsabs.harvard.edu/abs/2016MNRAS.459L..46M}
}

@ARTICLE{Pillepich2015,
       author = {{Pillepich}, Annalisa and {Madau}, Piero and {Mayer}, Lucio},
        title = "{Building Late-type Spiral Galaxies by In-situ and Ex-situ Star Formation}",
      journal = {\apj},
         year = "2015",
        month = "Feb",
       volume = {799},
       number = {2},
          eid = {184},
        pages = {184},
          doi = {10.1088/0004-637X/799/2/184},
archivePrefix = {arXiv},
       eprint = {1407.7855},
 primaryClass = {astro-ph.GA},
       adsurl = {https://ui.adsabs.harvard.edu/abs/2015ApJ...799..184P}
}

@ARTICLE{DSouza2018_MNRAS,
       author = {{D'Souza}, Richard and {Bell}, Eric F.},
        title = "{The masses and metallicities of stellar haloes reflect galactic merger histories}",
      journal = {\mnras},
         year = "2018",
        month = "Mar",
       volume = {474},
       number = {4},
        pages = {5300-5318},
          doi = {10.1093/mnras/stx3081},
archivePrefix = {arXiv},
       eprint = {1705.08442},
 primaryClass = {astro-ph.GA},
       adsurl = {https://ui.adsabs.harvard.edu/abs/2018MNRAS.474.5300D}
}

@ARTICLE{Anderson10,
   author = {{Anderson}, J. and {Bedin}, L.~R.},
    title = "{An Empirical Pixel-Based Correction for Imperfect CTE. I. HST's Advanced Camera for Surveys}",
  journal = {\pasp},
archivePrefix = "arXiv",
   eprint = {1007.3987},
 primaryClass = "astro-ph.IM",
     year = 2010,
    month = sep,
   volume = 122,
    pages = {1035-1064},
      doi = {10.1086/656399},
   adsurl = {http://adsabs.harvard.edu/abs/2010PASP..122.1035A}
}

@ARTICLE{Robin2007,
       author = {{Robin}, A.~C. and {Rich}, R.~M. and {Aussel}, H. and {Capak}, P. and
         {Tasca}, L.~A.~M. and {Jahnke}, K. and {Kakazu}, Y. and
         {Kneib}, J. -P. and {Koekemoer}, A. and {Leauthaud}, A.~C. and
         {Lilly}, S. and {Mobasher}, B. and {Scoville}, N. and {Taniguchi}, Y. and
         {Thompson}, D.~J.},
        title = "{The Stellar Content of the COSMOS Field as Derived from Morphological and SED-based Star/Galaxy Separation}",
      journal = {\apjs},
         year = "2007",
        month = "Sep",
       volume = {172},
       number = {1},
        pages = {545-559},
          doi = {10.1086/516600},
archivePrefix = {arXiv},
       eprint = {astro-ph/0612349},
 primaryClass = {astro-ph},
       adsurl = {https://ui.adsabs.harvard.edu/abs/2007ApJS..172..545R}
}

@INPROCEEDINGS{Thilker08,
       author = {{Thilker}, D.~A. and {Bianchi}, L. and {Meurer}, G. and
         {Gil de Paz}, A. and {Boissier}, S. and {Madore}, B.~F. and
         {Ferguson}, A.~N.~M. and {Hameed}, S. and {Neff}, S. and
         {Martin}, C.~D. and {Rich}, R.~M. and {Schiminovich}, D. and
         {Seibert}, M. and {Wyder}, T.},
        title = "{Resolved Stellar Populations Constituting Extended UV Disks in Nearby Galaxies}",
    booktitle = {Formation and Evolution of Galaxy Disks},
         year = "2008",
       editor = {{Funes}, J.~G. and {Corsini}, E.~M.},
       series = {Astronomical Society of the Pacific Conference Series},
       volume = {396},
        month = "Oct",
        pages = {223},
archivePrefix = {arXiv},
       eprint = {0712.3724},
 primaryClass = {astro-ph},
       adsurl = {https://ui.adsabs.harvard.edu/abs/2008ASPC..396..223T}
}

@ARTICLE{Muller2015,
       author = {{M{\"u}ller}, Oliver and {Jerjen}, Helmut and {Binggeli}, Bruno},
        title = "{New dwarf galaxy candidates in the Centaurus group}",
      journal = {\aap},
         year = "2015",
        month = "Nov",
       volume = {583},
          eid = {A79},
        pages = {A79},
          doi = {10.1051/0004-6361/201526748},
archivePrefix = {arXiv},
       eprint = {1509.04931},
 primaryClass = {astro-ph.GA},
       adsurl = {https://ui.adsabs.harvard.edu/abs/2015A&A...583A..79M}
}

@ARTICLE{Bertin_arnouts96,
   author = {{Bertin}, E. and {Arnouts}, S.},
    title = "{SExtractor: Software for source extraction.}",
  journal = {\aaps},
     year = 1996,
    month = jun,
   volume = 117,
    pages = {393-404},
   adsurl = {http://adsabs.harvard.edu/abs/1996A%26AS..117..393B}
}

@ARTICLE{Dolphin00,
   author = {{Dolphin}, A.~E.},
    title = "{WFPC2 Stellar Photometry with HSTPHOT}",
  journal = {\pasp},
   eprint = {arXiv:astro-ph/0006217},
     year = 2000,
    month = oct,
   volume = 112,
    pages = {1383-1396},
      doi = {10.1086/316630},
   adsurl = {http://adsabs.harvard.edu/abs/2000PASP..112.1383D}
}

@BOOK{Gonzaga12,
   author = {{Gonzaga}, S.},
    title = "{The DrizzlePac Handbook}",
booktitle = {The DrizzlePac Handbook, HST Data Handbook},
     year = 2012,
    month = jun,
   adsurl = {http://adsabs.harvard.edu/abs/2012drzp.book.....G}
}

@ARTICLE{Merritt2016,
       author = {{Merritt}, Allison and {van Dokkum}, Pieter and {Abraham}, Roberto and
         {Zhang}, Jielai},
        title = "{The Dragonfly nearby Galaxies Survey. I. Substantial Variation in the Diffuse Stellar Halos around Spiral Galaxies}",
      journal = {\apj},
         year = "2016",
        month = "Oct",
       volume = {830},
       number = {2},
          eid = {62},
        pages = {62},
          doi = {10.3847/0004-637X/830/2/62},
archivePrefix = {arXiv},
       eprint = {1606.08847},
 primaryClass = {astro-ph.GA},
       adsurl = {https://ui.adsabs.harvard.edu/abs/2016ApJ...830...62M}
}

@ARTICLE{Deason2016,
       author = {{Deason}, Alis J. and {Mao}, Yao-Yuan and {Wechsler}, Risa H.},
        title = "{The Eating Habits of Milky Way-mass Halos: Destroyed Dwarf Satellites and the Metallicity Distribution of Accreted Stars}",
      journal = {\apj},
         year = "2016",
        month = "Apr",
       volume = {821},
       number = {1},
          eid = {5},
        pages = {5},
          doi = {10.3847/0004-637X/821/1/5},
archivePrefix = {arXiv},
       eprint = {1601.07905},
 primaryClass = {astro-ph.GA},
       adsurl = {https://ui.adsabs.harvard.edu/abs/2016ApJ...821....5D}
}

@ARTICLE{Tully2013,
       author = {{Tully}, R. Brent and {Courtois}, H{\'e}l{\`e}ne M. and
         {Dolphin}, Andrew E. and {Fisher}, J. Richard and
         {H{\'e}raudeau}, Philippe and {Jacobs}, Bradley A. and
         {Karachentsev}, Igor D. and {Makarov}, Dmitry and {Makarova}, Lidia and
         {Mitronova}, Sofia and {Rizzi}, Luca and {Shaya}, Edward J. and
         {Sorce}, Jenny G. and {Wu}, Po-Feng},
        title = "{Cosmicflows-2: The Data}",
      journal = {\aj},
         year = "2013",
        month = "Oct",
       volume = {146},
       number = {4},
          eid = {86},
        pages = {86},
          doi = {10.1088/0004-6256/146/4/86},
archivePrefix = {arXiv},
       eprint = {1307.7213},
 primaryClass = {astro-ph.CO},
       adsurl = {https://ui.adsabs.harvard.edu/abs/2013AJ....146...86T}
}

@ARTICLE{vandenBergh1980,
       author = {{van den Bergh}, S.},
        title = "{The posteruptive galaxy NGC 5253.}",
      journal = {\pasp},
         year = "1980",
        month = "Apr",
       volume = {92},
        pages = {122-124},
          doi = {10.1086/130631},
       adsurl = {https://ui.adsabs.harvard.edu/abs/1980PASP...92..122V}
}

@ARTICLE{Calzetti1999,
       author = {{Calzetti}, Daniela and {Conselice}, Christopher J. and
         {Gallagher}, John S., III and {Kinney}, Anne L.},
        title = "{The Structure and Morphology of the Ionized Gas in Starburst Galaxies: NGC 5253/5236}",
      journal = {\aj},
         year = "1999",
        month = "Aug",
       volume = {118},
       number = {2},
        pages = {797-816},
          doi = {10.1086/300972},
archivePrefix = {arXiv},
       eprint = {astro-ph/9904428},
 primaryClass = {astro-ph},
       adsurl = {https://ui.adsabs.harvard.edu/abs/1999AJ....118..797C}
}

@ARTICLE{Chandar2010,
       author = {{Chandar}, Rupali and {Whitmore}, Bradley C. and {Kim}, Hwihyun and
         {Kaleida}, Catherine and {Mutchler}, Max and {Calzetti}, Daniela and
         {Saha}, Abhijit and {O'Connell}, Robert and {Balick}, Bruce and
         {Bond}, Howard and {Carollo}, Marcella and {Disney}, Michael and
         {Dopita}, Michael A. and {Frogel}, Jay A. and {Hall}, Donald and
         {Holtzman}, Jon A. and {Kimble}, Randy A. and {McCarthy}, Patrick and
         {Paresce}, Francesco and {Silk}, Joe and {Trauger}, John and
         {Walker}, Alistair R. and {Windhorst}, Rogier A. and {Young}, Erick},
        title = "{The Luminosity, Mass, and Age Distributions of Compact Star Clusters in M83 Based on Hubble Space Telescope/Wide Field Camera 3 Observations}",
      journal = {\apj},
         year = "2010",
        month = "Aug",
       volume = {719},
       number = {1},
        pages = {966-978},
          doi = {10.1088/0004-637X/719/1/966},
archivePrefix = {arXiv},
       eprint = {1007.5237},
 primaryClass = {astro-ph.GA},
       adsurl = {https://ui.adsabs.harvard.edu/abs/2010ApJ...719..966C}
}

@ARTICLE{Thilker2005,
       author = {{Thilker}, David A. and {Bianchi}, Luciana and {Boissier}, Samuel and
         {Gil de Paz}, Armando and {Madore}, Barry F. and
         {Martin}, D. Christopher and {Meurer}, Gerhardt R. and
         {Neff}, Susan G. and {Rich}, R. Michael and {Schiminovich}, David and
         {Seibert}, Mark and {Wyder}, Ted K. and {Barlow}, Tom A. and
         {Byun}, Yong-Ik and {Donas}, Jose and {Forster}, Karl and
         {Friedman}, Peter G. and {Heckman}, Timothy M. and
         {Jelinsky}, Patrick N. and {Lee}, Young-Wook and {Malina}, Roger F. and
         {Milliard}, Bruno and {Morrissey}, Patrick and
         {Siegmund}, Oswald H.~W. and {Small}, Todd and {Szalay}, Alex S. and
         {Welsh}, Barry Y.},
        title = "{Recent Star Formation in the Extreme Outer Disk of M83}",
      journal = {\apj},
         year = "2005",
        month = "Jan",
       volume = {619},
       number = {1},
        pages = {L79-L82},
          doi = {10.1086/425251},
archivePrefix = {arXiv},
       eprint = {astro-ph/0411306},
 primaryClass = {astro-ph},
       adsurl = {https://ui.adsabs.harvard.edu/abs/2005ApJ...619L..79T}
}

@ARTICLE{NibPear2025,
       author = {{Nibauer}, Jacob and {Pearson}, Sarah},
        title = "{Testing Dark Matter with Generative Models for Extragalactic Stellar Streams}",
      journal = {arXiv e-prints},
         year = 2025,
        month = aug,
          eid = {arXiv:2508.02666},
        pages = {arXiv:2508.02666},
archivePrefix = {arXiv},
       eprint = {2508.02666},
 primaryClass = {astro-ph.GA},
       adsurl = {https://ui.adsabs.harvard.edu/abs/2025arXiv250802666N}
}

@ARTICLE{Thatte2000,
       author = {{Thatte}, N. and {Tecza}, M. and {Genzel}, R.},
        title = "{Stellar dynamics observations of a double nucleus in M 83}",
      journal = {\aap},
         year = "2000",
        month = "Dec",
       volume = {364},
        pages = {L47-L53},
archivePrefix = {arXiv},
       eprint = {astro-ph/0009392},
 primaryClass = {astro-ph},
       adsurl = {https://ui.adsabs.harvard.edu/abs/2000A&A...364L..47T}
}

@ARTICLE{Harris2001,
       author = {{Harris}, Jason and {Calzetti}, Daniela and {Gallagher}, John S., III and
         {Conselice}, Christopher J. and {Smith}, Denise A.},
        title = "{Young Clusters in the Nuclear Starburst of M83}",
      journal = {\aj},
         year = "2001",
        month = "Dec",
       volume = {122},
       number = {6},
        pages = {3046-3064},
          doi = {10.1086/324230},
archivePrefix = {arXiv},
       eprint = {astro-ph/0109076},
 primaryClass = {astro-ph},
       adsurl = {https://ui.adsabs.harvard.edu/abs/2001AJ....122.3046H}
}

@ARTICLE{Monachesi2016_GHOSTS,
       author = {{Monachesi}, Antonela and {Bell}, Eric F. and {Radburn-Smith}, David J. and
         {Bailin}, Jeremy and {de Jong}, Roelof S. and {Holwerda}, Benne and
         {Streich}, David and {Silverstein}, Grace},
        title = "{The GHOSTS survey - II. The diversity of halo colour and metallicity profiles of massive disc galaxies}",
      journal = {\mnras},
         year = "2016",
        month = "Apr",
       volume = {457},
       number = {2},
        pages = {1419-1446},
          doi = {10.1093/mnras/stv2987},
archivePrefix = {arXiv},
       eprint = {1507.06657},
 primaryClass = {astro-ph.GA},
       adsurl = {https://ui.adsabs.harvard.edu/abs/2016MNRAS.457.1419M}
}

@ARTICLE{DSouza2018_M31,
       author = {{D'Souza}, Richard and {Bell}, Eric F.},
        title = "{The Andromeda galaxy's most important merger about 2 billion years ago as M32's likely progenitor}",
      journal = {Nature Astronomy},
         year = "2018",
        month = "Jul",
       volume = {2},
        pages = {737-743},
          doi = {10.1038/s41550-018-0533-x},
archivePrefix = {arXiv},
       eprint = {1807.08819},
 primaryClass = {astro-ph.GA},
       adsurl = {https://ui.adsabs.harvard.edu/abs/2018NatAs...2..737D}
}

@ARTICLE{Bell2017,
       author = {{Bell}, Eric F. and {Monachesi}, Antonela and {Harmsen}, Benjamin and
         {de Jong}, Roelof S. and {Bailin}, Jeremy and
         {Radburn-Smith}, David J. and {D'Souza}, Richard and
         {Holwerda}, Benne W.},
        title = "{Galaxies Grow Their Bulges and Black Holes in Diverse Ways}",
      journal = {\apj},
         year = "2017",
        month = "Mar",
       volume = {837},
       number = {1},
          eid = {L8},
        pages = {L8},
          doi = {10.3847/2041-8213/aa6158},
archivePrefix = {arXiv},
       eprint = {1702.06116},
 primaryClass = {astro-ph.GA},
       adsurl = {https://ui.adsabs.harvard.edu/abs/2017ApJ...837L...8B}
}

@ARTICLE{Somerville2015,
       author = {{Somerville}, Rachel S. and {Dav{\'e}}, Romeel},
        title = "{Physical Models of Galaxy Formation in a Cosmological Framework}",
      journal = {\araa},
         year = "2015",
        month = "Aug",
       volume = {53},
        pages = {51-113},
          doi = {10.1146/annurev-astro-082812-140951},
archivePrefix = {arXiv},
       eprint = {1412.2712},
 primaryClass = {astro-ph.GA},
       adsurl = {https://ui.adsabs.harvard.edu/abs/2015ARA&A..53...51S}
}

@ARTICLE{Bullock2001,
       author = {{Bullock}, James S. and {Kravtsov}, Andrey V. and {Weinberg}, David H.},
        title = "{Hierarchical Galaxy Formation and Substructure in the Galaxy's Stellar Halo}",
      journal = {\apj},
         year = "2001",
        month = "Feb",
       volume = {548},
       number = {1},
        pages = {33-46},
          doi = {10.1086/318681},
archivePrefix = {arXiv},
       eprint = {astro-ph/0007295},
 primaryClass = {astro-ph},
       adsurl = {https://ui.adsabs.harvard.edu/abs/2001ApJ...548...33B}
}

@ARTICLE{BullockJohnston2005,
       author = {{Bullock}, James S. and {Johnston}, Kathryn V.},
        title = "{Tracing Galaxy Formation with Stellar Halos. I. Methods}",
      journal = {\apj},
         year = "2005",
        month = "Dec",
       volume = {635},
       number = {2},
        pages = {931-949},
          doi = {10.1086/497422},
archivePrefix = {arXiv},
       eprint = {astro-ph/0506467},
 primaryClass = {astro-ph},
       adsurl = {https://ui.adsabs.harvard.edu/abs/2005ApJ...635..931B}
}

@ARTICLE{Cooper2010,
       author = {{Cooper}, A.~P. and {Cole}, S. and {Frenk}, C.~S. and {White}, S.~D.~M. and
         {Helly}, J. and {Benson}, A.~J. and {De Lucia}, G. and {Helmi}, A. and
         {Jenkins}, A. and {Navarro}, J.~F. and {Springel}, V. and {Wang}, J.},
        title = "{Galactic stellar haloes in the CDM model}",
      journal = {\mnras},
         year = "2010",
        month = "Aug",
       volume = {406},
       number = {2},
        pages = {744-766},
          doi = {10.1111/j.1365-2966.2010.16740.x},
archivePrefix = {arXiv},
       eprint = {0910.3211},
 primaryClass = {astro-ph.GA},
       adsurl = {https://ui.adsabs.harvard.edu/abs/2010MNRAS.406..744C}
}

@ARTICLE{walder2024,
       author = {{Walder}, Madison and {Erkal}, Denis and {Collins}, Michelle and {Martinez-Delgado}, David},
        title = "{Probing the dark matter haloes of external galaxies with stellar streams}",
      journal = {arXiv e-prints},
         year = 2024,
        month = feb,
          eid = {arXiv:2402.13314},
        pages = {arXiv:2402.13314},
          doi = {10.48550/arXiv.2402.13314},
archivePrefix = {arXiv},
       eprint = {2402.13314},
 primaryClass = {astro-ph.GA},
       adsurl = {https://ui.adsabs.harvard.edu/abs/2024arXiv240213314W}
}

@ARTICLE{Fardal2019,
       author = {{Fardal}, Mark A. and {van der Marel}, Roeland P. and {Law}, David R. and {Sohn}, Sangmo Tony and {Sesar}, Branimir and {Hernitschek}, Nina and {Rix}, Hans-Walter},
        title = "{Connecting the Milky Way potential profile to the orbital time-scales and spatial structure of the Sagittarius Stream}",
      journal = {\mnras},
         year = 2019,
        month = mar,
       volume = {483},
       number = {4},
        pages = {4724-4741},
          doi = {10.1093/mnras/sty3428},
archivePrefix = {arXiv},
       eprint = {1804.04995},
 primaryClass = {astro-ph.GA},
       adsurl = {https://ui.adsabs.harvard.edu/abs/2019MNRAS.483.4724F}
}

@ARTICLE{hendel2015,
       author = {{Hendel}, David and {Johnston}, Kathryn V.},
        title = "{Tidal debris morphology and the orbits of satellite galaxies}",
      journal = {\mnras},
         year = 2015,
        month = dec,
       volume = {454},
       number = {3},
        pages = {2472-2485},
          doi = {10.1093/mnras/stv2035},
archivePrefix = {arXiv},
       eprint = {1509.06369},
 primaryClass = {astro-ph.GA},
       adsurl = {https://ui.adsabs.harvard.edu/abs/2015MNRAS.454.2472H}
}

@ARTICLE{belokurov2014,
       author = {{Belokurov}, V. and {Koposov}, S.~E. and {Evans}, N.~W. and {Pe{\~n}arrubia}, J. and {Irwin}, M.~J. and {Smith}, M.~C. and {Lewis}, G.~F. and {Gieles}, M. and {Wilkinson}, M.~I. and {Gilmore}, G. and {Olszewski}, E.~W. and {Niederste-Ostholt}, M.},
        title = "{Precession of the Sagittarius stream}",
      journal = {\mnras},
         year = 2014,
        month = jan,
       volume = {437},
       number = {1},
        pages = {116-131},
          doi = {10.1093/mnras/stt1862},
archivePrefix = {arXiv},
       eprint = {1301.7069},
 primaryClass = {astro-ph.GA},
       adsurl = {https://ui.adsabs.harvard.edu/abs/2014MNRAS.437..116B}
}

@ARTICLE{Miyamoto1975,
       author = {{Miyamoto}, M. and {Nagai}, R.},
        title = "{Three-dimensional models for the distribution of mass in galaxies.}",
      journal = {\pasj},
         year = 1975,
        month = jan,
       volume = {27},
        pages = {533-543},
       adsurl = {https://ui.adsabs.harvard.edu/abs/1975PASJ...27..533M}
}

@ARTICLE{buist2015,
       author = {{Buist}, Hans J.~T. and {Helmi}, Amina},
        title = "{The evolution of streams in a time-dependent potential}",
      journal = {\aap},
         year = 2015,
        month = dec,
       volume = {584},
          eid = {A120},
        pages = {A120},
          doi = {10.1051/0004-6361/201526203},
archivePrefix = {arXiv},
       eprint = {1504.00008},
 primaryClass = {astro-ph.GA},
       adsurl = {https://ui.adsabs.harvard.edu/abs/2015A&A...584A.120B}
}

@ARTICLE{Law2009,
       author = {{Law}, David R. and {Majewski}, Steven R. and {Johnston}, Kathryn V.},
        title = "{Evidence for a Triaxial Milky Way Dark Matter Halo from the Sagittarius Stellar Tidal Stream}",
      journal = {\apjl},
         year = 2009,
        month = sep,
       volume = {703},
       number = {1},
        pages = {L67-L71},
          doi = {10.1088/0004-637X/703/1/L67},
archivePrefix = {arXiv},
       eprint = {0908.3187},
 primaryClass = {astro-ph.GA},
       adsurl = {https://ui.adsabs.harvard.edu/abs/2009ApJ...703L..67L}
}

@ARTICLE{Harmsen2017,
       author = {{Harmsen}, Benjamin and {Monachesi}, Antonela and {Bell}, Eric F. and
         {de Jong}, Roelof S. and {Bailin}, Jeremy and
         {Radburn-Smith}, David J. and {Holwerda}, Benne W.},
        title = "{Diverse stellar haloes in nearby Milky Way mass disc galaxies}",
      journal = {\mnras},
         year = "2017",
        month = "Apr",
       volume = {466},
       number = {2},
        pages = {1491-1512},
          doi = {10.1093/mnras/stw2992},
archivePrefix = {arXiv},
       eprint = {1611.05448},
 primaryClass = {astro-ph.GA},
       adsurl = {https://ui.adsabs.harvard.edu/abs/2017MNRAS.466.1491H}
}

@ARTICLE{Radburn-Smith2011,
       author = {{Radburn-Smith}, D.~J. and {de Jong}, R.~S. and {Seth}, A.~C. and
         {Bailin}, J. and {Bell}, E.~F. and {Brown}, T.~M. and {Bullock}, J.~S. and
         {Courteau}, S. and {Dalcanton}, J.~J. and {Ferguson}, H.~C. and
         {Goudfrooij}, P. and {Holfeltz}, S. and {Holwerda}, B.~W. and
         {Purcell}, C. and {Sick}, J. and {Streich}, D. and {Vlajic}, M. and
         {Zucker}, D.~B.},
        title = "{The GHOSTS Survey. I. Hubble Space Telescope Advanced Camera for Surveys Data}",
      journal = {\apjs},
         year = "2011",
        month = "Aug",
       volume = {195},
       number = {2},
          eid = {18},
        pages = {18},
          doi = {10.1088/0067-0049/195/2/18},
       adsurl = {https://ui.adsabs.harvard.edu/abs/2011ApJS..195...18R}
}

@ARTICLE{Ibata2014,
       author = {{Ibata}, Rodrigo A. and {Lewis}, Geraint F. and {McConnachie}, Alan W. and
         {Martin}, Nicolas F. and {Irwin}, Michael J. and
         {Ferguson}, Annette M.~N. and {Babul}, Arif and {Bernard}, Edouard J. and
         {Chapman}, Scott C. and {Collins}, Michelle and {Fardal}, Mark and
         {Mackey}, A.~D. and {Navarro}, Julio and {Pe{\~n}arrubia}, Jorge and
         {Rich}, R. Michael and {Tanvir}, Nial and {Widrow}, Lawrence},
        title = "{The Large-scale Structure of the Halo of the Andromeda Galaxy. I. Global Stellar Density, Morphology and Metallicity Properties}",
      journal = {\apj},
         year = "2014",
        month = "Jan",
       volume = {780},
       number = {2},
          eid = {128},
        pages = {128},
          doi = {10.1088/0004-637X/780/2/128},
archivePrefix = {arXiv},
       eprint = {1311.5888},
 primaryClass = {astro-ph.GA},
       adsurl = {https://ui.adsabs.harvard.edu/abs/2014ApJ...780..128I}
}

@ARTICLE{Brooks2025,
       author = {{Brooks}, Richard A.~N. and {Garavito-Camargo}, Nicol{\'a}s and {Johnston}, Kathryn V. and {Price-Whelan}, Adrian M. and {Sanders}, Jason L. and {Lilleengen}, Sophia},
        title = "{LMC Calls, Milky Way Halo Answers: Disentangling the Effects of the MW{\textendash}LMC Interaction on Stellar Stream Populations}",
      journal = {\apj},
         year = 2025,
        month = jan,
       volume = {978},
       number = {1},
          eid = {79},
        pages = {79},
          doi = {10.3847/1538-4357/ad93a7},
archivePrefix = {arXiv},
       eprint = {2410.02574},
 primaryClass = {astro-ph.GA},
       adsurl = {https://ui.adsabs.harvard.edu/abs/2025ApJ...978...79B}
}

@ARTICLE{Lilleengen2023,
       author = {{Lilleengen}, Sophia and {Petersen}, Michael S. and {Erkal}, Denis and {Pe{\~n}arrubia}, Jorge and {Koposov}, Sergey E. and {Li}, Ting S. and {Cullinane}, Lara R. and {Ji}, Alexander P. and {Kuehn}, Kyler and {Lewis}, Geraint F. and {Mackey}, Dougal and {Pace}, Andrew B. and {Shipp}, Nora and {Zucker}, Daniel B. and {Bland-Hawthorn}, Joss and {Hilmi}, Tariq and {S5 Collaboration}},
        title = "{The effect of the deforming dark matter haloes of the Milky Way and the Large Magellanic Cloud on the Orphan-Chenab stream}",
      journal = {\mnras},
         year = 2023,
        month = jan,
       volume = {518},
       number = {1},
        pages = {774-790},
          doi = {10.1093/mnras/stac3108},
archivePrefix = {arXiv},
       eprint = {2205.01688},
 primaryClass = {astro-ph.GA},
       adsurl = {https://ui.adsabs.harvard.edu/abs/2023MNRAS.518..774L}
}

@article{Zolotov2009,
	doi = {10.1088/0004-637x/702/2/1058},
	url = {https://doi.org/10.1088%2F0004-637x%2F702%2F2%2F1058},
	year = 2009,
	month = {aug},
	publisher = {{IOP} Publishing},
	volume = {702},
	number = {2},
	pages = {1058--1067},
	author = {Adi Zolotov and Beth Willman and Alyson M. Brooks and Fabio Governato and Chris B. Brook and David W. Hogg and Tom Quinn and Greg Stinson},
	title = {{THE} {DUAL} {ORIGIN} {OF} {STELLAR} {HALOS}},
	journal = {The Astrophysical Journal}
}

@ARTICLE{Bell2008,
       author = {{Bell}, Eric F. and {Zucker}, Daniel B. and {Belokurov}, Vasily and
         {Sharma}, Sanjib and {Johnston}, Kathryn V. and {Bullock}, James S. and
         {Hogg}, David W. and {Jahnke}, Knud and {de Jong}, Jelte T.~A. and
         {Beers}, Timothy C. and {Evans}, N.~W. and {Grebel}, Eva K. and
         {Ivezi{\'c}}, {\v{Z}}eljko and {Koposov}, Sergey E. and
         {Rix}, Hans-Walter and {Schneider}, Donald P. and
         {Steinmetz}, Matthias and {Zolotov}, Adi},
        title = "{The Accretion Origin of the Milky Way's Stellar Halo}",
      journal = {\apj},
         year = "2008",
        month = "Jun",
       volume = {680},
       number = {1},
        pages = {295-311},
          doi = {10.1086/588032},
archivePrefix = {arXiv},
       eprint = {0706.0004},
 primaryClass = {astro-ph},
       adsurl = {https://ui.adsabs.harvard.edu/abs/2008ApJ...680..295B}
}

@ARTICLE{HuchtmeierBohnenstengel1981,
       author = {{Huchtmeier}, W.~K. and {Bohnenstengel}, H. -D.},
        title = "{The extended H I-envelope of NGC 5236 (M 83).}",
      journal = {\aap},
         year = "1981",
        month = "Jul",
       volume = {100},
        pages = {72-78},
       adsurl = {https://ui.adsabs.harvard.edu/abs/1981A&A...100...72H}
}

@ARTICLE{Heald2016,
       author = {{Heald}, G. and {de Blok}, W.~J.~G. and {Lucero}, D. and {Carignan}, C. and
         {Jarrett}, T. and {Elson}, E. and {Oozeer}, N. and {Randriamampand
        ry}, T.~H. and {van Zee}, L.},
        title = "{Neutral hydrogen and magnetic fields in M83 observed with the SKA Pathfinder KAT-7}",
      journal = {\mnras},
         year = "2016",
        month = "Oct",
       volume = {462},
       number = {2},
        pages = {1238-1255},
          doi = {10.1093/mnras/stw1698},
archivePrefix = {arXiv},
       eprint = {1607.03365},
 primaryClass = {astro-ph.GA},
       adsurl = {https://ui.adsabs.harvard.edu/abs/2016MNRAS.462.1238H}
}

@ARTICLE{MalinHadley1997,
       author = {{Malin}, David and {Hadley}, Brian},
        title = "{HI in Shell Galaxies and Other Merger Remnants}",
      journal = {\pasa},
         year = "1997",
        month = "Apr",
       volume = {14},
       number = {1},
        pages = {52-58},
          doi = {10.1071/AS97052},
       adsurl = {https://ui.adsabs.harvard.edu/abs/1997PASA...14...52M}
}

@ARTICLE{Carrillo2017,
       author = {{Carrillo}, Andreia and {Bell}, Eric F. and {Bailin}, Jeremy and
         {Monachesi}, Antonela and {de Jong}, Roelof S. and
         {Harmsen}, B2enjamin and {Slater}, Colin T.},
        title = "{Characterizing dw1335-29, a recently discovered dwarf satellite of M83}",
      journal = {\mnras},
         year = "2017",
        month = "Mar",
       volume = {465},
       number = {4},
        pages = {5026-5032},
          doi = {10.1093/mnras/stw3025},
archivePrefix = {arXiv},
       eprint = {1611.10361},
 primaryClass = {astro-ph.GA},
       adsurl = {https://ui.adsabs.harvard.edu/abs/2017MNRAS.465.5026C}
}

@ARTICLE{Barnes2014,
   author = {{Barnes}, K.~L. and {van Zee}, L. and {Dale}, D.~A. and {Staudaher}, S. and 
	{Bullock}, J.~S. and {Calzetti}, D. and {Chandar}, R. and {Dalcanton}, J.~J.
	},
    title = "{New Insights on the Formation and Assembly of M83 from Deep Near-infrared Imaging}",
  journal = {\apj},
     year = 2014,
    month = jul,
   volume = 789,
      eid = {126},
    pages = {126},
      doi = {10.1088/0004-637X/789/2/126},
   adsurl = {http://adsabs.harvard.edu/abs/2014ApJ...789..126B}
}

@ARTICLE{navarro1997,
       author = {{Navarro}, Julio F. and {Frenk}, Carlos S. and {White}, Simon D.~M.},
        title = "{A Universal Density Profile from Hierarchical Clustering}",
      journal = {\apj},
         year = 1997,
        month = dec,
       volume = {490},
       number = {2},
        pages = {493-508},
          doi = {10.1086/304888},
archivePrefix = {arXiv},
       eprint = {astro-ph/9611107},
 primaryClass = {astro-ph},
       adsurl = {https://ui.adsabs.harvard.edu/abs/1997ApJ...490..493N}
}

@ARTICLE{Johnston2001,
       author = {{Johnston}, Kathryn V. and {Sackett}, Penny D. and {Bullock}, James S.},
        title = "{Interpreting Debris from Satellite Disruption in External Galaxies}",
      journal = {\apj},
         year = "2001",
        month = "Aug",
       volume = {557},
       number = {1},
        pages = {137-149},
          doi = {10.1086/321644},
archivePrefix = {arXiv},
       eprint = {astro-ph/0101543},
 primaryClass = {astro-ph},
       adsurl = {https://ui.adsabs.harvard.edu/abs/2001ApJ...557..137J}
}

@ARTICLE{Bresolin2009,
       author = {{Bresolin}, Fabio and {Ryan-Weber}, Emma and {Kennicutt}, Robert C. and
         {Goddard}, Quinton},
        title = "{The Flat Oxygen Abundance Gradient in the Extended Disk of M83}",
      journal = {\apj},
         year = "2009",
        month = "Apr",
       volume = {695},
       number = {1},
        pages = {580-595},
          doi = {10.1088/0004-637X/695/1/580},
archivePrefix = {arXiv},
       eprint = {0901.1127},
 primaryClass = {astro-ph.GA},
       adsurl = {https://ui.adsabs.harvard.edu/abs/2009ApJ...695..580B}
}

@ARTICLE{KormendyKennicutt2004,
       author = {{Kormendy}, John and {Kennicutt}, Robert C., Jr.},
        title = "{Secular Evolution and the Formation of Pseudobulges in Disk Galaxies}",
      journal = {\araa},
         year = "2004",
        month = "Sep",
       volume = {42},
       number = {1},
        pages = {603-683},
          doi = {10.1146/annurev.astro.42.053102.134024},
archivePrefix = {arXiv},
       eprint = {astro-ph/0407343},
 primaryClass = {astro-ph},
       adsurl = {https://ui.adsabs.harvard.edu/abs/2004ARA&A..42..603K}
}
\bibliographystyle{aasjournalv7}



\end{document}